**Title: Evaluation of machine-learning and deep-learning SPEI drought forecasts against persistence across timescales and lead times in Bangladesh**

**Authors: Mahir Ahnaf[1*], Showmitra Kumar Sarkar[2]**

[1]Graduate Student, Department of Urban and Regional Planning, Khulna University of Engineering & Technology (KUET), Khulna-9203, Bangladesh.

[ahnf.mahir@gmail.com] (https://orcid.org/0009-0007-0992-9560)

[2]Assistant Professor, Department of Urban and Regional Planning, Khulna University of Engineering & Technology (KUET), Khulna-9203, Bangladesh.

[mail4dhrubo@gmail.com] (https://orcid.org/0000-0001-5928-7577)

[*]Corresponding author

**Corresponding author details**

Email: ahnf.mahir@gmail.com

ORCID: https://orcid.org/0009-0007-0992-9560

# Evaluation of machine-learning and deep-learning SPEI drought forecasts against persistence across timescales and lead times in Bangladesh

**Abstract:**

Bangladesh is home to nearly 178 million people, many of whom depend on agriculture, making the country highly vulnerable to recurring droughts that threaten food security and rural livelihoods. Machine-learning and deep-learning models are increasingly used to forecast drought indices such as the Standardized Precipitation Evapotranspiration Index (SPEI). Most studies, however, report accuracy without a persistence reference, although accumulated indices are strongly autocorrelated, and seldom test drought detection, longer leads or variation between training runs. This study forecast SPEI at 3-, 6- and 12-month timescales one to three months ahead for the eight divisions of Bangladesh, using ERA5 reanalysis checked against 35 Bangladesh Meteorological Department stations. Five models, XGBoost, CNN, Bi-LSTM, CNN-BiLSTM and CEEMDAN-BiLSTM, were trained with five random seeds each, scored on accuracy, skill against persistence and drought detection, and interpreted with Shapley values. At a one-month lead, the CNN-BiLSTM and Bi-LSTM beat persistence clearly at SPEI-3 and SPEI-6 (skill 0.408 and 0.350), while at SPEI-12 a Nash-Sutcliffe efficiency of 0.872 barely exceeded persistence (0.850). Skill and detection could disagree: at SPEI-3 and a three-month lead, XGBoost had the highest skill in the study (0.439) yet detected none of the 119 drought months. Only at SPEI-6 did the forecasts beat persistence significantly and match its detection at all three leads, with fewer false alarms. At SPEI-12, gains were small or absent and varied between seeds. The forecasts reproduced the spatial pattern of the worst drought month (spatial correlation 0.854-0.932), kept 76-80% of their correlation against station-derived SPEI, and relied mainly on precipitation and lagged SPEI. Forecasts should therefore be judged on skill against persistence and detection across several seeds, not on accuracy alone. In Bangladesh, SPEI-6 forecasts can support early warning up to three months ahead and SPEI-3 forecasts one month ahead, while SPEI-12 is better suited to monitoring.

**Keywords:** Drought forecasting, Standardized Precipitation Evapotranspiration Index, Persistence baseline, Forecast verification, Deep learning, Lead time.

# 1. Introduction

Drought builds over weeks and is often recognized only after the damage has reached crops and water supplies (Wilhite & Glantz, 1985). It develops as a chain, from a deficit of precipitation to deficits of soil moisture, streamflow and storage, and finally to losses for the households that depend on them (Raposo et al., 2023). Meteorological drought is the first link in that chain, and therefore the stage at which a warning can still be issued while there is time to act. In Bangladesh, regional droughts affect about 2.5 million ha of agricultural land in the wet season and 1.2 million ha in the dry season in a typical year (Osmani et al., 2022), much of the rice crop depends on rainfall (Rahman et al., 2022), and drought vulnerability is high in several regions, from the north-west to the eastern hills (Sarkar et al., 2024). A forecast that gives reliable notice of a coming drought, even a month or two ahead, would therefore be of direct use for crop and water planning.

The Standardized Precipitation Evapotranspiration Index (SPEI) is well suited to this task. It measures the climatic water balance rather than precipitation alone, so it responds to the rising evaporative demand that a rainfall-only index misses (Vicente-Serrano et al., 2010). This matters because meteorological droughts defined by precipitation show no clear global trend over the past 120 years, while the severity of agricultural and ecological drought has increased with evaporative demand (Vicente-Serrano et al., 2022). Since 2015, SPEI has become the leading index among the three most widely used meteorological drought indices, with much of that work in Asia (Kim et al., 2023). Like the Standardized Precipitation Index (McKee et al., 1993), SPEI is computed over accumulation windows of several months, and these windows give the index a strong memory: consecutive values of a k-month index share $k - 1$ months, so simply repeating the last observed value is already a good forecast at long windows. With warming expected to reduce the dynamic predictability of agricultural drought (Wu et al., 2025), it is important to know how much of the skill of a drought forecast is genuinely predicted and how much is inherited from this memory. Drought forecasting has moved from statistical time-series models to machine learning and deep learning. In Bangladesh, seasonal autoregressive (SARIMA), PROPHET and ARMA-GARCH models (Hossain, Rahman, et al., 2025; Mahmud et al., 2026) and classical machine-learning models such as random forests (Elbeltagi et al., 2022; Yaseen et al., 2021), model trees (Al Mamun et al., 2024) and extreme learning machines (Akter et al., 2025; Hossain, Begum, Akhtar, Talukder, et al., 2025) have been used to estimate or forecast drought indices. Reviews of the wider field identify SPI and SPEI as the most used targets and note a shift towards hybrid and deep-learning models (Ahmed Osman et al., 2025; Oyarzabal et al., 2025). Recurrent models based on long short-term memory (LSTM) are the usual choice for drought indices (Dikshit, Pradhan, & Huete, 2021; Luz et al., 2026), and hybrids with a convolutional neural network (CNN) (Danandeh Mehr et al., 2023; Wang et al., 2026), with statistical or decomposition components (Ding et al., 2022; Xu et al., 2022) and with attention mechanisms (Y. Chen et al., 2025) have been reported to improve accuracy further, and stacked LSTM models have been applied to leads of several months (Dikshit, Pradhan, & Alamri, 2021).

Several gaps limit what this body of work can tell a forecaster. Most studies report accuracy through the root mean square error (RMSE), the coefficient of determination ($R^2$) or the Nash-Sutcliffe efficiency (NSE) and compare models only with one another; a systematic review found that 20% to 25% of deep-learning drought studies report no explicit baseline (Luz et al., 2026). Without a persistence reference, high accuracy at long windows cannot be separated from the memory of the index, and accuracy does indeed rise when the input lags span the accumulation window (Poudel et al., 2024). Accuracy metrics also say nothing about whether a forecast warns when a drought arrives, which is what an early-warning system needs, and even seasonal climate models detect fewer than 30% of drought onsets (Yuan & Wood, 2013). Long forecast horizons are explored far less than short ones (Luz et al., 2026). Most results come from a single training run, although random initialization alone can change the outcome of a model comparison (Bouthillier et al., 2021), and decomposition applied to the full series before the data are split lets future information enter the training inputs (Yang et al., 2024). Reviews for South Asia accordingly list missing benchmarks and missing uncertainty analysis among the main gaps (Al Reshan et al., 2025). The same review identifies the lack of explainable and interpretable models as a further gap, because deep-learning models are often treated as black boxes that give little insight into the factors driving a drought forecast, and few studies have addressed this (Al Reshan et al., 2025; Dikshit & Pradhan, 2021). In Bangladesh, earlier studies cover single stations or regions, mostly one index and a single lead, in some cases with random splits of the record, and deep learning has been applied to drought classification, daily SPI and water-scarcity mapping (Hossain, Begum, Akhtar, Hossain, et al., 2025; Mridul et al., 2025; Sifat et al., 2025). To our knowledge, bidirectional and convolutional or decomposition-based hybrid deep-learning models have not been applied to SPEI forecasting in the country, and no study has reported how forecast quality changes from one to three months ahead.

This study addresses these gaps for the eight divisions of Bangladesh. SPEI is computed from ERA5 reanalysis at 3-, 6- and 12-month windows, after testing the reanalysis against 35 stations of the Bangladesh Meteorological Department. Extreme gradient boosting (XGBoost), a CNN, a bidirectional LSTM (Bi-LSTM) and two hybrids, CNN-BiLSTM and CEEMDAN-BiLSTM (complete ensemble empirical mode decomposition with adaptive noise), are trained through one leakage-controlled pipeline with five random seeds each. Every forecast is evaluated with block-bootstrap intervals and Diebold-Mariano tests, and the selected models are re-scored against an independent station-derived target. This study aims to develop machine-learning and deep-learning models that forecast drought at the 3-, 6- and 12-month timescales one to three months ahead, and to interpret the selected models with SHAP to identify the factors that most influence the forecasts. It further aims to assess whether accuracy, skill against persistence and event detection agree across these timescales and leads, to identify the timescale that keeps genuine forecast value beyond one month and test whether that horizon can be anticipated before any model is trained, and to quantify the robustness of the skill estimates to the random initialization of the models and to sampling. In doing so, the study fills several of these gaps. To our knowledge, it provides the first SPEI forecasts for all eight divisions of Bangladesh at three timescales and one to three months ahead, the first application of bidirectional and hybrid deep-learning models to SPEI forecasting in the country, and the first evaluation of such forecasts against persistence and as drought-event detectors. It demonstrates that the three views of forecast quality can disagree, to the point that a skill score can reward a forecast that issues no drought warning.

## 2. Materials and methods

### 2.1 Study area and data

Bangladesh covers about 147,570 km² of the lower Ganges-Brahmaputra-Meghna delta (Fig. 1). Most of the country is a flat plain below 10 m above sea level, with hills only in the east. Its eight administrative divisions are the spatial units of this study. The climate is tropical monsoon, with most of the annual rainfall between June and September (Bushra et al., 2026) and annual totals that rise from about 1,400 mm in the west to more than 4,300 mm in the east (Rakib, 2019). Temperature peaks in May and is lowest in January (Shahid, 2010), so the climatic water balance moves from a large monsoon surplus to a clear deficit in the dry and pre-monsoon months, when largely rain-fed agriculture is most exposed (Rahman et al., 2022).

The main dataset is the ERA5 monthly reanalysis on single levels (Hersbach et al., 2020, 2023) at 0.25° resolution for January 1980 to December 2024 (540 months). ERA5 was chosen because the stations of Bangladesh are sparse and their records have gaps, and because it has already served as the observational reference for the country in a benchmark of CMIP6 models (Kamruzzaman et al., 2022). Nine variables were retrieved (Table S1; Text S1): total precipitation, mean, maximum, minimum and dew-point temperature at 2 m, surface solar radiation, the two 10 m wind components and total cloud cover. Fig. 2 maps the 1980-2024 trend of each variable, with the two wind components combined into wind speed. The fields were masked to the national boundary (274 grid cells), and the cells of each division were averaged into one monthly series per variable. An independent compiled record of 35 Bangladesh Meteorological Department (BMD) stations (Diptho et al., 2026), which overlaps ERA5 from 1980 to 2022, was used to test the reanalysis. The compilation stores missing data as zero rainfall, so station-years with a zero annual total and zero monthly totals in May-September were set to missing (Text S2). The stations of each division were averaged into one divisional series, which gives 4,128 division-month pairs (8 divisions × 516 months).

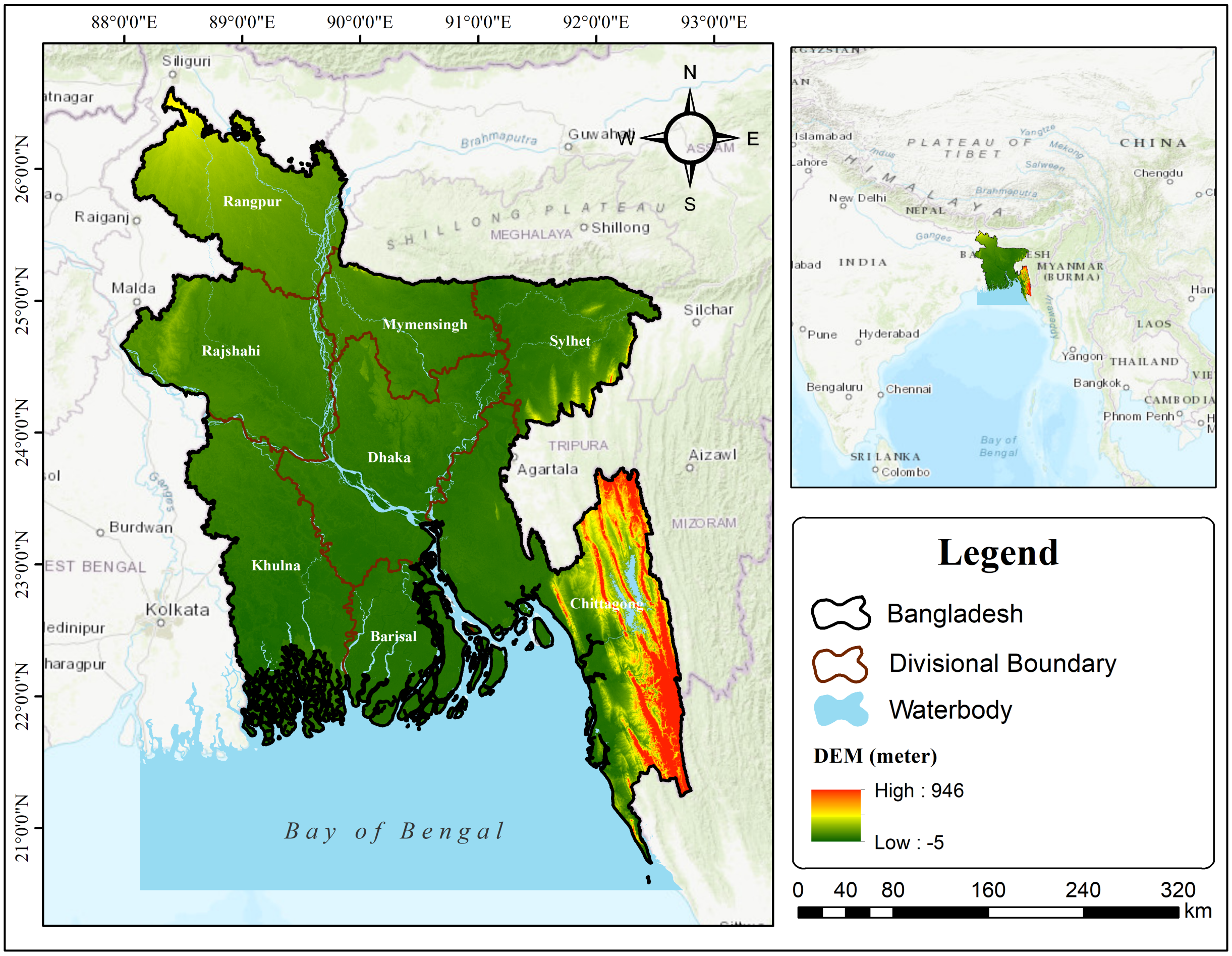

**Fig. 1.** Study area: the eight administrative divisions of Bangladesh, with elevation (−5 to 946 m) from the HydroSHEDS digital elevation model (Lehner et al., 2008) and the regional setting in the inset. Base map: Esri.

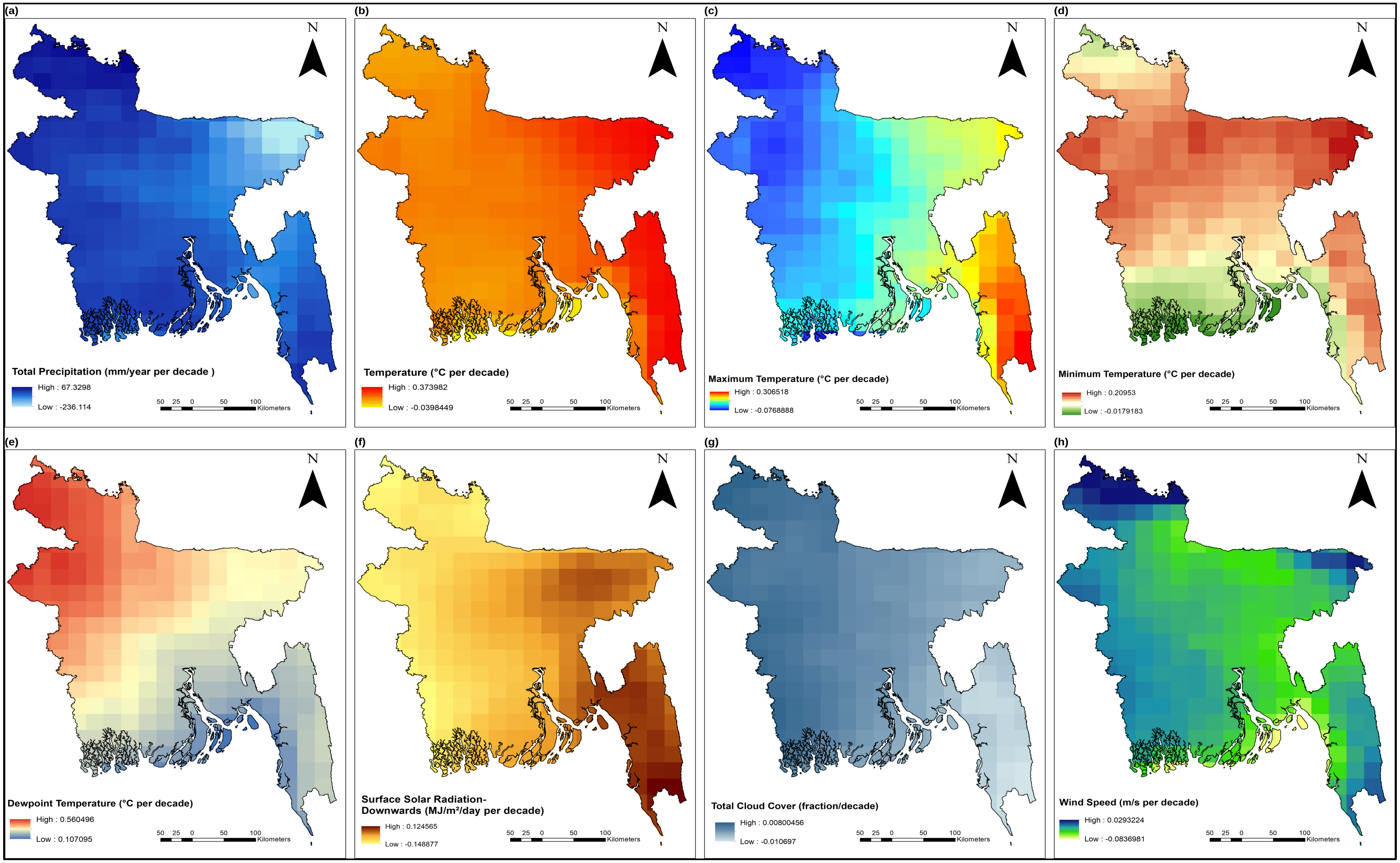

**Fig. 2.** Theil-Sen trends (Sen, 1968) of the annual series of the eight ERA5 climatic indicators over 1980-2024, computed at each 0.25° grid cell: (a) total precipitation (mm yr$^{-1}$ per decade), (b) mean 2 m temperature, (c) maximum temperature, (d) minimum temperature and (e) dew-point temperature (°C per decade), (f) surface solar radiation (MJ m$^{-2}$ day$^{-1}$ per decade), (g) total cloud cover (fraction per decade) and (h) 10 m wind speed (m s$^{-1}$ per decade). Each panel has its own color scale, with the highest and lowest cell values in the legend. Mean temperature was removed from the predictors after the VIF test (Table S7) and is shown for completeness.

## 2.2 Research framework

The workflow is summarized in Fig. 3 and described step by step in Sections 2.3 to 2.10.

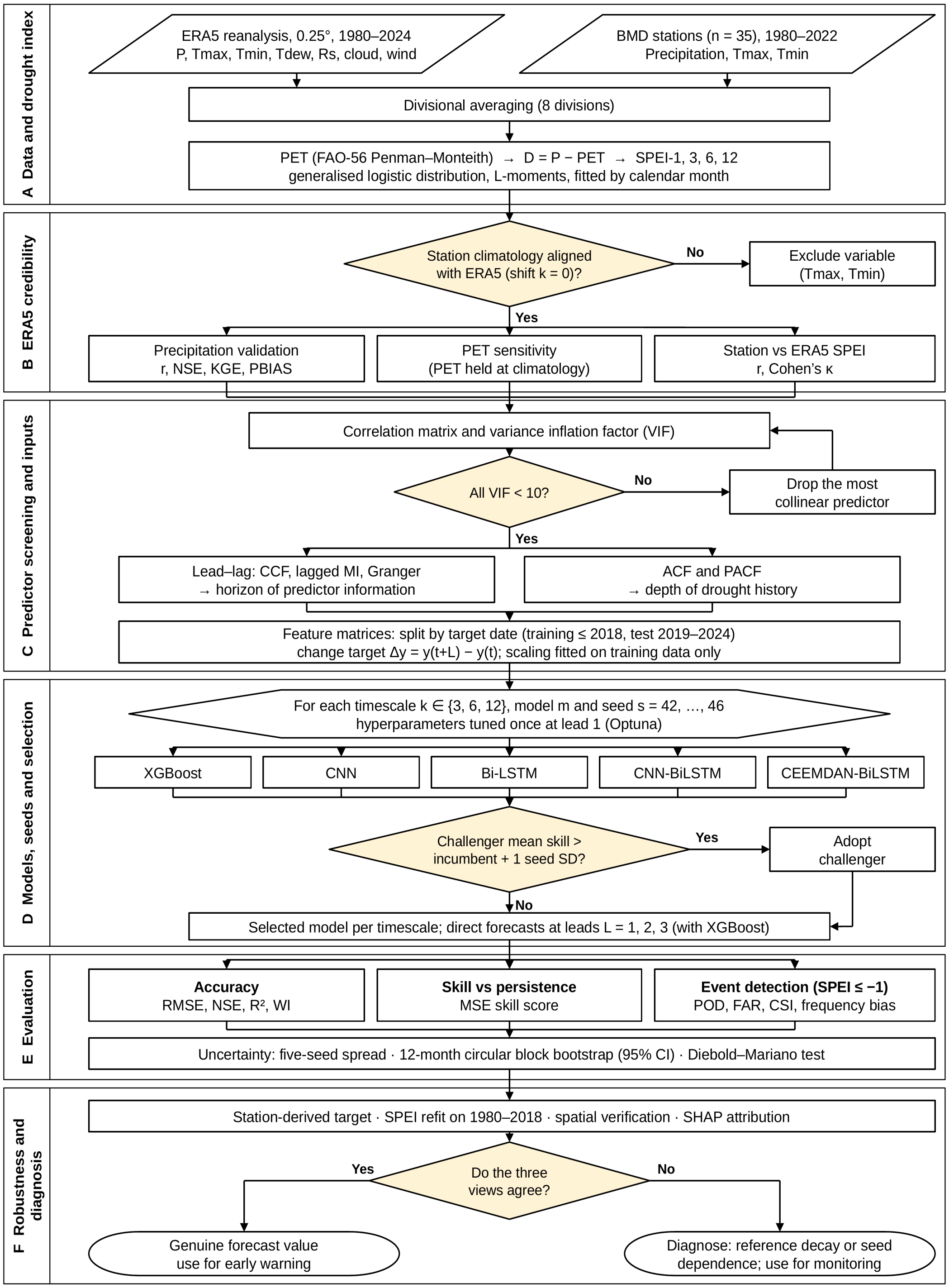

**Fig. 3.** Methodological framework: (A) data and drought index; (B) credibility of ERA5; (C) predictor screening and inputs; (D) model training with five seeds and model selection; (E) evaluation; (F) robustness checks and outcome.

Parallelograms denote data, rectangles processes, diamonds decisions, the hexagon the training loop and rounded shapes outcomes. P, precipitation; PET, potential evapotranspiration; D, climatic water balance; Tdew, dew-point temperature; Rs, surface solar radiation.

## 2.3 Drought index

Potential evapotranspiration (PET) was computed with the FAO-56 Penman-Monteith method (Allen et al., 1998), whose radiative and aerodynamic inputs ERA5 supplies in full. Temperature-only methods such as Thornthwaite (Thornthwaite, 1948) were not used in the main analysis, because they can misstate evaporative demand under warming (Li et al., 2022). Reference evapotranspiration (mm day$^{-1}$) is given by Eq. (1),

$$ET_0 = \frac{0.408\,\Delta\,(R_n - G) + \gamma\,\frac{900}{T + 273}\,u_2\,(e_s - e_a)}{\Delta + \gamma\,(1 + 0.34\,u_2)} \tag{1}$$

where T is the mean air temperature (°C), Δ is the slope of the saturation vapor pressure curve, $e_s$ and $e_a$ are the saturation and actual vapor pressures derived from maximum, minimum and dew-point temperature, $R_n$ is net radiation, G the soil heat flux, γ the psychrometric constant and $u_2$ the wind speed at 2 m. The auxiliary FAO-56 relations, including the psychrometric constant from the mean elevation of each division in the HydroSHEDS digital elevation model (Lehner et al., 2008), are given in Text S1. The climatic water balance of month i (mm) is then given by Eq. (2),

$$D_i = \left(P_i - ET_{0,i}\right) \times n_i \tag{2}$$

where $P_i$ and $ET_{0,i}$ are daily rates and $n_i$ is the number of days in the month.

SPEI was computed from D (Vicente-Serrano et al., 2010) following the procedure of Beguería et al. (2014). The water balance was accumulated over k months following Eq. (3),

$$D_n^{(k)} = \sum_{i=0}^{k-1} D_{n-i} \tag{3}$$

and the generalized logistic distribution of Eq. (4),

$$F(x) = \left\{1 + \left[1 - \frac{\kappa\,(x - \xi)}{\alpha}\right]^{1/\kappa}\right\}^{-1} \tag{4}$$

in which ξ, α and κ are the location, scale and shape parameters, was fitted separately for each calendar month by L-moments (Hosking & Wallis, 1997), so that the index expresses departures from the normal for that month rather than the monsoon cycle itself. The cumulative probability was converted to a standard normal value with the rational approximation of Abramowitz and Stegun (1965) (Text S1). SPEI was computed at 1-, 3-, 6- and 12-month windows. SPEI-3, SPEI-6 and SPEI-12 are the forecast targets, and SPEI-1 enters only the screening and the credibility tests. A month with SPEI $\leq -1$ was treated as a drought month throughout, following the conventional classes (McKee et al., 1993) listed in Table S2. The distributions were fitted on the full 1980-2024 record, which includes the test period. To test this choice, they were refitted on 1980-2018 with training-year climatologies for the predictors, and the Bi-LSTM at SPEI-6 and lead 1 was retrained with all five seeds; the main pipeline was to be kept if mean skill changed by less than one seed standard deviation, a rule fixed before the run (Text S6; Table S18; Fig. S5).

**2.4 Credibility of ERA5**

The reanalysis was tested against the stations in four steps, independently of the forecasting pipeline (Text S2). First, each station variable was screened by finding the circular month shift that best aligns its climatology with the ERA5 climatology of its division, which separates a uniform alignment error from a station-specific defect (Table S4; Section 3.1). The compilation contains precipitation and temperature only, so the humidity, wind and radiation inputs of the Penman-Monteith method could not be checked against stations. Second, precipitation, the variable that passed the screening, was compared with ERA5 for each division and for the pooled record using the correlation of the raw and de-seasonalized series, the root mean square error (RMSE), the Nash-Sutcliffe efficiency (NSE) (Nash & Sutcliffe, 1970), the percent bias (PBIAS) and the Kling-Gupta efficiency (KGE) (Gupta et al., 2009), defined in Eqs. (5) and (6),

$$\mathrm{PBIAS} = 100 \times \frac{\sum_i (P_i - O_i)}{\sum_i O_i} \quad (5)$$

$$\mathrm{KGE} = 1 - \sqrt{(r-1)^2 + \left(\frac{\sigma_P}{\sigma_O} - 1\right)^2 + \left(\frac{\mu_P}{\mu_O} - 1\right)^2} \quad (6)$$

where O and P are the station and reanalysis values, r their correlation, σ the standard deviation and μ the mean. Agreement was rated with the classes of Moriasi et al. (2007), in which NSE > 0.75 together with |PBIAS| < 10% counts as very good.

Third, SPEI was recomputed with PET held at its monthly climatology and correlated with the full index, so that $1 - r^2$ gives the share of SPEI variance attributable to interannual PET variation (Table S5). Fourth, SPEI was computed separately from station and ERA5 precipitation with the same Thornthwaite PET derived from ERA5 temperature, so that any difference between the two comes from precipitation alone. Agreement was scored with r, RMSE, NSE, the probability of detection, the false alarm ratio and Cohen's κ (Cohen, 1960), read with the bands of Landis and Koch (1977) (Table S6; Fig. S1).

### 2.5 Predictor screening

The eight candidate variables (Fig. 2) were reduced to an informative, non-redundant set (Text S3). Pearson correlations among the predictors, PET, the water balance and SPEI were computed on same-month values pooled over the divisions (4,320 division-months), as in the concurrent screening used for Bangladesh by Al Mamun et al. (2024). Multicollinearity was assessed with the variance inflation factor (VIF) of Eq. (7),

$$\mathrm{VIF}_j = \frac{1}{1 - R_j^2} \tag{7}$$

where $R_j^2$ is the coefficient of determination of predictor j regressed on all the others (O'Brien, 2007). VIF was computed on de-seasonalized anomalies, because the shared annual cycle of the raw series inflates it, and a value above 10 was treated as serious. The removal was confirmed by a greedy procedure that repeatedly drops the variable with the highest VIF (Dormann et al., 2013) (Table S7). PET and the water balance were excluded from the predictors, because they are built from the other variables and would leak the construction of the target.

The lead-lag analysis measured how far ahead each de-seasonalized predictor X carries information about the index S, through the lagged correlation of Eq. (8),

$$r_k = \frac{\sum_t (X_{t-k} - \bar{X})\,(S_t - \bar{S})}{\sqrt{\sum_t (X_{t-k} - \bar{X})^2}\,\sqrt{\sum_t (S_t - \bar{S})^2}} \tag{8}$$

where $\bar{X}$ and $\bar{S}$ are the series means, for k = 1-12 months. It was checked with the lagged mutual information from the k-nearest-neighbor estimator (Kraskov et al., 2004; Zare et al., 2026) and with Granger causality (Granger, 1969) combined across divisions by Fisher's method (Fisher, 1932). The memory of the index was described by the autocorrelation function (ACF) of Eq. (9),

$$\rho_k = \frac{\sum_{t=1}^{n-k}(S_t - \bar{S})\,(S_{t+k} - \bar{S})}{\sum_{t=1}^{n}(S_t - \bar{S})^2} \tag{9}$$

and by the partial autocorrelation function (PACF), estimated with the Yule-Walker equations (Box et al., 2015), for lags 0-24. Both were averaged over divisions and judged against the 95% white-noise bound ±1.96/√n, with n the mean series length (538, 535 and 529 months for SPEI-3, SPEI-6 and SPEI-12). All screening statistics used the full 1980-2024 record. They decided only which predictors enter the models and how many lags of the index are kept, and no model parameter was fitted to them; we nevertheless note this use of the test years as a limitation (Section 4).

### 2.6 Inputs, partition and leakage controls

Each row of the feature matrix is one division at one forecast origin t. It holds the seven screened predictors over the three most recent months up to t, de-seasonalized by the calendar-month mean of each division; the lagged index, from SPEI(t) back to SPEI(t − 7) at SPEI-3 and to SPEI(t − 13) at SPEI-6 and SPEI-12; the calendar month as a sine and cosine pair; and the division as indicator columns. The depth of the lagged index is the largest lag with a significant PACF (Section 3.2), so it always includes the lag one month beyond the accumulation window. The persistence forecast, SPEI(t), is known at the forecast origin and serves as both the reference forecast and the base of the deep-learning target. XGBoost receives each row as a flat vector, and the deep-learning models receive monthly sequences with the division as a static input (Text S3). The sequence length is 7 months at SPEI-3 and 13 at SPEI-12. At SPEI-6 the deep-learning models read only the last 6 of the 13 months; this length was chosen in a lag-depth test during model development that was scored on the test period. It is the only design choice informed by test-period scores, it affects only the four SPEI-6 deep-learning models, and the central SPEI-6 result beyond lead 1 is shared by XGBoost, which uses all 14 lags (Section 3.6). The record was split by the date of the target month, not by the date of the row. Target months from 1980 to 2018 form the training set and those from 2019 to 2024 the test set, which holds 72 months per division and 576 division-months in total. The split was recomputed for each lead, and the code checks that every training target precedes every test target. Hyperparameter search and early stopping used only the last training years (2014 - 2018 for the deep-learning models, 2017 - 2018 for XGBoost). The climate anomalies were standardized with a scaler fitted on training rows, and sequence windows end at the forecast origin and span only contiguous months. Because a long-window index is strongly autocorrelated, a model trained on the index level can reach a low error while learning little beyond persistence. The deep-learning models were therefore trained on the change from persistence over the lead L, defined in Eq. (10),

$$\Delta y_{t+L} = y_{t+L} - y_t \tag{10}$$

and their output was added back to the persistence forecast through Eq. (11),

$$\hat{y}_{t+L} = \Delta\hat{y}_{t+L} + y_t \tag{11}$$

XGBoost was trained on the index level. Both families are scored on the reconstructed index.

### 2.7 Forecasting models, training and seeds

Five models were compared under the same inputs, partition, tuning protocol and evaluation code. Extreme Gradient Boosting (XGBoost) (T. Chen & Guestrin, 2016) is the machine-learning reference, since tree ensembles are the most common drought-modeling technique in South Asia (Al Reshan et al., 2025). It adds M regression trees under the regularized objective of Eq. (12),

$$\begin{gathered}\hat{y}_i = \sum_{m=1}^{M} f_m(\mathbf{x}_i),\\ \mathcal{L} = \sum_i l(y_i, \hat{y}_i) + \sum_{m=1}^{M} (\gamma T_m + 1/2\,\lambda \parallel \mathbf{w}_m \parallel^2 + \alpha \parallel \mathbf{w}_m \parallel_1)\end{gathered} \tag{12}$$

where $\hat{y}_i$ is the prediction for row i with input vector $\mathbf{x}_i$, $f_m$ is the $m^{th}$ tree, $\mathcal{L}$ is the objective minimized in training, l is the loss between $y_i$ and $\hat{y}_i$, $T_m$ is the number of leaves of tree m, $\mathbf{w}_m$ its leaf weights, γ the minimum loss reduction required for a split, and λ and α the L2 and L1 penalties. The CNN (LeCun et al., 1998) applies two one-dimensional convolution blocks with batch normalization (Ioffe & Szegedy, 2015) and ReLU activation, followed by global average pooling over time, as summarized in Eq. (13),

$$\mathbf{h} = \mathrm{GAP}\left(\phi\left(\mathrm{BN}\left(W_2 * \phi\left(\mathrm{BN}(W_1 * X)\right)\right)\right)\right) \tag{13}$$

where X is the input sequence, $W_1$ and $W_2$ are the convolution kernels, $*$ denotes convolution, BN is batch normalization, φ is the ReLU activation, GAP is global average pooling and **h** is the pooled representation. The Bi-LSTM reads the input window forwards and backwards and concatenates the final states (Gers et al., 2000; Hochreiter & Schmidhuber, 1997; Schuster & Paliwal, 1997). The window lies wholly before the forecast month, so the backward pass uses no future information. The CNN-BiLSTM places one convolution block in front of the recurrent layer, as in comparable drought-index hybrids (Wang et al., 2026). The CEEMDAN-BiLSTM decomposes the SPEI channel of each input window into intrinsic mode functions with CEEMDAN (Huang et al., 1998; Torres et al., 2011; Wu & Huang, 2009). The decomposition is applied inside each window and never to the full series, which avoids the leakage that arises when decomposition precedes the split (Yang et al., 2024). A raw stream and a mode-function stream each pass through a Bi-LSTM and are pooled by additive attention (Bahdanau et al., 2014) according to Eq. (14),

$$\mathbf{a}^{(s)} = \mathrm{softmax}\left(\mathbf{v}^{\top}\tanh\left(WH^{(s)} + \mathbf{b}\right)\right), \qquad \mathbf{p}^{(s)} = \mathrm{LN}\left(\sum_t a_t^{(s)}\,\mathbf{h}_t^{(s)}\right) \tag{14}$$

where $H^{(s)}$ holds the hidden states $\mathbf{h}_t^{(s)}$ of stream s at time steps t, W, **v** and **b** are learned weights, $\mathbf{a}^{(s)}$ are the attention weights, LN is layer normalization and $\mathbf{p}^{(s)}$ is the pooled representation of the stream. All four deep-learning models end in the same dense head, which joins the pooled representation with the division vector and outputs the predicted change. The gate and head equations are given in Text S4, the architectures in Fig. 4 and the tuned values in Tables S8 and S9. Hyperparameters were tuned once, at lead 1, with the Optuna tree-structured Parzen estimator (Akiba et al., 2019), using 20 trials per model and timescale. The deep-learning models were trained with the Adam optimizer (Kingma & Ba, 2014) on a Huber loss with threshold 1.0 (Huber, 1964), with dropout between layers (Srivastava et al., 2014) and early stopping on the validation years, and XGBoost used up to 3,000 trees with early stopping. The remaining settings are listed in Table S8, and a check of tuned against untuned settings is given in Table S10. Each model was retrained exactly as it had been tuned, including the order of the training batches (Text S4).

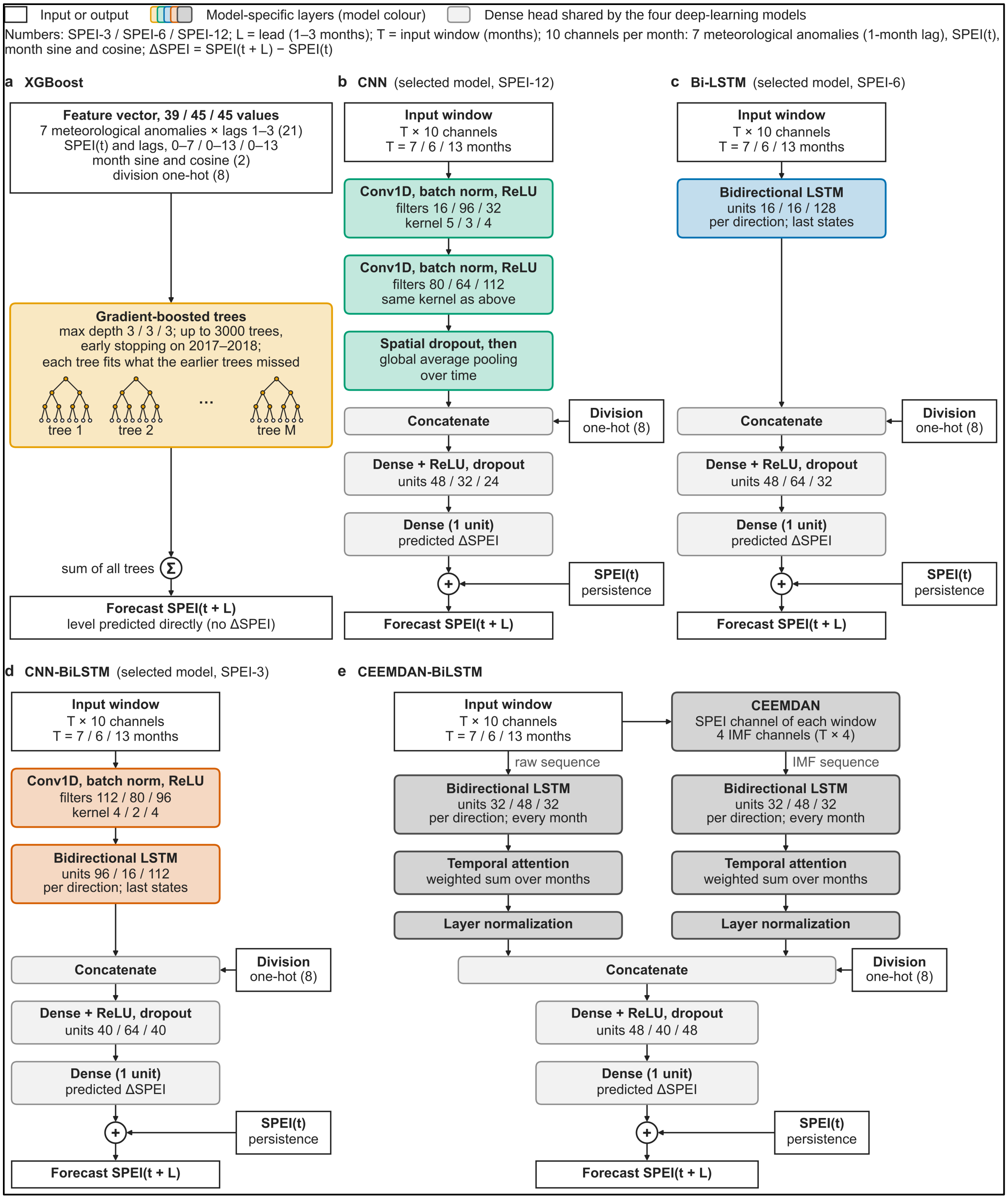

**Fig. 4.** Architectures of the five models: (a) XGBoost, (b) CNN, (c) Bi-LSTM, (d) CNN-BiLSTM and (e) CEEMDAN-BiLSTM. Numbers in the boxes are the values at SPEI-3, SPEI-6 and SPEI-12. White boxes are inputs and outputs, colored boxes are the layers specific to each model and light gray boxes form the dense head shared by the four deep-learning models. The deep-learning models predict the change ΔSPEI = SPEI(t + L) − SPEI(t), to which the persistence forecast SPEI(t) is added, whereas XGBoost predicts the SPEI level directly. Dropout rates and learning rates are listed in Table S8.

Random initialization alone can change the outcome of a model comparison (Bouthillier et al., 2021), so every configuration was trained with five seeds (42-46). The seed sets the weight initialization, dropout masks and batch order of the deep-learning models and the subsampling of XGBoost. In total, 135 runs were made: 75 at lead 1 (five models, three timescales, five seeds) and 60 at leads 2 and 3 (the selected deep-learning model and XGBoost at each of the three timescales, two leads, five seeds), plus the five runs of the reference-period test. Seed 42 reproduces the earlier single-seed results (Table S12).

The model carried forward at each timescale was chosen by a rule fixed before the five-seed runs were made. The score is the five-seed mean lead-1 skill against persistence. The incumbents from the earlier single-seed comparison (CNN-BiLSTM at SPEI-3, Bi-LSTM at SPEI-6 and CNN at SPEI-12) are kept unless a challenger's mean exceeds the incumbent's by more than one seed standard deviation, taking the larger of the two. If no challenger clears this margin, the leading models are reported as indistinguishable and the incumbent is carried forward as their representative. Event detection is reported but not used for selection, and the rule was applied once.

### 2.8 Multi-lead forecasts

Forecasts at leads of two and three months were made directly, with a separate model for each lead trained on the target $y_{t+L}$, so that one-step forecasts are never chained and errors do not compound (Ben Taieb et al., 2012). The persistence forecast remains $y_t$ at every lead, so the reference itself weakens as the lead grows. The partition was re-derived from the shifted target dates, and the test period remains the 2019-2024 target months, with 576 forecast-observation pairs at every lead. Only the selected model of each timescale and XGBoost were extended. The lead-1 hyperparameters were reused, since the inputs are identical across leads and the gain from tuning is of the same size as the variation between seeds (Table S10).

### 2.9 Evaluation

Continuous accuracy was measured with five metrics over the n test pairs of observations $y_i$ and forecasts $\hat{y}_i$, given in Eqs. (15)-(17),

$$\mathrm{RMSE} = \sqrt{\frac{1}{n}\sum_{i=1}^{n}(y_i - \hat{y}_i)^2}, \qquad \mathrm{MAE} = \frac{1}{n}\sum_{i=1}^{n}|\, y_i - \hat{y}_i| \tag{15}$$

$$\mathrm{NSE} = 1 - \frac{\sum_i (y_i - \hat{y}_i)^2}{\sum_i (y_i - \bar{y})^2}, \qquad R^2 = \left[\frac{\sum_i (y_i - \bar{y})\left(\hat{y}_i - \bar{\hat{y}}\right)}{\sqrt{\sum_i (y_i - \bar{y})^2}\sqrt{\sum_i \left(\hat{y}_i - \bar{\hat{y}}\right)^2}}\right]^2 \tag{16}$$

$$\mathrm{WI} = 1 - \frac{\sum_i (y_i - \hat{y}_i)^2}{\sum_i (|\hat{y}_i - \bar{y}| + |y_i - \bar{y}|)^2} \tag{17}$$

where MAE is the mean absolute error, WI is the Willmott index (Willmott, 1981) and the bars denote means over the test period, so that NSE is computed against the test-period mean. Skill against persistence is the mean squared error (MSE) skill score of Eq. (18),

$$SS = 1 - \frac{\mathrm{MSE_{model}}}{\mathrm{MSE_{persistence}}} = 1 - \frac{1 - \mathrm{NSE}}{1 - \mathrm{NSE}_p} \tag{18}$$

where $\mathrm{NSE_p}$ is the NSE of persistence on the same test set. A positive value means that the model beats carrying the last observed value forward. The second form of Eq. (18) shows that, on a fixed test set, skill is an exact transformation of NSE, so differences in NSE are multiplied by $1/(1 - \mathrm{NSE_p})$ when expressed as skill. Normalized Taylor diagrams (Taylor, 2001) summarize the correlation and the standard deviation of each forecast relative to the observations, with persistence as a reference point.

For event detection, each observed and forecast value was classed as drought when SPEI $\leq -1$. With H hits, M misses and F false alarms, Eq. (19) gives the probability of detection, false alarm ratio, success ratio, critical success index and frequency bias (Wilks, 2019),

$$\mathrm{POD} = \frac{H}{H + M}, \quad \mathrm{FAR} = \frac{F}{H + F}, \quad \mathrm{SR} = 1 - \mathrm{FAR}, \quad \mathrm{CSI} = \frac{H}{H + M + F}, \quad B = \frac{H + F}{H + M} \tag{19}$$

which are displayed together on performance diagrams (Roebber, 2009). Because CSI is not linear in POD and the success ratio, the CSI isoline under a seed-mean marker can differ slightly from the mean CSI. FAR and SR are undefined for a forecast that issues no drought month.

Every score carries two kinds of uncertainty. The seed uncertainty is reported as the mean ± standard deviation over the five seeds, and the sampling uncertainty was estimated with a circular block bootstrap (Text S5) (Politis & Romano, 1992). The 72 test months were resampled 2,000 times in blocks of 12 consecutive months, with all eight divisions of a month kept together, which preserves both the serial and the spatial correlation of the 576 pairs. The same resampled months were used for every model and for persistence, which gives paired intervals for differences, and scores were pooled over seeds, so the 95% intervals cover both sampling and seed variation. Panels that must show a single forecast series use each model's median-skill seed; the third of the five seeds ranked by lead-1 skill. Whether a model beats persistence in squared error was tested with the Diebold-Mariano test (Diebold & Mariano, 1995). The loss differential $d_t$ of each test month is the mean over the eight divisions of the difference in squared error between the model and persistence, which gives T = 72 values, and the test statistic is given by Eq. (20),

$$DM = \frac{\bar{d}}{\sqrt{\hat{\sigma}^2_{\mathrm{LR}}/T}} \tag{20}$$

where $\bar{d}$ is the mean loss differential and $\hat{\sigma}^2_{LR}$ its long-run variance, estimated with the Newey-West estimator (Newey & West, 1987) using Bartlett weights and a bandwidth of 3 (Text S5). The statistic was multiplied by the small-sample correction of Harvey et al. (1997) and compared with Student's t distribution with $T - 1$ degrees of freedom (two-sided). The test was applied to each seed and to the seed-ensemble mean forecast.

## 2.10 Spatial verification, station target and interpretation

The spatial pattern of the selected forecasts was checked in two test months per timescale, the worst observed drought month and a near-normal month. For each seed we computed the spatial correlation and the ratio of spatial standard deviations between forecast and observed divisional values, the forecast national mean relative to the observed one, the driest division and the number of sign reversals (Fig. S3; Table S16).

To test whether the models learned drought behavior rather than features specific to the reanalysis, the selected models were re-scored, without retraining, against SPEI computed from BMD station precipitation with the Thornthwaite PET of Section 2.4. Only the reference series changes; the forecasts and the persistence forecast are those used everywhere else. The common period of the test set and the station record is 2019-2022, which gives 384 division-months. The difference in PET method is acceptable here because interannual PET variation explains only about 4-5% of SPEI variance at the forecast timescales (Section 3.1).

The selected models were interpreted with SHapley Additive exPlanations (SHAP) (Lundberg & Lee, 2017), which attribute a prediction to its inputs through Shapley values from cooperative game theory (Shapley, 1953). Attributions were computed with the gradient-based explainer for every seed of the three selected models, using a background sample of 150 training windows, and summed over the time steps of each predictor. The importance of a predictor is its mean absolute SHAP value over the test windows, reported as the mean ± standard deviation over seeds. The sign of the mean signed SHAP value is reported only where all five seeds agree.

# 3. Results

All scores refer to the independent test period, January 2019 to December 2024, which contains 576 division-months, of which 119, 156 and 204 are drought months (SPEI $\leq -1$) at SPEI-3, SPEI-6 and SPEI-12. Model scores are the mean ± standard deviation (SD) over the five seeds (per-seed values in Table S14), and persistence, being deterministic, has no seed spread. Bracketed intervals are 95% confidence intervals (CI) from the circular block bootstrap (Table S13), and Diebold-Mariano (DM) tests compare the seed-ensemble mean forecast with persistence (Section 2.9).

## 3.1 ERA5 as a reference for SPEI in Bangladesh

Station precipitation passed the quality screening, but the station temperature records did not. The month shift that best aligns each station's precipitation climatology with the ERA5 climatology of its division is zero at all 35 stations, with climatology correlations of 0.955-0.999 (Table S4; Fig. 5a, b). The station temperatures, by contrast, have almost no seasonal cycle. Their annual range is 0.5 °C for maximum and 0.8 °C for minimum temperature, against 8.3 °C and 12.6 °C in ERA5, and the best-aligning shift for maximum temperature varies between stations from 0 to 11 months, with correlations of only 0.511-0.677 (Fig. 5c, d). Only precipitation was therefore used as a validation reference.

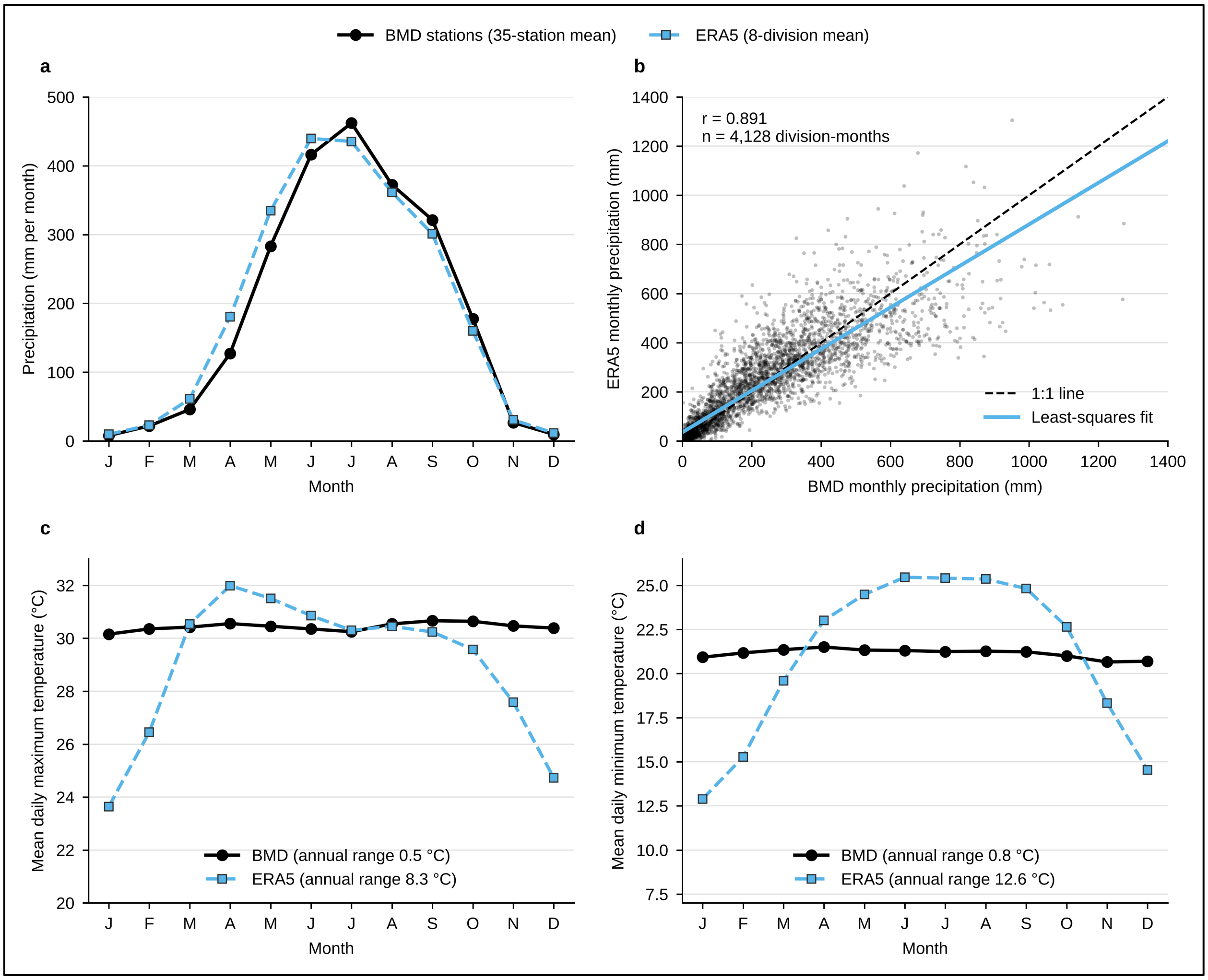

**Fig. 5.** Screening of the BMD station compilation against ERA5, 1980-2022. (a) Mean monthly precipitation of the 35-station mean and of the eight-division ERA5 mean. (b) ERA5 against BMD monthly precipitation for all 4,128 division-months, with the 1:1 line (dashed) and the least-squares line. (c, d) Climatologies of mean daily maximum and minimum temperature, with the annual range of each record in the legend. BMD values are drawn in black and ERA5 values in blue. Station values stored as zero for missing data are excluded (Text S2).

Pooled over the eight divisions and 4,128 division-months of 1980-2022, ERA5 and station precipitation correlate at 0.891 for the raw series and 0.696 for the de-seasonalized anomalies, with an NSE of 0.790, a KGE of 0.875 and a percent bias of +3.4% (Table 1A). Agreement is weakest in Rangpur for the anomaly correlation (0.564), in Mymensingh for NSE (0.658) and in Chittagong for bias (−24.1%) and KGE (0.597) (Table S3).

The drought index depends only weakly on the evaporative term that could not be checked against stations. With PET held at its monthly climatology, the recomputed index correlates with the full index at r = 0.973, 0.979 and 0.977 for SPEI-3, SPEI-6 and SPEI-12 (Table 1B).

Interannual PET variation therefore accounts for 5.3%, 4.2% and 4.6% of the variance of the three forecast targets, and for 12.6% at SPEI-1. The division-level values are given in Table S5.

SPEI computed from station and from ERA5 precipitation agrees best at short windows (Table 1C; Fig. S1). The correlation falls from 0.809 at SPEI-1 to 0.766, 0.712 and 0.647 at SPEI-3, SPEI-6 and SPEI-12, and Cohen's κ falls from 0.660 to 0.579, 0.468 and 0.422. At SPEI-3 agreement is moderate, and ERA5 flags 65.2% of the station drought months (POD = 0.652).

**Table 1.** Credibility of ERA5 as the reference for SPEI. (A) ERA5 against BMD station precipitation, pooled over the divisions, 1980-2022; $r_a$ is the correlation of the de-seasonalized anomalies. (B) Correlation of SPEI with time-varying and with climatological PET, and the share of SPEI variance due to interannual PET variation. (C) SPEI from station against ERA5 precipitation, with the same Thornthwaite PET; drought is SPEI ≤ −1. Division values are given in Tables S3 and S5.

**(A) Precipitation, ERA5 against stations (pooled)**

| n | r | $r_a$ | PBIAS (%) | RMSE (mm) | NSE | KGE |
|---|---|---|---|---|---|---|
| 4,128 | 0.891 | 0.696 | +3.4 | 94.6 | 0.790 | 0.875 |

**(B) Sensitivity of SPEI to PET**

| Timescale | r (full PET vs climatological PET) | Variance attributable to PET (%) |
|---|---|---|
| SPEI-1 | 0.935 | 12.6 |
| SPEI-3 | 0.973 | 5.3 |
| SPEI-6 | 0.979 | 4.2 |
| SPEI-12 | 0.977 | 4.6 |

**(C) SPEI from station precipitation against SPEI from ERA5 precipitation**

| Timescale | n | r | RMSE | NSE | POD | FAR | Cohen's κ |
|---|---|---|---|---|---|---|---|
| SPEI-1 | 4,128 | 0.809 | 0.635 | 0.622 | 0.713 | 0.270 | 0.660 |
| SPEI-3 | 4,112 | 0.766 | 0.689 | 0.534 | 0.652 | 0.345 | 0.579 |
| SPEI-6 | 4,088 | 0.712 | 0.755 | 0.427 | 0.557 | 0.436 | 0.468 |
| SPEI-12 | 4,040 | 0.647 | 0.825 | 0.294 | 0.522 | 0.482 | 0.422 |

## 3.2 Predictor screening and the horizon of meteorological information

Two groups of candidate predictors are redundant (Fig. 6a). Mean temperature correlates at 0.93-0.98 with maximum, minimum and dew-point temperature, and precipitation correlates at 0.85 with cloud cover and at 0.99 with the water balance. Maximum, minimum and mean temperature exceed the VIF threshold of 10, with values of 11.5, 10.9 and 11.3, and removing mean temperature alone brings every remaining VIF below 10, the largest being 8.8 (Table S7). Seven predictors were therefore retained: precipitation, maximum and minimum temperature, dew-point temperature, solar radiation, cloud cover and wind speed (trends in Fig. 2).

The lead-lag analysis shows how long a predictor observed before the target month stays informative (Fig. 6b-d). De-seasonalized precipitation, solar radiation, cloud cover and maximum temperature carry the strongest signal, and each curve stays roughly level until the lead reaches the accumulation length and then collapses to near zero. At SPEI-3, precipitation correlates at 0.51 and 0.54 with the index one and two months later but at only 0.06 three months later. At SPEI-6 the precipitation correlation stays between 0.38 and 0.45 up to a lead of five months and drops to 0.07 at six, and at SPEI-12 it stays between 0.29 and 0.33 up to eleven months and drops to 0.04 at twelve. For a forecast issued at month t, the newest predictors are therefore informative at leads 1 and 2 but not at lead 3 for SPEI-3, and at all three leads for SPEI-6 and SPEI-12.

The autocorrelation function (ACF) and the partial autocorrelation function (PACF) of the index show the same structure (Fig. 7). The lag-1 autocorrelation is 0.64, 0.80 and 0.93 at SPEI-3, SPEI-6 and SPEI-12, and the autocorrelation stays above the 95% white-noise bound up to lags 2, 5 and 10. The PACF is dominated by lag 1 but has a positive secondary peak one month beyond the window length (0.21 at lag 4, 0.27 at lag 7 and 0.40 at lag 13). Significant partial autocorrelations extend to lags 7, 13 and 13, which sets the depth of the drought-history input (Section 2.6).

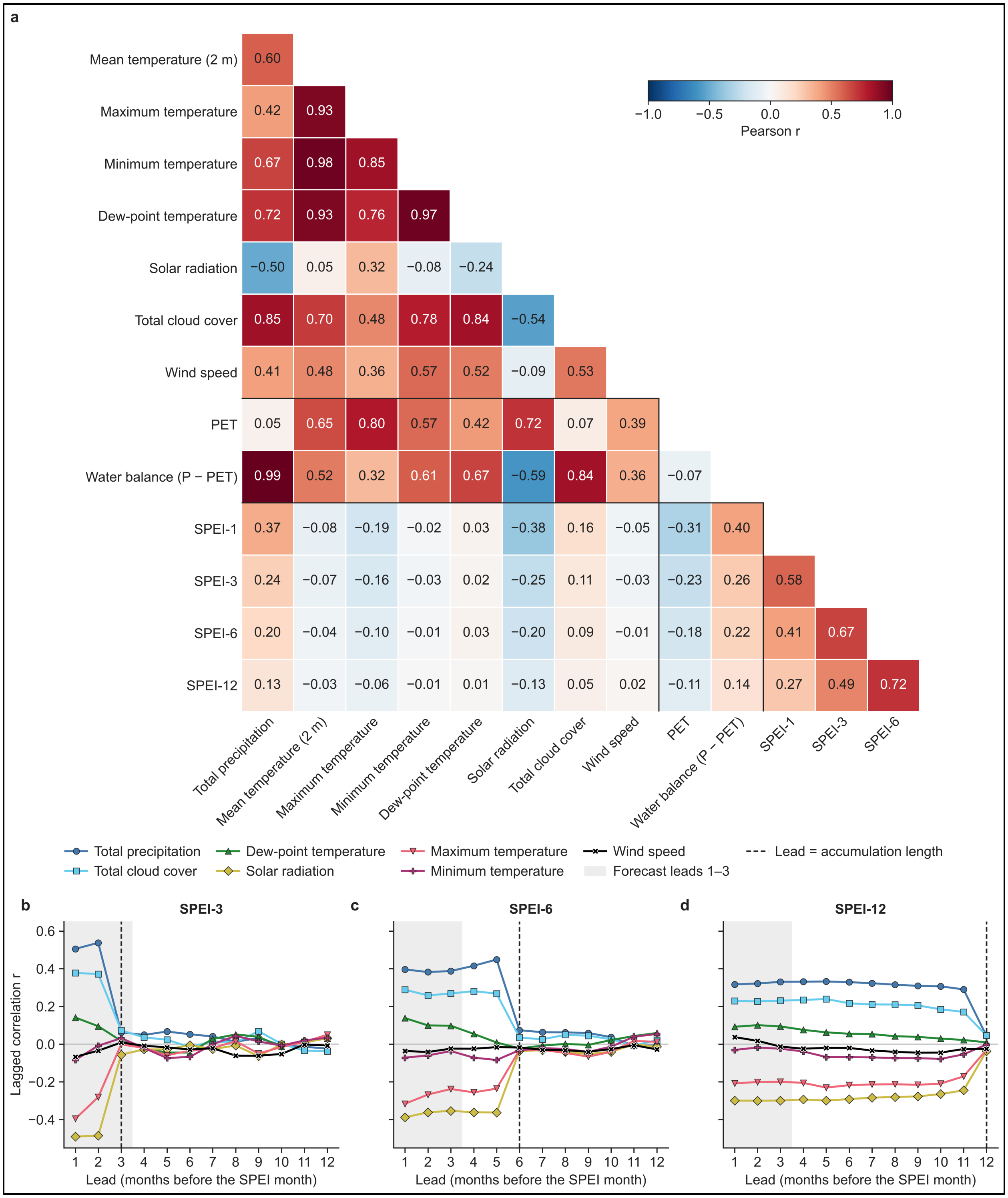

**Fig. 6.** Predictor screening. (a) Pearson correlations among the eight candidate ERA5 predictors, PET, the water balance (P − PET) and SPEI at four timescales (1980-2024, pooled over divisions). (b-d) Correlation of each de-seasonalized predictor, observed k months before the target month, with (b) SPEI-3, (c) SPEI-6 and (d) SPEI-12. The gray band marks lead 1-3 and the dashed line a lead equal to the accumulation length.

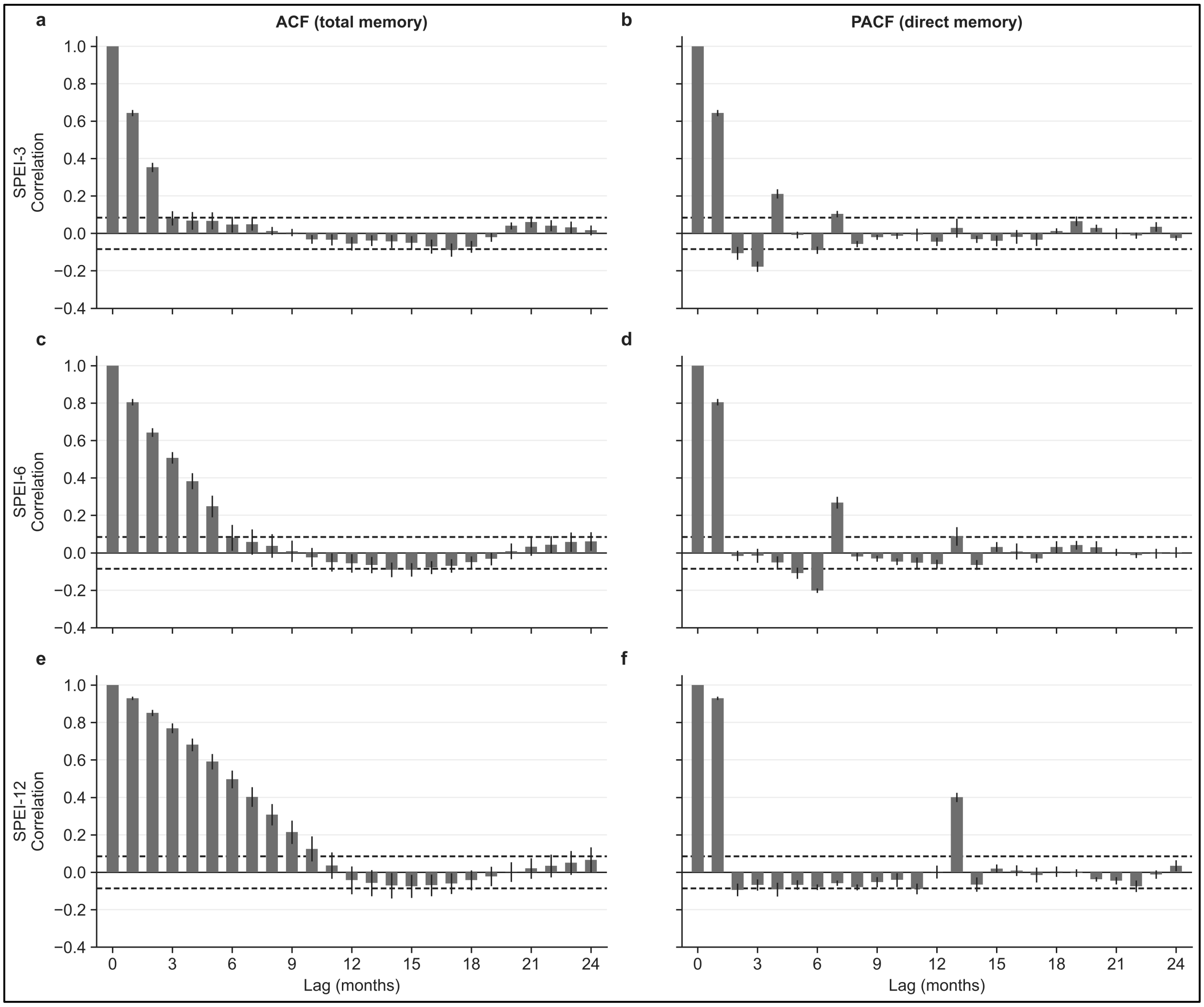

**Fig. 7.** Autocorrelation (a, c, e) and partial autocorrelation (b, d, f) functions of SPEI-3, SPEI-6 and SPEI-12 for lags of 0-24 months, 1980-2024. Bars are the mean over the eight divisions and whiskers their standard deviation. Dashed lines are the 95% white-noise bounds $\pm 1.96/\sqrt{n}$, with n the mean series length (538, 535 and 529 months).

## 3.3 Lead-1 comparison and model selection

At lead 1 every model is more accurate than persistence at SPEI-3 and SPEI-6, whereas at SPEI-12 two of the five models are less accurate than persistence (Fig. 8; Table 2). Persistence itself becomes much stronger as the window lengthens, with NSE values of 0.315, 0.565 and 0.850 at SPEI-3, SPEI-6 and SPEI-12.

At SPEI-3 the CNN-BiLSTM is the best model on all six continuous metrics, with an RMSE of 0.627 ± 0.009, an NSE of 0.595 ± 0.011 and a skill of 0.408 ± 0.016 [0.34, 0.47]. The Bi-LSTM (0.375 ± 0.012) and XGBoost (0.373 ± 0.006) follow closely, and the CNN is the weakest (0.255

± 0.031). The skill advantage of the CNN-BiLSTM over XGBoost, +0.035 [−0.030, +0.090], is not significant.

At SPEI-6 the Bi-LSTM is the best model on all six metrics, with an RMSE of 0.491 ± 0.017, an NSE of 0.717 ± 0.020 and a skill of 0.350 ± 0.045 [0.26, 0.46]. The CNN is close behind (0.327 ± 0.059), while XGBoost (0.144 ± 0.019) and the CEEMDAN-BiLSTM (0.111 ± 0.060) gain far less over persistence. Here the advantage of the selected model over XGBoost is significant, +0.206 [0.063, 0.368].

SPEI-12 gives the highest accuracy and the least skill. NSE ranges from 0.836 to 0.872 across the five models, against 0.850 for persistence. Skill is 0.148 ± 0.042 for the CNN, 0.141 ± 0.118 for the Bi-LSTM, 0.110 ± 0.059 for the CNN-BiLSTM, −0.034 ± 0.105 for the CEEMDAN-BiLSTM and −0.093 ± 0.038 for XGBoost, and none of the five intervals excludes zero (CNN [−0.04, 0.30]). The CNN's skill nevertheless exceeds that of XGBoost by +0.241 [0.042, 0.392].

The selection rule of Section 2.7 left the incumbent models in place: the CNN-BiLSTM at SPEI-3, the Bi-LSTM at SPEI-6 and the CNN at SPEI-12 (Table S10). No challenger exceeded an incumbent by more than one seed SD. Under the same rule the CNN is indistinguishable from the Bi-LSTM at SPEI-6, and the Bi-LSTM and the CNN-BiLSTM are indistinguishable from the CNN at SPEI-12, so at these two timescales the selected model represents a group of equivalent models. The CEEMDAN-BiLSTM is not the best model on any metric at any timescale.

These results do not depend on the reference period of the SPEI fit. Refitting the distribution on 1980-2018 leaves the test-period index almost unchanged (correlations of 0.993-0.997; Fig. S5) and changes the lead-1 skill of the Bi-LSTM at SPEI-6 from 0.350 ± 0.045 to 0.354 ± 0.059, well within one seed SD (Table S18; Text S6), so the full-record index was kept.

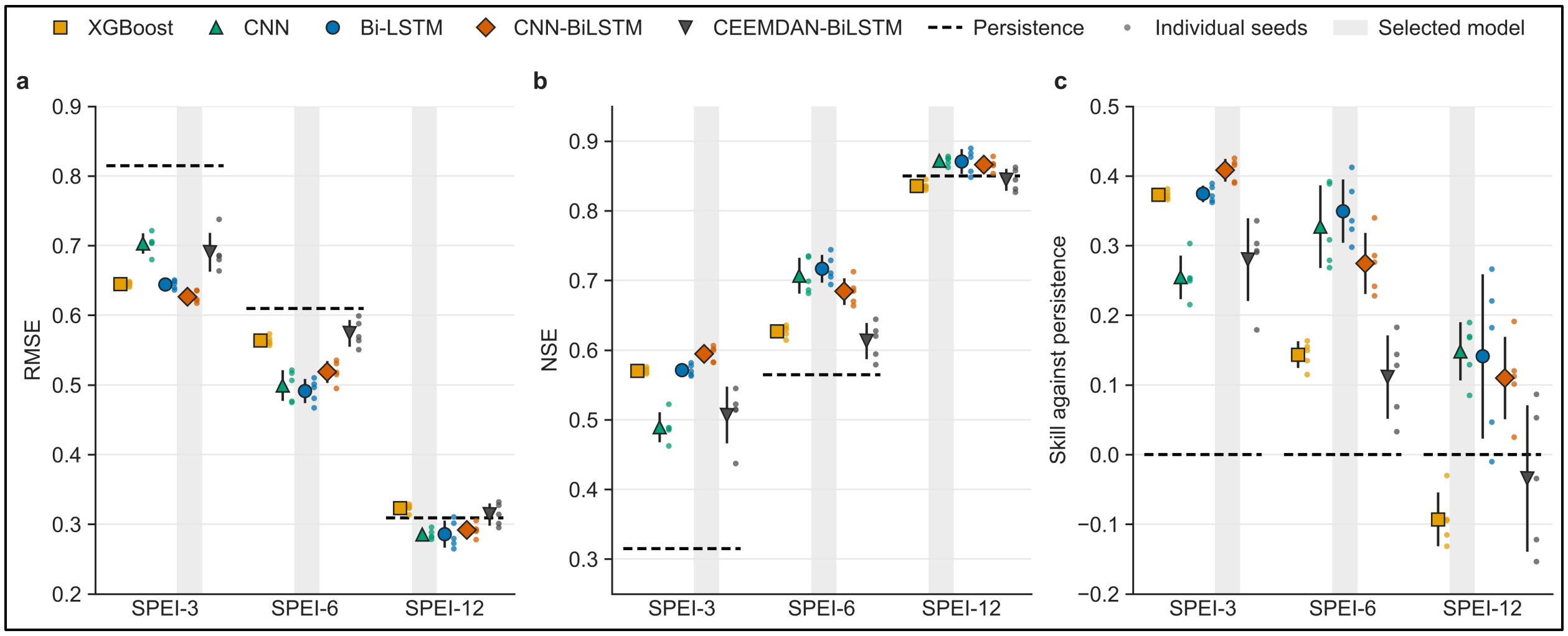

**Fig. 8.** Lead-1 performance of the five models over the test period (2019-2024): (a) RMSE, (b) NSE and (c) skill against persistence. Large markers are five-seed means, whiskers ±1 SD and small dots individual seeds; the dashed tick marks persistence and the gray band the selected model. Values are given in Table 2.

**Table 2.** Lead-1 scores of the five models and of persistence over the test period (2019-2024, 576 division-months). Model values are five-seed means ± SD, and bold rows mark the selected model of each timescale. Skill is against persistence (Eq. (18)) and is zero for persistence by definition. WI, Willmott index. Bootstrap intervals are given in Table S13 and per-seed values in Table S14.

| SPEI | Model | RMSE | MAE | NSE | WI | $R^2$ | Skill | Skill range over seeds | Seeds with skill > 0 |
|---|---|---|---|---|---|---|---|---|---|
| 3 | XGBoost | 0.645 ± 0.003 | 0.518 ± 0.001 | 0.571 ± 0.004 | 0.838 ± 0.001 | 0.576 ± 0.006 | 0.373 ± 0.006 | +0.366 to +0.381 | 5/5 |
| 3 | CNN | 0.703 ± 0.015 | 0.564 ± 0.009 | 0.489 ± 0.021 | 0.818 ± 0.011 | 0.498 ± 0.020 | 0.255 ± 0.031 | +0.215 to +0.303 | 5/5 |
| 3 | Bi-LSTM | 0.644 ± 0.006 | 0.512 ± 0.006 | 0.572 ± 0.008 | 0.857 ± 0.002 | 0.577 ± 0.004 | 0.375 ± 0.012 | +0.362 to +0.390 | 5/5 |
| **3** | **CNN-BiLSTM** | **0.627 ± 0.009** | **0.499 ± 0.008** | **0.595 ± 0.011** | **0.861 ± 0.004** | **0.595 ± 0.011** | **0.408 ± 0.016** | **+0.390 to +0.426** | **5/5** |
| 3 | CEEMDAN-BiLSTM | 0.691 ± 0.028 | 0.555 ± 0.018 | 0.507 ± 0.041 | 0.837 ± 0.014 | 0.519 ± 0.035 | 0.280 ± 0.059 | +0.179 to +0.336 | 5/5 |
| 3 | Persistence | 0.815 | 0.660 | 0.315 | 0.808 | 0.432 | 0 | n/a | n/a |
| 6 | XGBoost | 0.564 ± 0.006 | 0.427 ± 0.006 | 0.627 ± 0.008 | 0.880 ± 0.003 | 0.635 ± 0.008 | 0.144 ± 0.019 | +0.115 to +0.164 | 5/5 |
| 6 | CNN | 0.499 ± 0.022 | 0.370 ± 0.014 | 0.707 ± 0.026 | 0.912 ± 0.009 | 0.711 ± 0.023 | 0.327 ± 0.059 | +0.269 to +0.392 | 5/5 |
| **6** | **Bi-LSTM** | **0.491 ± 0.017** | **0.356 ± 0.012** | **0.717 ± 0.020** | **0.921 ± 0.006** | **0.729 ± 0.019** | **0.350 ± 0.045** | **+0.298 to +0.413** | **5/5** |
| 6 | CNN-BiLSTM | 0.519 ± 0.016 | 0.368 ± 0.011 | 0.684 ± 0.019 | 0.909 ± 0.007 | 0.701 ± 0.017 | 0.274 ± 0.044 | +0.228 to +0.340 | 5/5 |
| 6 | CEEMDAN-BiLSTM | 0.574 ± 0.019 | 0.418 ± 0.015 | 0.613 ± 0.026 | 0.890 ± 0.007 | 0.640 ± 0.019 | 0.111 ± 0.060 | +0.033 to +0.183 | 5/5 |
| 6 | Persistence | 0.609 | 0.450 | 0.565 | 0.882 | 0.611 | 0 | n/a | n/a |
| 12 | XGBoost | 0.323 ± 0.006 | 0.247 ± 0.005 | 0.836 ± 0.006 | 0.954 ± 0.002 | 0.859 ± 0.005 | −0.093 ± 0.038 | −0.131 to −0.030 | 0/5 |
| **12** | **CNN** | **0.285 ± 0.007** | **0.207 ± 0.007** | **0.872 ± 0.006** | **0.965 ± 0.002** | **0.874 ± 0.006** | **0.148 ± 0.042** | **+0.085 to +0.190** | **5/5** |
| 12 | Bi-LSTM | 0.286 ± 0.020 | 0.202 ± 0.011 | 0.871 ± 0.018 | 0.965 ± 0.005 | 0.878 ± 0.015 | 0.141 ± 0.118 | −0.010 to +0.266 | 4/5 |
| 12 | CNN-BiLSTM | 0.292 ± 0.010 | 0.204 ± 0.007 | 0.866 ± 0.009 | 0.963 ± 0.002 | 0.879 ± 0.005 | 0.110 ± 0.059 | +0.025 to +0.191 | 5/5 |
| 12 | CEEMDAN-BiLSTM | 0.314 ± 0.016 | 0.217 ± 0.008 | 0.845 ± 0.016 | 0.959 ± 0.004 | 0.858 ± 0.012 | −0.034 ± 0.105 | −0.154 to +0.087 | 2/5 |
| 12 | Persistence | 0.309 | 0.220 | 0.850 | 0.961 | 0.855 | 0 | n/a | n/a |

### 3.4 Forecast fidelity: high correlation, compressed spread

The models gain over persistence mainly through correlation (Fig. 9a-c). The correlation of the seed-mean forecasts with the observations is 0.706-0.772 at SPEI-3, 0.797-0.854 at SPEI-6 and 0.926-0.938 at SPEI-12, against 0.657, 0.781 and 0.925 for persistence.

Every model lies inside the arc of the observed SD. The ratio of forecast to observed SD ranges from 0.69 to 0.96 across models and timescales, whereas persistence sits on the arc (1.000, 0.993 and 0.995). For the selected models the ratio is 0.775, 0.941 and 0.939, and the least-squares slope of forecast on observation is 0.598, 0.804 and 0.878 (Table S15). In all fifteen scatter panels the fitted line is flatter than the 1:1 line and the largest anomalies of either sign are under-predicted (Fig. 9d-r).

The selected forecasts also stay close to persistence, with which they correlate at 0.883, 0.930 and 0.971, and their predicted month-to-month change (the forecast minus persistence) reaches only 0.581, 0.556 and 0.618 of the observed amplitude (Table S15). The national mean series in Fig. 10 shows the same behavior over time: the forecasts follow the observed index but do not reach its largest dry and wet anomalies. The divisional series of the selected models are given in Fig. S2.

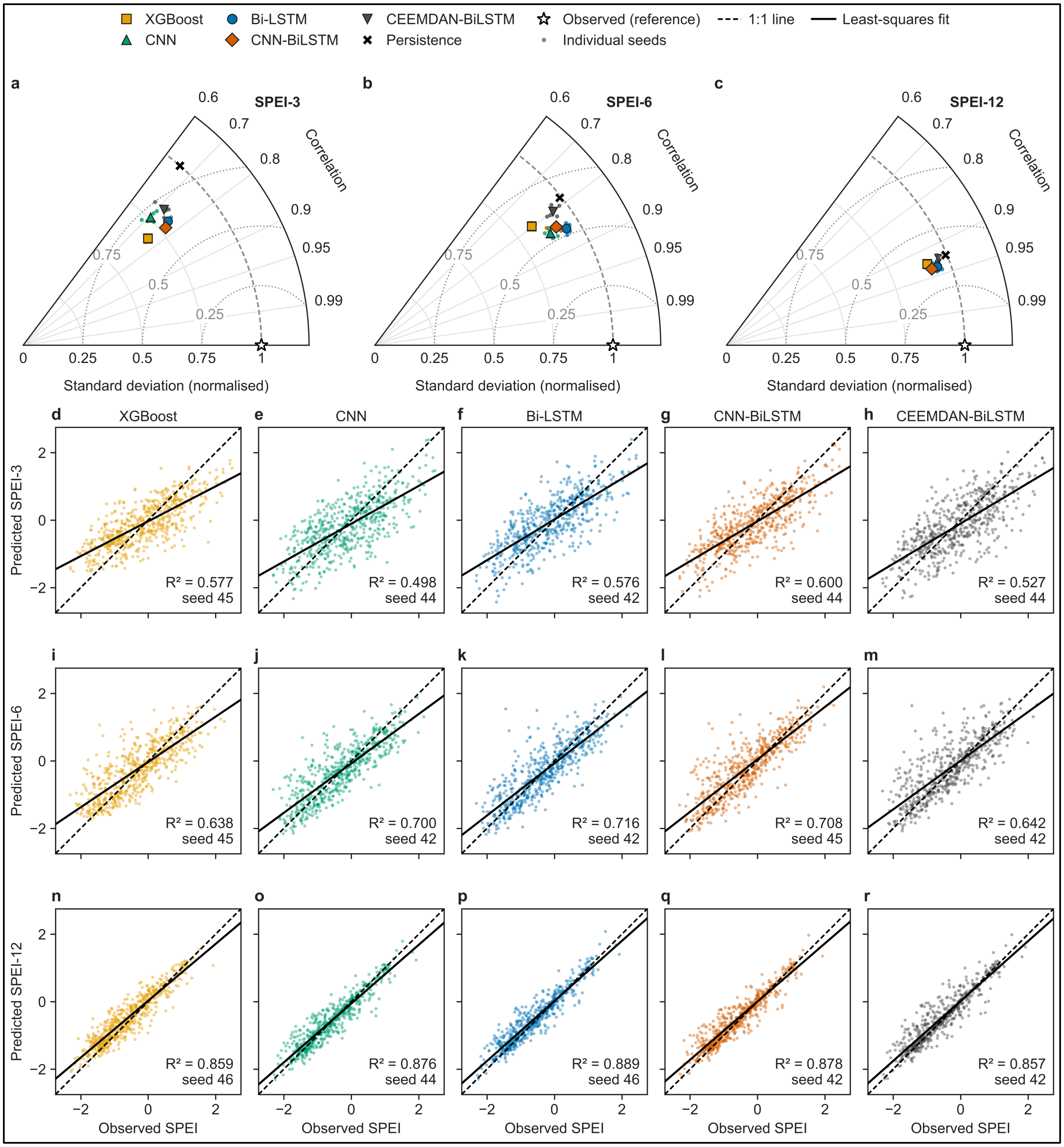

**Fig. 9.** Lead-1 forecast fidelity. (a-c) Normalized Taylor diagrams for SPEI-3, SPEI-6 and SPEI-12 (radius, SD ratio; angle, correlation; arcs, centered RMSE), with seed means (large markers), seeds (dots), persistence (cross) and observations (star). (d–r) Predicted against observed SPEI for the median-skill seed of each model, defined as the third of its five seeds ranked by lead-1 skill (Section 2.9), so the plotted seed differs between panels and is printed in each, with the 1:1 line (dashed) and the least-squares fit (solid).

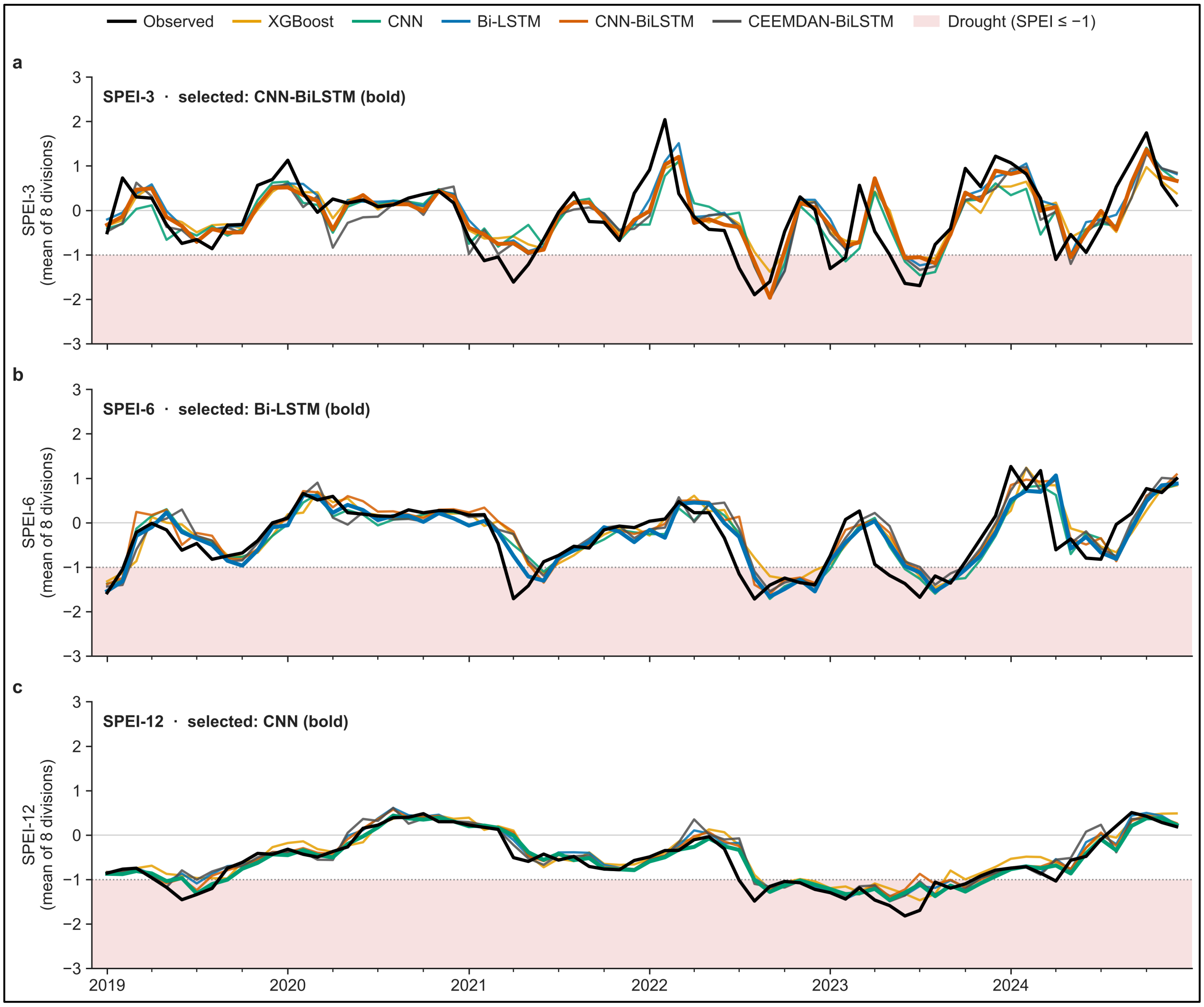

**Fig. 10.** National mean SPEI (unweighted mean of the eight divisions) over the test period, January 2019 to December 2024, at (a) SPEI-3, (b) SPEI-6 and (c) SPEI-12. The observed index (black) is shown with the lead-1 forecast of the median-skill seed of each model, and the selected model is drawn in bold. The shaded band marks drought (SPEI ≤ −1).

## 3.5 At lead 1, drought detection does not follow skill

Detection of drought months improves as the accumulation window lengthens (Fig. 11; Table S11). Across the five models, CSI ranges from 0.252 to 0.346 at SPEI-3, from 0.515 to 0.615 at SPEI-6 and from 0.664 to 0.735 at SPEI-12, and POD rises from 0.284-0.425 to 0.606-0.714 and 0.715-0.810.

Persistence detects more drought months than any model at every timescale, with POD values of 0.555, 0.769 and 0.853. Its CSI of 0.379 at SPEI-3 is above the point estimate of every model, although the difference is significant only for XGBoost, and its CSI at SPEI-6 (0.606) and SPEI-12 (0.734) is level with that of the best models, the Bi-LSTM (0.615 ± 0.023) and the CNN (0.735 ± 0.014). The models raise fewer false alarms than persistence, with FAR values of 0.310-0.399 against 0.455 at SPEI-3 and 0.077-0.111 against 0.159 at SPEI-12, but they miss more droughts. Every model has a frequency bias below 1, at most 0.966, and misses outnumber false alarms for every model at every timescale.

The ranking by detection also differs from the ranking by skill. At SPEI-3 the Bi-LSTM is the best detector (POD 0.425 ± 0.066, CSI 0.346 ± 0.041), ahead of the selected CNN-BiLSTM (CSI 0.298 ± 0.044), and XGBoost, which matches the Bi-LSTM in skill, has the lowest CSI (0.252 ± 0.022). Relative to persistence, the CSI deficit of the CNN-BiLSTM, −0.081 [−0.231, +0.038], is not significant, whereas that of XGBoost, −0.127 [−0.274, −0.016], is. At SPEI-6 and SPEI-12 the selected models are the best detectors among the five models and are level with persistence, with CSI differences of +0.009 [−0.087, +0.085] and +0.001 [−0.043, +0.049].

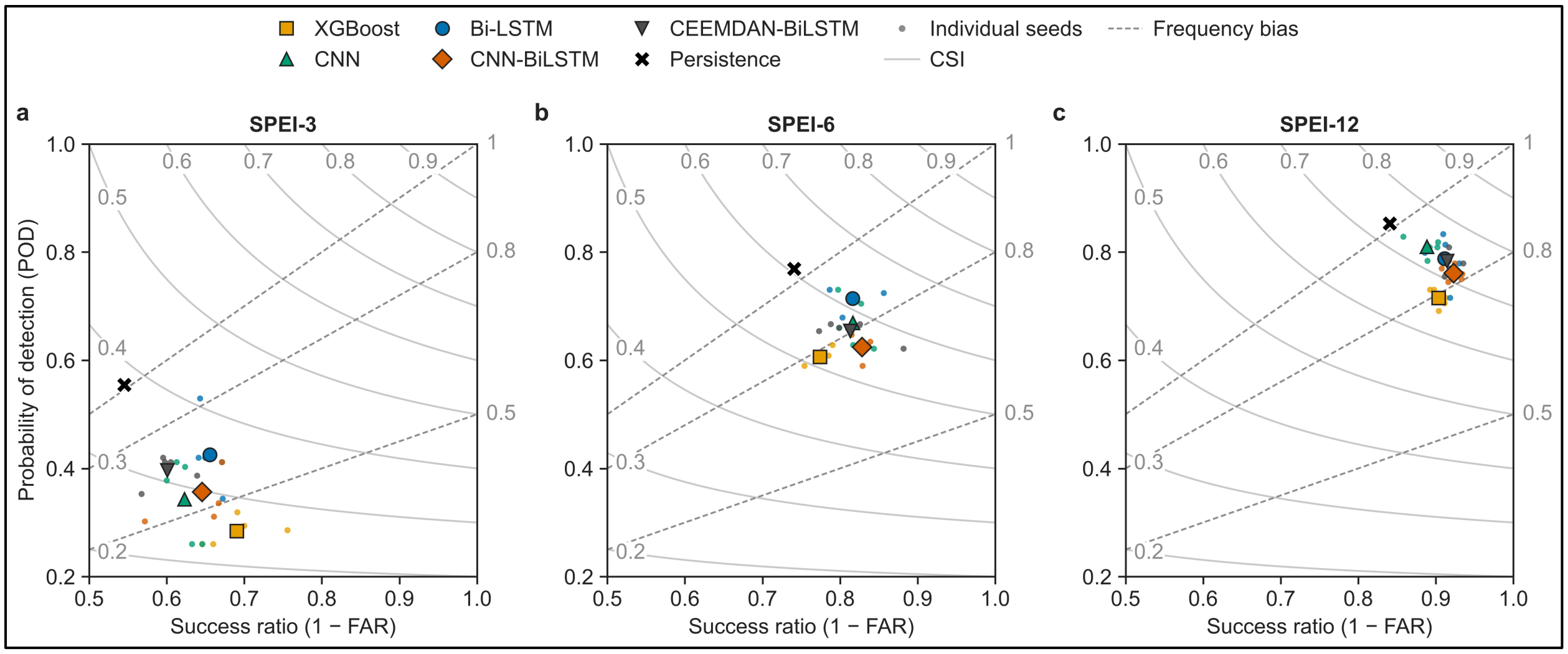

**Fig. 11.** Lead-1 drought detection (SPEI ≤ −1) on performance diagrams for (a) SPEI-3, (b) SPEI-6 and (c) SPEI-12. The axes are the success ratio and POD; solid curves are lines of equal CSI and dashed lines frequency bias (0.5, 0.8 and 1). Markers as in Fig. 8, with persistence as a cross. Contingency counts are given in Table S11.

### 3.6 Accuracy, skill and detection diverge with lead time

Extending the lead from one to three months degrades every forecast, including the reference (Fig. 12; Table 3). The RMSE of persistence grows by 70%, 75% and 89% at SPEI-3, SPEI-6 and SPEI-12, and its NSE falls from 0.315 to −0.989, from 0.565 to −0.340 and from 0.850 to 0.462.

At SPEI-3, skill stays almost flat while detection collapses. The skill of the CNN-BiLSTM is 0.408 ± 0.016, 0.394 ± 0.010 and 0.355 ± 0.004 at leads 1-3, and its squared error remains significantly lower than that of persistence at every lead (ensemble DM $p \leq 0.0018$, and $p < 0.05$ in all five single seeds). Over the same leads its NSE falls from 0.595 to 0.159 and then to −0.283, and its RMSE rises from 0.627 to 1.115. The model catches on average 42.4 of the 119 drought months at lead 1, 7.2 at lead 2 and 0.2 at lead 3, and its CSI falls from 0.298 to 0.054 and 0.002, significantly below persistence at leads 2 and 3 (CSI differences of −0.128 [−0.259, −0.018] and −0.094 [−0.134, −0.045]). Persistence still catches 66, 38 and 22 of the same drought months.

XGBoost shows the divergence in its clearest form. At lead 3 it has the highest mean skill in the study, 0.439 ± 0.025 [0.35, 0.53], significantly higher than that of the CNN-BiLSTM at the same lead (CNN-BiLSTM minus XGBoost: −0.085 [−0.166, −0.015]), yet it detects none of the 119 drought months in any seed. Four of its five seeds issue no drought forecast at lead 3, and the fifth issues a single false alarm.

At SPEI-6, the forecast keeps its value at all three leads. The skill of the Bi-LSTM is 0.350 ± 0.045, 0.367 ± 0.070 and 0.367 ± 0.038, the intervals stay above zero ([0.26, 0.46], [0.23, 0.51] and [0.19, 0.50]), and the DM test is significant at every lead (ensemble $p \leq 0.0008$, and $p < 0.05$ in all five single seeds). NSE declines from 0.717 to 0.445 and 0.152, and RMSE rises from 0.491 to 0.851. In detection the Bi-LSTM is level with persistence at every lead, with CSI values of 0.615, 0.423 and 0.313 against 0.606, 0.392 and 0.272 and no CSI-difference interval excluding zero (+0.009, +0.031 and +0.041). It reaches this level with fewer false alarms (FAR 0.184, 0.276 and 0.294 against 0.259, 0.455 and 0.591) and still catches 56.2 of the 156 drought months at lead 3. XGBoost is level with the Bi-LSTM in skill at leads 2 and 3 (0.322 and 0.372) and with persistence in CSI, but at lead 1 its CSI is significantly below persistence (−0.091 [−0.169, −0.027]).

At SPEI-12, no lead shows robust skill. The CNN's skill is 0.148 ± 0.042, −0.025 ± 0.261 and −0.129 ± 0.339 at leads 1-3, and all three intervals include zero ([−0.04, 0.30], [−0.95, 0.27] and [−1.16, 0.26]). The DM test of the seed-ensemble forecast is significant only at lead 1 ($p = 0.006$), where only two of the five single seeds pass it, and gives p values of 0.25 and 0.71 at leads 2 and 3. NSE is 0.872 at lead 1 against 0.850 for persistence, and 0.652 and 0.393 at leads 2 and 3 against 0.661 and 0.462. CSI equals that of persistence at lead 1 (0.735 against 0.734) and is lower, though not significantly, at leads 2 and 3 (0.596 against 0.624, and 0.438 against 0.539). XGBoost never beats persistence at this timescale (DM p = 0.45-0.91), and its CSI is significantly below persistence at every lead.

Across the three views, SPEI-6 is thus the only timescale at which the selected model beats persistence significantly in squared error and stays level with it in detection at leads 2 and 3.

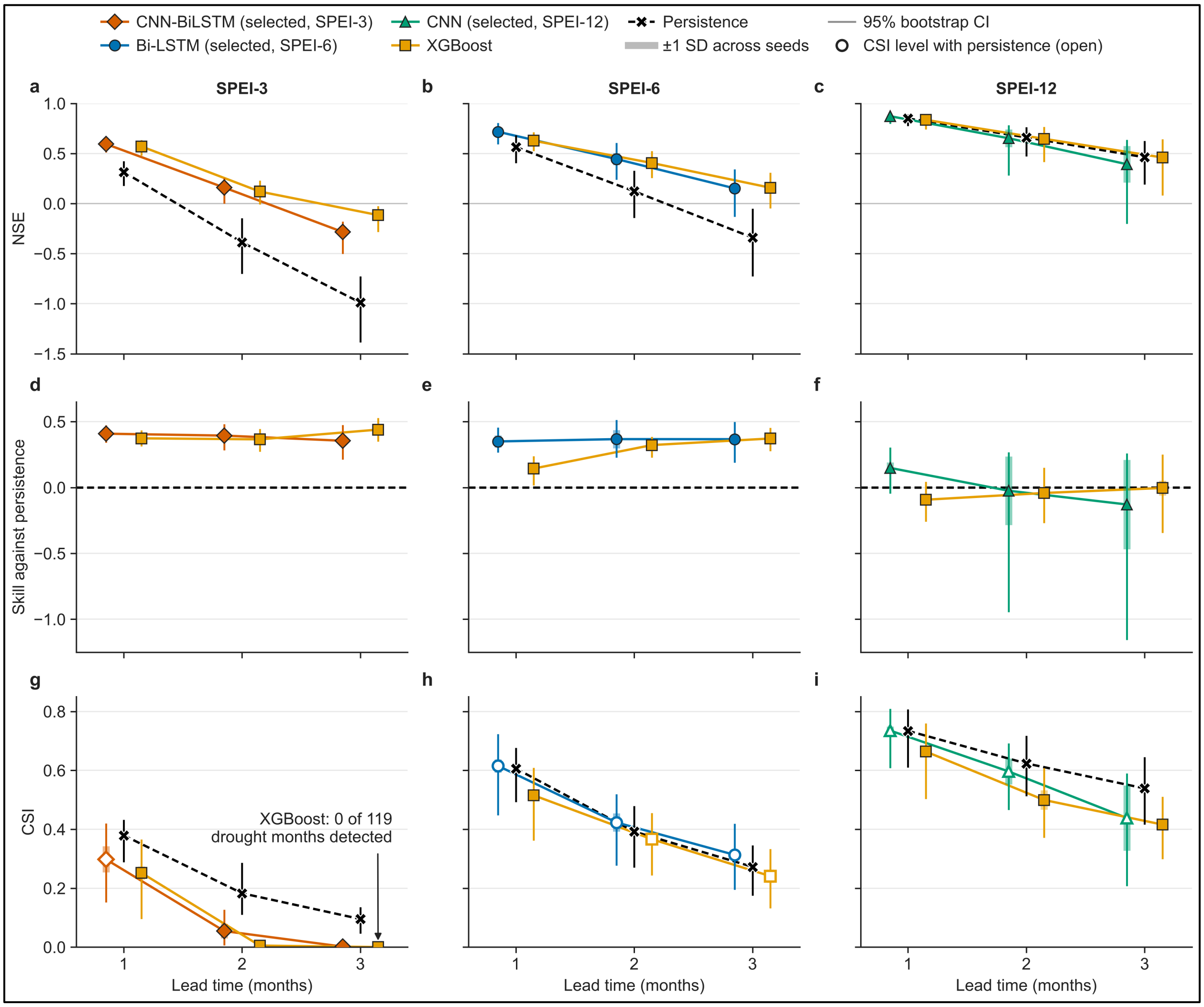

**Fig. 12.** NSE (a-c), skill against persistence (d-f) and CSI (g-i) against lead time for the selected model of each timescale, XGBoost and persistence. Markers are seed means, thick bars ±1 SD and thin bars 95% bootstrap intervals; in the CSI row, filled markers differ significantly from persistence. The annotation in (g) marks XGBoost at SPEI-3 and lead 3, which detects none of the 119 drought months. Values are given in Table 3.

**Table 3.** Multi-lead performance over the test period (2019-2024) of the selected model at each timescale, XGBoost and persistence. Model values are five-seed means ± SD; skill is against persistence, with its 95% bootstrap interval. H / M / FA are mean hits, misses and false alarms for drought (SPEI ≤ −1), and the number of seeds with a defined FAR is given in parentheses when it is below five. ΔCSI is the difference from persistence, and DM p is the Diebold-Mariano test of the seed-ensemble forecast, with the number of significant single seeds. n/a, not applicable.

| SPEI | Lead | Model | RMSE | NSE | Skill [95% CI] | POD | FAR | CSI | H / M / FA | ΔCSI [95% CI] | DM p; seeds |
|---|---|---|---|---|---|---|---|---|---|---|---|
| 3 | 1 | CNN-BiLSTM | 0.627 ± 0.009 | 0.595 ± 0.011 | 0.408 ± 0.016 [0.34, 0.47] | 0.356 ± 0.056 | 0.354 ± 0.042 | 0.298 ± 0.044 | 42.4 / 76.6 / 23.2 | −0.081 [−0.231, +0.038] | <0.0001; 5/5 |
| 3 | 1 | XGBoost | 0.645 ± 0.003 | 0.571 ± 0.004 | 0.373 ± 0.006 [0.31, 0.43] | 0.284 ± 0.025 | 0.310 ± 0.043 | 0.252 ± 0.022 | 33.8 / 85.2 / 15.2 | −0.127 [−0.274, −0.016] | <0.0001; 5/5 |
| 3 | 1 | Persistence | 0.815 | 0.315 | 0 | 0.555 | 0.455 | 0.379 | 66 / 53 / 55 | n/a | n/a |
| 3 | 2 | CNN-BiLSTM | 0.902 ± 0.007 | 0.159 ± 0.014 | 0.394 ± 0.010 [0.28, 0.48] | 0.061 ± 0.029 | 0.641 ± 0.118 | 0.054 ± 0.025 | 7.2 / 111.8 / 12.6 | −0.128 [−0.259, −0.018] | 0.0001; 5/5 |
| 3 | 2 | XGBoost | 0.923 ± 0.005 | 0.120 ± 0.010 | 0.366 ± 0.007 [0.27, 0.44] | 0.005 ± 0.005 | 0.250 ± 0.433 (3) | 0.005 ± 0.005 | 0.6 / 118.4 / 0.6 | −0.178 [−0.287, −0.102] | 0.0001; 5/5 |
| 3 | 2 | Persistence | 1.159 | −0.387 | 0 | 0.319 | 0.701 | 0.183 | 38 / 81 / 89 | n/a | n/a |
| 3 | 3 | CNN-BiLSTM | 1.115 ± 0.004 | −0.283 ± 0.008 | 0.355 ± 0.004 [0.21, 0.47] | 0.002 ± 0.004 | 0.889 ± 0.192 (3) | 0.002 ± 0.004 | 0.2 / 118.8 / 1.2 | −0.094 [−0.134, −0.045] | 0.0018; 5/5 |
| 3 | 3 | XGBoost | 1.039 ± 0.023 | −0.115 ± 0.049 | 0.439 ± 0.025 [0.35, 0.53] | 0.000 ± 0.000 | 1.000 (1) | 0.000 ± 0.000 | 0.0 / 119.0 / 0.2 | −0.095 [−0.136, −0.046] | <0.0001; 5/5 |
| 3 | 3 | Persistence | 1.388 | −0.989 | 0 | 0.185 | 0.836 | 0.095 | 22 / 97 / 112 | n/a | n/a |
| 6 | 1 | Bi-LSTM | 0.491 ± 0.017 | 0.717 ± 0.020 | 0.350 ± 0.045 [0.26, 0.46] | 0.714 ± 0.020 | 0.184 ± 0.026 | 0.615 ± 0.023 | 111.4 / 44.6 / 25.2 | +0.009 [−0.087, +0.085] | 0.0001; 5/5 |
| 6 | 1 | XGBoost | 0.564 ± 0.006 | 0.627 ± 0.008 | 0.144 ± 0.019 [0.02, 0.24] | 0.606 ± 0.014 | 0.226 ± 0.014 | 0.515 ± 0.016 | 94.6 / 61.4 / 27.6 | −0.091 [−0.169, −0.027] | 0.0730; 1/5 |
| 6 | 1 | Persistence | 0.609 | 0.565 | 0 | 0.769 | 0.259 | 0.606 | 120 / 36 / 42 | n/a | n/a |
| 6 | 2 | Bi-LSTM | 0.687 ± 0.039 | 0.445 ± 0.061 | 0.367 ± 0.070 [0.23, 0.51] | 0.506 ± 0.046 | 0.276 ± 0.041 | 0.423 ± 0.031 | 79.0 / 77.0 / 30.6 | +0.031 [−0.067, +0.111] | 0.0003; 5/5 |
| 6 | 2 | XGBoost | 0.712 ± 0.006 | 0.405 ± 0.010 | 0.322 ± 0.012 [0.23, 0.39] | 0.433 ± 0.017 | 0.297 ± 0.014 | 0.366 ± 0.011 | 67.6 / 88.4 / 28.6 | −0.026 [−0.101, +0.056] | 0.0024; 5/5 |
| 6 | 2 | Persistence | 0.865 | 0.123 | 0 | 0.583 | 0.455 | 0.392 | 91 / 65 / 76 | n/a | n/a |
| 6 | 3 | Bi-LSTM | 0.851 ± 0.025 | 0.152 ± 0.050 | 0.367 ± 0.038 [0.19, 0.50] | 0.360 ± 0.018 | 0.294 ± 0.035 | 0.313 ± 0.015 | 56.2 / 99.8 / 23.6 | +0.041 [−0.045, +0.149] | 0.0008; 5/5 |
| 6 | 3 | XGBoost | 0.847 ± 0.010 | 0.158 ± 0.019 | 0.372 ± 0.014 [0.28, 0.45] | 0.273 ± 0.034 | 0.314 ± 0.071 | 0.241 ± 0.022 | 42.6 / 113.4 / 20.4 | −0.031 [−0.119, +0.061] | 0.0004; 5/5 |
| 6 | 3 | Persistence | 1.069 | −0.340 | 0 | 0.449 | 0.591 | 0.272 | 70 / 86 / 101 | n/a | n/a |
| 12 | 1 | CNN | 0.285 ± 0.007 | 0.872 ± 0.006 | 0.148 ± 0.042 [−0.04, 0.30] | 0.810 ± 0.016 | 0.111 ± 0.018 | 0.735 ± 0.014 | 165.2 / 38.8 / 20.8 | +0.001 [−0.043, +0.049] | 0.0062; 2/5 |
| 12 | 1 | XGBoost | 0.323 ± 0.006 | 0.836 ± 0.006 | −0.093 ± 0.038 [−0.26, 0.04] | 0.715 ± 0.017 | 0.096 ± 0.009 | 0.664 ± 0.012 | 145.8 / 58.2 / 15.6 | −0.070 [−0.138, −0.021] | 0.4489; 0/5 |
| 12 | 1 | Persistence | 0.309 | 0.850 | 0 | 0.853 | 0.159 | 0.734 | 174 / 30 / 33 | n/a | n/a |
| 12 | 2 | CNN | 0.468 ± 0.056 | 0.652 ± 0.088 | −0.025 ± 0.261 [−0.95, 0.27] | 0.638 ± 0.063 | 0.093 ± 0.039 | 0.596 ± 0.043 | 130.2 / 73.8 / 14.0 | −0.027 [−0.151, +0.063] | 0.2500; 0/5 |
| 12 | 2 | XGBoost | 0.474 ± 0.010 | 0.647 ± 0.015 | −0.042 ± 0.043 [−0.27, 0.15] | 0.536 ± 0.039 | 0.120 ± 0.017 | 0.499 ± 0.033 | 109.4 / 94.6 / 15.0 | −0.124 [−0.208, −0.040] | 0.8805; 0/5 |
| 12 | 2 | Persistence | 0.465 | 0.661 | 0 | 0.779 | 0.243 | 0.624 | 159 / 45 / 51 | n/a | n/a |
| 12 | 3 | CNN | 0.617 ± 0.088 | 0.393 ± 0.182 | −0.129 ± 0.339 [−1.16, 0.26] | 0.473 ± 0.139 | 0.112 ± 0.055 | 0.438 ± 0.110 | 96.4 / 107.6 / 14.0 | −0.101 [−0.343, +0.068] | 0.7110; 1/5 |
| 12 | 3 | XGBoost | 0.586 ± 0.016 | 0.460 ± 0.030 | −0.004 ± 0.056 [−0.34, 0.25] | 0.443 ± 0.013 | 0.129 ± 0.023 | 0.416 ± 0.015 | 90.4 / 113.6 / 13.4 | −0.123 [−0.196, −0.029] | 0.9067; 0/5 |
| 12 | 3 | Persistence | 0.585 | 0.462 | 0 | 0.716 | 0.315 | 0.539 | 146 / 58 / 67 | n/a | n/a |

### 3.7 At SPEI-12, the sign of skill depends on the seed

At SPEI-12 the seed spread is large enough to change the conclusion (Fig. 13). At lead 1 the five CNN runs agree on a small positive skill, between +0.085 and +0.190, although the gain is not robust (Section 3.6). The other architectures do not agree on the sign of skill even at lead 1 (Fig. 13c). The Bi-LSTM ranges from −0.010 to +0.266 (positive in four of five seeds), the CEEMDAN-BiLSTM from −0.154 to +0.087 (two of five) and XGBoost from −0.131 to −0.030 (none), while the CNN-BiLSTM is positive in all five seeds (+0.025 to +0.191).

At leads 2 and 3, one seed in five produces a badly degraded CNN: seed 43 at lead 2 (NSE 0.497, skill −0.483) and seed 46 at lead 3 (NSE 0.084, skill −0.702). The number of seeds with positive skill falls from five at lead 1 to four at lead 2 and three at lead 3, and the skill range widens to −0.483 to +0.140 and −0.702 to +0.170, so a single training run could have reported a skill of either sign at these leads.

By Eq. (18), any difference in NSE between two runs is multiplied by $1/(1 - NSE_p)$ when it is expressed as skill. This factor is 6.65 at SPEI-12 and lead 1, and 2.95 and 1.86 at leads 2 and 3, compared with 1.46 at SPEI-3 and 2.30 at SPEI-6 at lead 1. The seed SD of the CNN's NSE (0.006, 0.088 and 0.182 at leads 1-3) thus becomes a skill SD of 0.042, 0.261 and 0.339.

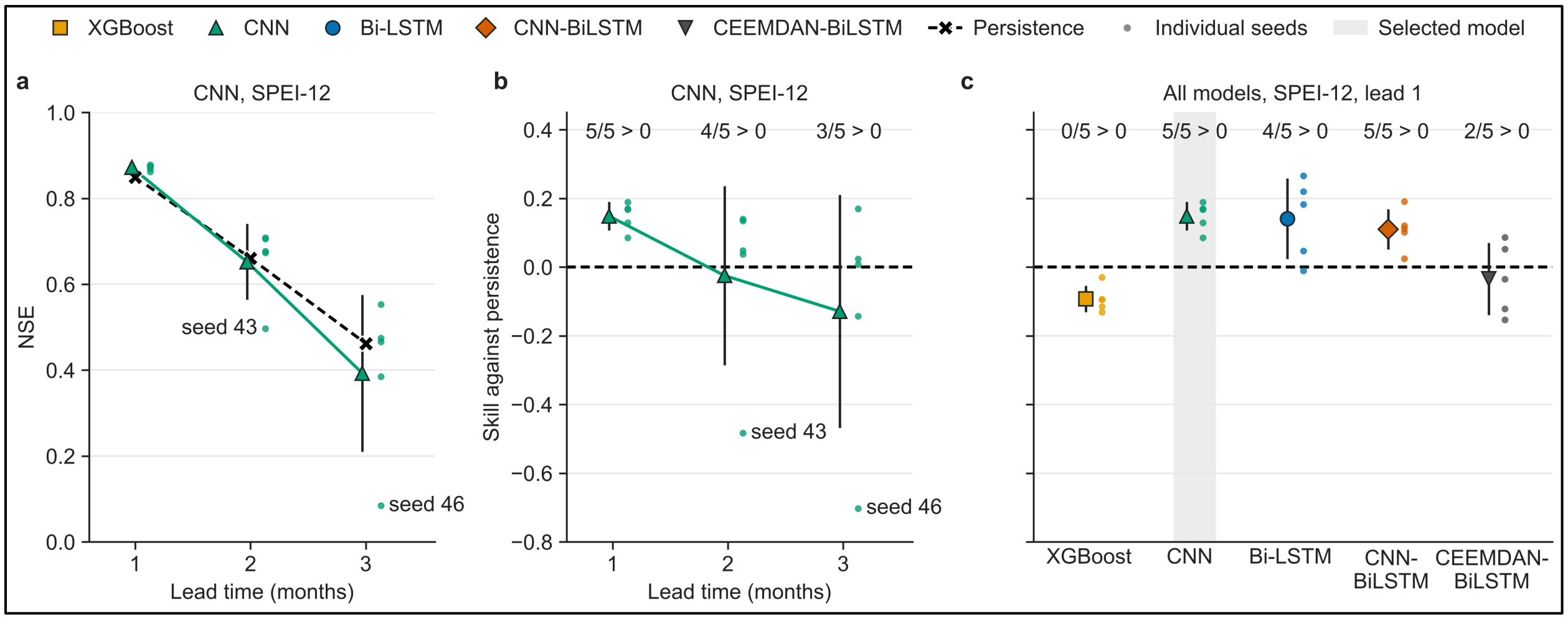

**Fig. 13.** Seed dependence at SPEI-12: (a) NSE and (b) skill of the CNN at leads 1-3, and (c) lead-1 skill of all five models. Dots are seeds 42-46 and markers with bars the mean ± 1 SD; the two degraded runs are labeled, and "n/5 > 0" gives the number of seeds with positive skill. Persistence is shown by crosses in (a) and by the zero line in (b) and (c).

### 3.8 Spatial pattern and the independent station target

The selected models reproduce the spatial pattern of drought across the eight divisions in the two test months examined (Figs. 14 and S3; Table S16). The spatial correlation between the forecast and observed fields is 0.932 ± 0.007, 0.912 ± 0.016 and 0.854 ± 0.054 at SPEI-3, SPEI-6 and SPEI-12 in the worst observed drought month (August 2022, and June 2023 at SPEI-12), and 0.927-0.970 in the near-normal month. The severity of the worst month is under-forecast, with the forecast national mean reaching 60%, 72% and 75% of the observed value. The placement of the driest division and the few sign reversals are given in Table S16.

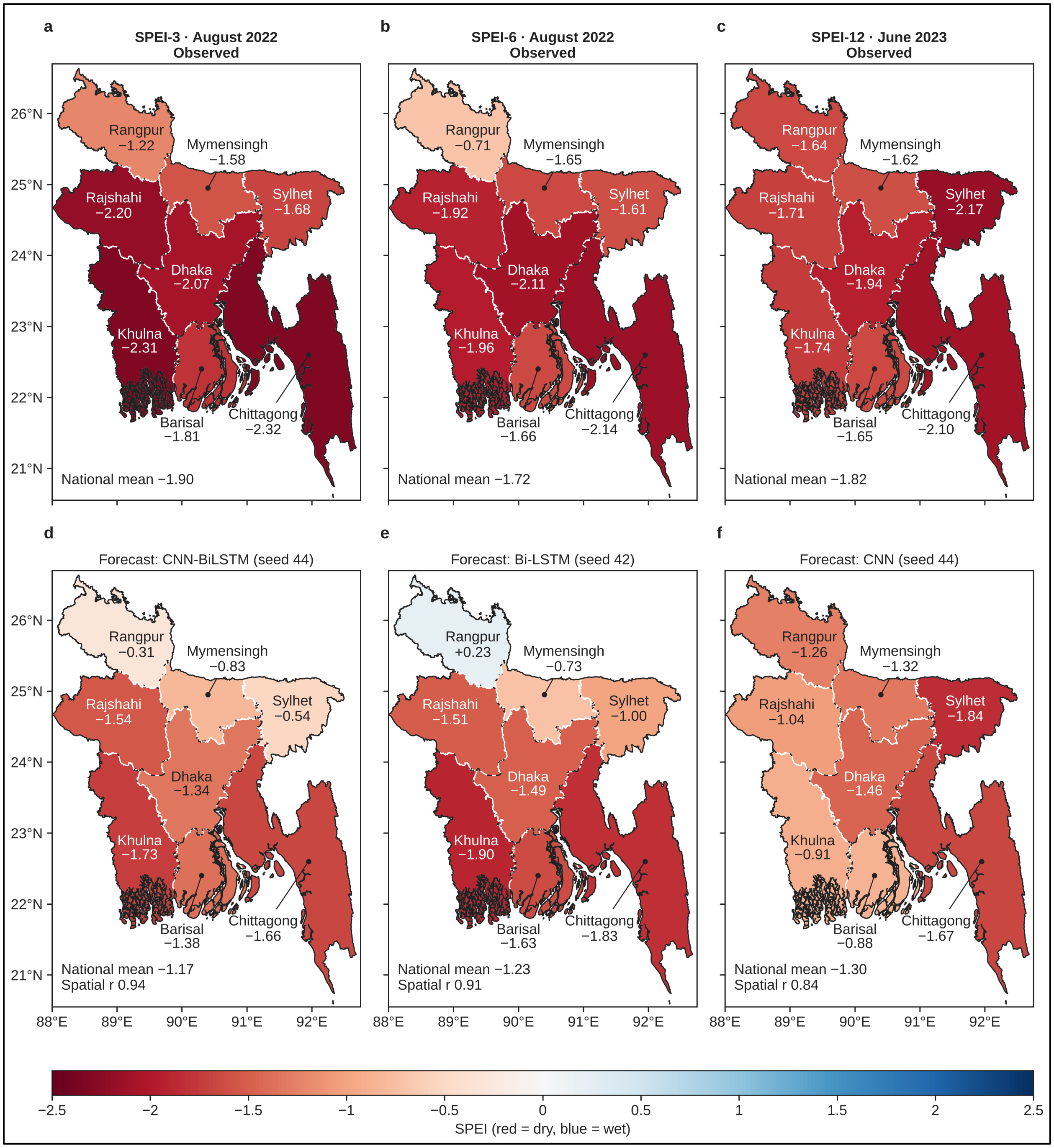

**Fig. 14.** Spatial verification in the worst observed drought month of the test period: (a, d) SPEI-3, August 2022; (b, e) SPEI-6, August 2022; (c, f) SPEI-12, June 2023. The top row shows the observed index and the bottom row the lead-1 forecast of the selected model (median-skill seed). Each division is labeled with its value, the national mean is printed in every panel and the spatial correlation with the observed field in the forecast panels. Red denotes dry and blue wet conditions. The summary over five seeds is given in Table S16.

Re-scored without retraining against SPEI derived entirely from station precipitation, the selected models keep most of their correlation (Table 4). Over the common period 2019-2022 (384 division-months), the correlation falls from 0.797, 0.893 and 0.938 against the ERA5-derived target to 0.639, 0.698 and 0.710 against the station-derived target, so the forecasts retain 80%, 78% and 76% of their correlation with a target they never saw during training. NSE falls from 0.633-0.877 to 0.405-0.471, and skill falls from 0.419, 0.374 and 0.238 to 0.298, 0.121 and 0.095, with the largest loss at the long windows.

**Table 4.** Validation of the selected models against an independent station-derived target, 2019-2022 (384 division-months), without retraining. Model values are five-seed means ± SD; skill is measured against the same ERA5 persistence forecast for both targets, and drought is SPEI ≤ −1.

| Timescale | Target | Forecast | r | RMSE | NSE | Skill | POD | FAR |
|---|---|---|---|---|---|---|---|---|
| SPEI-3 | ERA5-derived | CNN-BiLSTM | 0.797 ± 0.007 | 0.564 ± 0.008 | 0.633 ± 0.011 | 0.419 ± 0.017 | 0.369 ± 0.042 | 0.256 ± 0.054 |
| | | Persistence | 0.685 | 0.740 | 0.369 | 0 | 0.583 | 0.432 |
| | Station-derived | CNN-BiLSTM | 0.639 ± 0.008 | 0.791 ± 0.008 | 0.405 ± 0.012 | 0.298 ± 0.014 | 0.272 ± 0.027 | 0.323 ± 0.012 |
| | | Persistence | 0.538 | 0.945 | 0.152 | 0 | 0.483 | 0.419 |
| SPEI-6 | ERA5-derived | Bi-LSTM | 0.893 ± 0.005 | 0.410 ± 0.011 | 0.790 ± 0.011 | 0.374 ± 0.034 | 0.782 ± 0.016 | 0.165 ± 0.020 |
| | | Persistence | 0.833 | 0.518 | 0.665 | 0 | 0.775 | 0.210 |
| | Station-derived | Bi-LSTM | 0.698 ± 0.006 | 0.740 ± 0.009 | 0.459 ± 0.013 | 0.121 ± 0.021 | 0.642 ± 0.031 | 0.289 ± 0.021 |
| | | Persistence | 0.661 | 0.790 | 0.385 | 0 | 0.604 | 0.360 |
| SPEI-12 | ERA5-derived | CNN | 0.938 ± 0.005 | 0.278 ± 0.013 | 0.877 ± 0.011 | 0.238 ± 0.070 | 0.766 ± 0.028 | 0.081 ± 0.010 |
| | | Persistence | 0.919 | 0.319 | 0.839 | 0 | 0.824 | 0.156 |
| | Station-derived | CNN | 0.710 ± 0.008 | 0.627 ± 0.005 | 0.471 ± 0.008 | 0.095 ± 0.013 | 0.613 ± 0.020 | 0.299 ± 0.005 |
| | | Persistence | 0.685 | 0.659 | 0.416 | 0 | 0.681 | 0.336 |

### 3.9 SHAP-Based Interpretation of the Selected Models

Fig. 15a-c shows the mean absolute SHAP values of the three selected models over the five seeds. At SPEI-3 and SPEI-6 the same three predictors occupy the top three ranks in every seed. At SPEI-12 only the first rank is fixed, with lagged SPEI first in all five seeds and precipitation second in four; lower ranks change between seeds and are not interpreted further (Table S17).

The ranking moves from meteorological inputs to drought memory as the window lengthens. At SPEI-3, precipitation (0.176 ± 0.013) and lagged SPEI (0.173 ± 0.017) are tied, followed by solar radiation (0.084 ± 0.012). At SPEI-6, precipitation leads in all five seeds (0.114 ± 0.008), and lagged SPEI (0.070 ± 0.009) and solar radiation (0.069 ± 0.003) share second place. At SPEI-12, lagged SPEI is clearly first (0.128 ± 0.043), ahead of precipitation (0.066 ± 0.013). The shares of the meteorological predictors alone are shown in Fig. S4.

Where all five seeds agree on the sign (colors in Fig. 15a-c), lagged SPEI and precipitation have a positive mean effect at every timescale, and maximum temperature, wind speed and cloud cover a negative one at SPEI-12. The beeswarm plots of the median-skill seeds (Fig. 15d-f) show that, at all three timescales, high lagged SPEI values push the forecast down and low values push it up. The mean effect of lagged SPEI is largest at SPEI-12 (+0.107 ± 0.036).

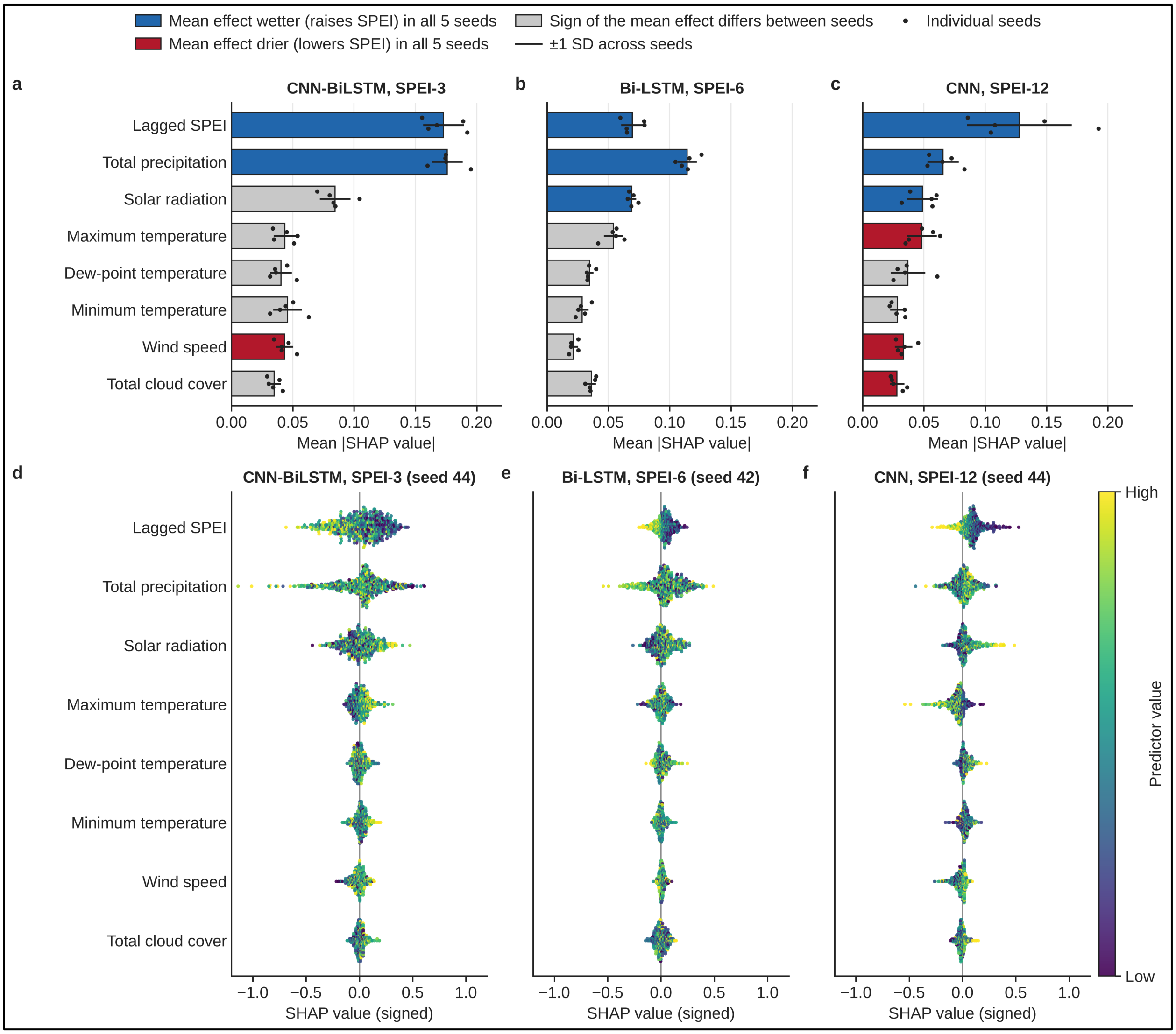

**Fig. 15.** SHAP interpretation of the selected models: CNN-BiLSTM at SPEI-3 (a, d), Bi-LSTM at SPEI-6 (b, e) and CNN at SPEI-12 (c, f). (a-c) Mean absolute SHAP values; bars are means over five seeds, whiskers ±1 SD and dots seeds, and colors give the sign of the mean signed value where all seeds agree (blue, wetter; red, drier; gray, the sign differs between seeds). (d-f) Beeswarm plots of the median-skill seeds (44, 42 and 44). Each dot is one test division-month, placed at its signed SHAP value summed over the input window and colored by the window mean of the predictor, scaled between its 5th and 95th percentiles. Positive values move the forecast toward wetter conditions.

# 4. Discussion

This study forecast SPEI-3, SPEI-6 and SPEI-12 one to three months ahead for the eight divisions of Bangladesh using one machine-learning model (XGBoost) and four deep-learning models (CNN, Bi-LSTM, CNN-BiLSTM and CEEMDAN-BiLSTM). Each forecast was evaluated by accuracy (NSE), skill against persistence and drought detection (POD, CSI and FAR). The central result of this study is that accuracy, skill against persistence and drought detection do not give the same verdict on the same forecasts, and that the disagreement grows with lead time. NSE ranks SPEI-12 first at every lead, whereas skill at lead 1 reverses the order. At SPEI-3 the skill of the selected model (CNN-BiLSTM) barely changes over three leads while its detection collapses, and XGBoost reaches the highest skill in the study (0.439) without detecting any of the 119 drought months. Only at SPEI-6 does a forecast keep a significant squared-error gain over persistence and remain level with it in detection at leads of two and three months.

The divergence follows from what a squared-error criterion rewards. When the information available to a model weakens, the forecast that minimizes the mean squared error moves towards the mean, and its spread shrinks with its correlation to the observations (Gupta et al., 2009; Murphy, 1988). The selected models (CNN-BiLSTM at SPEI-3, Bi-LSTM at SPEI-6 and CNN at SPEI-12) behave in this way, with SD ratios of 0.775 to 0.941 and a predicted month-to-month change of only 0.56 to 0.62 of the observed amplitude. Persistence does the opposite. It repeats an observed value and keeps the full variance, but its error grows quickly as the autocorrelation of the index decays with lead (RMSE up by 70% to 89% from lead 1 to lead 3), so a skill score measured against it can rise simply because the reference gets worse. The case is clearest at SPEI-3 and lead 3, where both models have a negative NSE (−0.283 for the CNN-BiLSTM and −0.115 for XGBoost). A constant forecast equal to the test-period mean would have a lower squared error than either model and, by Eq. (18), a skill of 0.497, higher than any forecast in the study, so a positive skill score at this lead is no evidence of predictive information. At SPEI-6 and lead 3 the same constant forecast would score 0.253, whereas the Bi-LSTM reaches 0.367 with a positive NSE of 0.152, so there the gain is genuine. The contraction also explains the detection results: a forecast pulled towards the mean crosses the drought threshold less often, so every model forecast fewer drought months than were observed, and persistence, which does not contract, keeps the highest POD at every timescale and lead.

Where this contraction becomes decisive is set by the accumulation window. A predictor observed k months before the target month shares months with the target's window only while k is shorter than the window, so the lead at which the predictor correlations collapse is fixed by the construction of the index rather than by the climate (Section 3.2). At SPEI-3 the newest predictors therefore carry information at leads 1 and 2 but none at lead 3. At that lead the CSI of the CNN-BiLSTM is 0.002, after a fall to 0.054 at lead 2 as the shared part of the window shrinks from two months to one, while its skill score, held up by the failing reference, shows no corresponding change. XGBoost loses its detection at the same leads (CSI 0.005 and 0.000), so the limit comes from the index rather than from the models. At SPEI-6 the window still overlaps the predictors at all three leads, and this is where both skill and detection hold. The useful horizon of a forecast can thus be bounded before any model is trained, by comparing the lead with the accumulation length; the trained models confirmed this bound rather than revealed it.

At SPEI-12 the sign of skill changes with the seed, and this instability is itself a result: it shows that little predictable signal remains beyond persistence, a dependence that only a multi-seed design can reveal. Persistence already explains 85% of the variance at lead 1, and the identity that links skill to NSE magnifies any difference in NSE between runs by a factor of 6.65 (Section 3.7), so ordinary differences between training runs become large differences in skill. Because the deep-learning models predict the change from persistence, a SHAP value describes how an input moves the forecast away from the last observed SPEI, and the lagged index acts mainly as a pull towards normal conditions. It leads clearly only at SPEI-12 (0.128 against 0.066 for precipitation), where its positive mean effect reflects upward corrections in a test period with 204 drought months out of 576. At the long window the models have largely learned how fast an anomaly fades, which agrees with their compressed spread.

Persistence is a strong baseline that costs nothing, and, to our knowledge, earlier accuracy figures for Bangladesh were reported without it. On our test period it reaches an $R^2$ of 0.432, 0.611 and 0.855 at SPEI-3, SPEI-6 and SPEI-12 (Table 2). Although data and test periods differ, this exceeds the best value of Al Mamun et al. (2024) at SPEI-6 (0.512) and nearly matches it at SPEI-12 (0.871), and their estimate lies well above persistence only at SPEI-3 (0.602), from a random split that estimated rather than forecast SPEI. Our lead-1 forecasts reach 0.595, 0.729 and 0.874. Hossain, Begum, Akhtar, Talukder, et al. (2025) reported $R^2$ of 0.914 to 0.965 in central Bangladesh with an 80/20 split, and other studies in the country worked at the scale of a single region or city, including the south-west with PROPHET and SARIMA models (Hossain, Rahman, et al., 2025), a semi-arid region where a hybrid model outperformed individual models (Mahmud et al., 2026), Rangpur with machine-learning prediction of SPI and SPEI (Akter et al., 2025), and Dhaka with traditional and custom models (Mridul et al., 2025). Yaseen et al. (2021)

and Elbeltagi et al. (2022) forecast or estimated a single index, SPI and SPEI respectively, with data-driven models. Without a persistence reference, however, an $R^2$ value cannot show how much a model adds to the memory of the index.

The same pattern appears beyond Bangladesh. Poudel et al. (2024) obtained $R^2$ above 0.9 for SPEI-3 when the input lags spanned the accumulation window, and needed 6- and 12-month lags at SPEI-6 and SPEI-12, as the overlap structure of the index predicts. Luz et al. (2026) found that models tuned on average error metrics tend to smooth extremes, and the compressed spread and under-forecast droughts found here are that smoothing in measurable form. Detection is difficult elsewhere too: climate models detected fewer than 30% of global drought onsets (Yuan & Wood, 2013), against 28% to 43% of drought months for our SPEI-3 models at lead 1. Decomposition hybrids are often reported to outperform single models (Ding et al., 2022), but decomposing the whole series before the split lets future values enter the training inputs (Yang et al., 2024); kept inside each input window, as here, the decomposition brought no measurable advantage.

Taken together, the study fills four gaps. The first is regional: to our knowledge, these are the first SPEI forecasts for all eight divisions of Bangladesh at three timescales and one to three months ahead, the first from bidirectional and hybrid deep-learning models (Bi-LSTM, CNN-BiLSTM and CEEMDAN-BiLSTM) in the country, and the first judged against persistence and as drought-event detectors. The shared, leakage-controlled pipeline makes the model comparison fair, and it shows that the value of deep learning depends on the timescale: the selected models, the CNN-BiLSTM at SPEI-3, the Bi-LSTM at SPEI-6 and the CNN at SPEI-12, beat XGBoost clearly at SPEI-6 (+0.206 in skill) and SPEI-12 (+0.241), but not at SPEI-3, so no single architecture was best at every timescale. The second gap is diagnostic. Studies that report accuracy alone, or compare models only with one another (Al Reshan et al., 2025; Luz et al., 2026), cannot tell a forecast that adds information from one that inherits it from the index, whereas scoring all three views together shows where they disagree. The third gap concerns lead time. Longer horizons are seldom explored (Luz et al., 2026), and no study had reported how forecast quality in Bangladesh changes from one to three months ahead; here only SPEI-6 keeps genuine value beyond one month, at a horizon the window-overlap rule anticipates before training. The fourth gap is uncertainty. Single training runs and full-series decomposition can both make a forecast look better than it is (Bouthillier et al., 2021; Yang et al., 2024). Here five seeds, block-bootstrap intervals, Diebold-Mariano tests, a selection rule fixed in advance and closed or tested routes of leakage reveal that SPEI-12 skill depends on the seed, which no single-run study could show, and make the remaining gains at SPEI-3 and SPEI-6 credible.

For forecasting practice, the main implication is that a drought forecast should be reported through all three views together with the scores of the reference forecast. A positive skill score with a negative NSE signals that the reference has failed, not that the model has skill, and skill at different leads should not be compared without the persistence scores beside it. Differences between architectures were often smaller than the spread between training runs of one model (0.350 ± 0.045 for the Bi-LSTM against 0.327 ± 0.059 for the CNN at SPEI-6, and 0.148 ± 0.042 for the CNN against 0.141 ± 0.118 for the Bi-LSTM at SPEI-12), so choosing the highest mean would select a favorable draw rather than a better model. For this reason, the selection rule was fixed before the five-seed runs: a challenger had to exceed the earlier choice by more than one seed SD, a margin that could not be adjusted once the results were known. No challenger cleared the margin, which shows that the leading architectures are equivalent at SPEI-6 and SPEI-12 and that the timescale and the lead matter more for forecast value than the choice among these models.

For drought early warning in Bangladesh, the results suggest a clear division of roles. SPEI-6 forecasts carry genuine information one to three months ahead and match persistence in detection with far fewer false alarms (FAR 0.184 to 0.294 against 0.259 to 0.591), so they are the natural basis for advance notice. SPEI-3 forecasts carry warning value at a one-month lead only; at lead 2 they catch 7.2 of the 119 drought months on average. At SPEI-12 no lead showed skill that held up under both the bootstrap and the seeds, so SPEI-12 is better treated as a monitoring index than as a forecast target. An operational system should adopt a model only if it beats persistence in detection as well as in squared error. Used in this way, divisional SPEI forecasts can support early warning and adaptive capacity under SDG 13 (targets 13.1 and 13.3), drought-resilient agriculture under SDG 2 (target 2.4), water-allocation planning under SDG 6 (target 6.4) and the management of drought-affected land under SDG 15 (target 15.3) (United Nations General Assembly, 2015), as well as the Sendai Framework target of wider access to multi-hazard early warning systems by 2030 (United Nations Office for Disaster Risk Reduction, 2015).

Several limitations qualify these results. The most direct concerns the test set: the predictor screening and the SPEI distribution used the full 1980-2024 record, which includes the test years. The screening selected inputs and fitted no model parameter, and refitting the distribution on the training years with training-period climatologies changed the lead-1 skill of the Bi-LSTM at SPEI-6 only from 0.350 to 0.354, well inside its seed SD of 0.045 (Table S18), so the full-record fit does not drive the results. Among the modeling choices, one bears directly on the central result. The forecasts are deterministic and were trained with a Huber loss (Huber, 1964), which is quadratic for errors smaller than one index unit and therefore rewards the same contraction towards the mean that separates accuracy, skill and detection (Gupta et al., 2009; Murphy, 1988). A different loss would change how the models trade spread for squared error, but not the diagnosis, because the skill that a constant forecast earns against a failing persistence reference (0.497 at SPEI-3 and lead 3) does not depend on how any model is trained. Hyperparameters were tuned at lead 1 and reused at leads 2 and 3; because the gain from tuning is of the same size as the variation between seeds (Table S10), the cost of reusing the lead-1 settings is of the same order as the seed spread already reported. The target itself is only partly verified. Because the station temperature records failed screening (Section 3.1), PET rests on ERA5 alone, whose near-surface temperature is reproduced better than its precipitation (Liu et al., 2024), and the station-derived target, which covers 2019-2022, checks only the precipitation side of the index. The forecast conclusions do not depend on this, because interannual variation in PET explains only 4.2% to 5.3% of the variance of the forecast targets, close to the 6% to 13% reported for another basin (Zheng et al., 2022), so the index is governed by precipitation, the variable that passed validation. Finally, the test window spans 72 months. This limits every skill estimate, above all at SPEI-12, where the intervals are widest, and it reflects a single sequence of wet and dry years. That is why the conclusions are framed as patterns across timescale and lead rather than as exact skill values: the patterns rest on the window overlap and on the strength of persistence, which follow mainly from the construction of the index rather than from the six test years. The study also covers one country, an index derived from reanalysis and one test period. The specific skill values may not transfer to other regions, but the window-overlap rule should, because it follows from how an accumulated index is built.

These limits point to the next steps. Training that preserves the observed spread or targets the drought threshold directly, and probabilistic forecasts built from seed ensembles, may recover the detection lost to contraction, and they should be judged through the same three views. Longer leads will need drivers that precede the index by several months. Large-scale climate models such as the El Niño-Southern Oscillation and the Indian Ocean Dipole were outside the scope of this study and are the natural candidates. The forecasts were designed for the eight divisions, the scale at which the 35 stations support validation; more stations with complete meteorological records would allow the evaporative term to be validated and the forecasts to be extended to districts. Attention-based architectures can then be tested within the same evaluation framework.

## 5. Conclusion

This study forecast the Standardized Precipitation Evapotranspiration Index for the eight divisions of Bangladesh at three timescales and one to three months ahead, with five models trained under five seeds and judged by accuracy, skill against persistence and drought detection. At a one-month lead the CNN-BiLSTM performed best at SPEI-3, the Bi-LSTM at SPEI-6 and the CNN at SPEI-12, although at the two longer timescales other deep-learning models were statistically equivalent. To our knowledge, these are the first such forecasts for the country, and the first in which bidirectional and hybrid deep-learning models are judged against persistence and as drought detectors. The three views of forecast quality do not agree. NSE ranks SPEI-12 first, whereas skill at lead 1 reverses the order (0.408, 0.350 and 0.148 for SPEI-3, SPEI-6 and SPEI-12), and XGBoost at SPEI-3 and lead 3 reaches a skill of 0.439 without detecting any of the 119 drought months, while a constant forecast would score 0.497. SPEI-6 is the only timescale at which the forecasts keep a significant gain over persistence at all three leads and match it in detection with fewer false alarms, a horizon set by the overlap between the accumulation window and the predictors and therefore known before any model is trained. At SPEI-12 the sign of skill changes with the seed, which shows that little predictable signal remains beyond persistence. The main results survive two independent checks: the forecasts keep 76% to 80% of their correlation against a station-derived target they never saw, and refitting the index on the training years changes the lead-1 skill at SPEI-6 only from 0.350 to 0.354. In practice, SPEI-6 is the basis for one to three months of drought notice in Bangladesh, SPEI-3 for warnings one month ahead, and SPEI-12 for monitoring rather than forecasting. Drought forecasts should report accuracy, skill and detection together with the scores of persistence, and models should be compared across several seeds before one is chosen.

## Declarations

**Funding:** None.

**Conflicts of interest:** The authors declare that they have no known competing financial interests or personal relationships that could have appeared to influence the work reported in this paper.

**Ethical approval:** No ethical approval was required for this study.

**Data and code availability:** ERA5 monthly data are available from the Copernicus Climate Change Service Climate Data Store (Hersbach et al., 2023). The BMD station compilation is available from Mendeley Data (Diptho et al., 2026). The processed data and model outputs of this study are available on Zenodo under the CC BY 4.0 license (https://doi.org/10.5281/zenodo.24115493). The code used in this study is available on GitHub (https://github.com/mahirahnaf1727/bangladesh-spei-drought-forecasting) and archived on Zenodo under the MIT license (https://doi.org/10.5281/zenodo.24120523).

### CRediT authorship contribution statement

Mahir Ahnaf: Conceptualization, Methodology, Software, Formal analysis, Data curation, Visualization, Writing - original draft.

Showmitra Kumar Sarkar: Conceptualization, Supervision, Writing - review and editing.

### Declaration of generative AI use

During the preparation of this work, the authors used Claude (Anthropic) for language editing and for assistance with code debugging. The authors reviewed, verified, and edited all AI-assisted output and take full responsibility for the content of the published article.

## Supplementary Information

# Evaluation of machine-learning and deep-learning SPEI drought forecasts against persistence across timescales and lead times in Bangladesh

Mahir Ahnaf[1*], Showmitra Kumar Sarkar[2]

[1]Graduate Student, Department of Urban and Regional Planning, Khulna University of Engineering & Technology (KUET), Khulna-9203, Bangladesh

[2]Assistant Professor, Department of Urban and Regional Planning, Khulna University of Engineering & Technology (KUET), Khulna-9203, Bangladesh

[*]Corresponding author: Mahir Ahnaf (ahnf.mahir@gmail.com)

## Contents of this file

## Introduction

This file gives the methodological details, figures and tables that support the main text. Equations, figures and tables numbered with the prefix S belong to this file, and those without it refer to the main text. Abbreviations follow the main text. Model scores refer to the test period, January 2019 to December 2024, unless stated otherwise, and model values are the mean ± standard deviation (SD) over seeds 42-46.

**List of contents**

**Text S1. Computation of the drought index**

The nine ERA5 variables and the unit conversions applied to them are listed in Table S1. Wind speed at 10 m was computed from the two horizontal components as

$$U_{10} = \sqrt{u_{10}^2 + v_{10}^2} \quad \text{(S1)}$$

where $u_{10}$ and $v_{10}$ are the zonal and meridional components (m $s^{-1}$). The converted fields were clipped to the national boundary with a polygon mask that set the cells outside the border to missing without removing them from the array, which leaves 274 grid cells with valid data. Each cell was assigned to the division whose polygon contains its center, and the cells of a division were averaged into one monthly series per variable. Working at the divisional scale also keeps strongly correlated neighboring cells from falling on opposite sides of the split between training and test data.

Potential evapotranspiration was computed with the FAO-56 Penman-Monteith equation (Eq. (1) of the main text) following Allen et al. (1998), and every intermediate term was kept as an output column so that each value can be traced. The saturation vapor pressure (kPa) at air temperature T (°C) is

$$e^\circ(T) = 0.6108 \exp\left(\frac{17.27\,T}{T + 237.3}\right) \quad \text{(S2)}$$

The saturation vapor pressure of the air was taken as the mean of the values at the maximum and minimum temperatures, and the actual vapor pressure was taken from the dew-point temperature,

$$e_s = \frac{e^\circ(T_{\max}) + e^\circ(T_{\min})}{2}, \qquad e_a = e^\circ(T_{dew}) \quad \text{(S3)}$$

Because monthly averaging can produce small non-physical values, $e_a$ was capped at $e_s$, so that the vapor pressure deficit $e_s - e_a$ is never negative. The slope of the saturation vapor pressure curve (kPa °$C^{-1}$) at the mean temperature is

$$\Delta = \frac{4098\, e^\circ(T)}{(T + 237.3)^2} \quad \text{(S4)}$$

Atmospheric pressure P (kPa) was computed from the mean elevation z (m) of each division, taken from the HydroSHEDS digital elevation model (Lehner et al., 2008), rather than assumed at sea level, and the psychrometric constant $\gamma$ (kPa °$C^{-1}$) follows from it,

$$P = 101.3\left(\frac{293 - 0.0065\,z}{293}\right)^{5.26}, \qquad \gamma = 0.665 \times 10^{-3}\,P \tag{S5}$$

Wind speed was reduced from the 10 m height of ERA5 to the 2 m reference height with the logarithmic wind profile,

$$u_2 = U_{10}\,\frac{4.87}{\ln(67.8 \times 10 - 5.42)} \tag{S6}$$

and was floored at 0.5 m $s^{-1}$ for numerical stability. Extraterrestrial radiation $R_a$ (MJ $m^{-2}$ $day^{-1}$) was computed for the latitude φ of each division (rad) and the day of the year J at the middle of each month,

$$R_a = \frac{24\,(60)}{\pi}\,G_{sc}\,d_r[\omega_s \sin\varphi \sin\delta + \cos\varphi \cos\delta \sin\omega_s] \tag{S7}$$

where $G_{sc}$ = 0.0820 MJ $m^{-2}$ $min^{-1}$ is the solar constant, $d_r = 1 + 0.033\cos(2\pi J/365)$ the inverse relative distance between the Earth and the Sun, $\delta = 0.409\sin(2\pi J/365 - 1.39)$ the solar declination and $\omega_s = \arccos(-\tan\varphi\tan\delta)$ the sunset hour angle. Clear-sky radiation is $R_{so} = (0.75 + 2 \times 10^{-5}\,z)\,R_a$, so the elevation of the division enters here as well. Net radiation was computed with the full longwave balance rather than approximated by the net shortwave term,

$$R_n = (1 - 0.23)\,R_s - \sigma\,\frac{T_{\max,K}^4 + T_{\min,K}^4}{2}\left(0.34 - 0.14\sqrt{e_a}\right)\left(1.35\,\frac{R_s}{R_{so}} - 0.35\right) \tag{S8}$$

where $R_s$ is the ERA5 surface solar radiation (MJ $m^{-2}$ $day^{-1}$), 0.23 the albedo of the grass reference surface, $\sigma = 4.903 \times 10^{-9}$ MJ $K^{-4}$ $m^{-2}$ $day^{-1}$ the Stefan-Boltzmann constant, and $T_{max,K}$ and $T_{min,K}$ the maximum and minimum temperatures in kelvin. The soil heat flux was not set to zero but was computed in the monthly form of FAO-56 from the change in mean air temperature between consecutive months,

$$G_i = 0.14\,(T_i - T_{i-1}) \tag{S9}$$

where $G_i$ is in MJ $m^{-2}$ $day^{-1}$ and $T_i$ is the mean temperature of month i. Negative values of $ET_0$, which have no physical meaning, were set to zero.

The accumulated water balance (Eq. (3) of the main text) was fitted separately for each calendar month with the generalized logistic distribution (Eq. (4) of the main text), whose parameters were estimated by the method of L-moments (Hosking & Wallis, 1997). The three-parameter log-logistic distribution can represent only right-skewed samples, whereas left-skewed calendar-month samples are common in the dry season, when the evaporative term dominates the water balance. The generalized logistic distribution covers both directions of skew, and where its shape parameter is negative it takes the same form as the log-logistic distribution. The cumulative probability F(x) was converted to a standard normal value with the rational approximation of Abramowitz and Stegun (1965),

$$W = \sqrt{-2\ln p}, \qquad p = 1 - F(x) \tag{S10}$$

$$\text{SPEI} = W - \frac{c_0 + c_1 W + c_2 W^2}{1 + d_1 W + d_2 W^2 + d_3 W^3} \tag{S11}$$

where p is the probability of exceeding the accumulated water balance and the constants are $c_0$ = 2.515517, $c_1$ = 0.802853, $c_2$ = 0.010328, $d_1$ = 1.432788, $d_2$ = 0.189269 and $d_3$ = 0.001308. Where p exceeds 0.5, the transformation is applied to 1 − p and the sign of the result is reversed. The index therefore has a mean of zero and a standard deviation of one, so that dry conditions in different months and divisions can be compared directly. The drought classes used in this study are listed in Table S2.

**Text S2. Protocol of the ERA5 credibility tests**

The credibility tests were written as a self-contained recomputation from the raw ERA5 and station files, so they inherit no intermediate product of the forecasting pipeline. The station record is a compiled monthly dataset of 35 BMD stations (Diptho et al., 2026) that overlaps ERA5 from January 1980 to December 2022. The compilation stores missing data as zero rainfall. Five stations have a zero total in every year before their record begins (Ambagan until 2014, Chuadanga and Syedpur until 2002, Mongla until 1990 and Tangail until 1986), Feni and Sandwip each have one such year (2001 and 2003), and Hatiya (May-August 1993) and Kutubdia (September 1980) report 0 mm in months of the May-September rainy season. These 101 station-years and five station-months, out of 1,505 station-years in 1980-2022, were set to missing, so each divisional mean uses only the stations that reported in that month. Each station was assigned to the division that contains it, and the stations of a division were averaged into one divisional series. A point gauge and a 0.25° grid cell do not describe the same area, and averaging both records to the divisional scale makes them comparable. This gives 4,128 division-month pairs (8 divisions × 516 months).

*Screening.* The monthly climatology of each station variable was compared with the corresponding ERA5 climatology in phase and amplitude. To locate any mismatch, the circular shift k (0-11 months) that maximizes the correlation between the two climatologies was found for each station, together with the correlation reached at that shift (Table S4). A record without defects needs k = 0 and a high correlation. Precipitation meets this condition at every station, but the temperature columns do not, since their optimal shift changes from station to station (Fig. 5 of the main text). A global search over row offsets of the source file, from −420 to +420 rows, recovered the seasonal amplitude of temperature at no offset, so the defect is not a uniform misalignment that could be repaired. Only precipitation was therefore used as a reference.

*Precipitation.* Agreement for each division and for the pooled record was scored with the Pearson correlation of the raw monthly series (r), the correlation of the de-seasonalized anomalies ($r_a$), the percent bias, RMSE, NSE and KGE (Eqs. (5) and (6) of the main text; Table S3). The anomalies were obtained by subtracting the calendar-month mean of each division. The anomaly correlation is reported alongside the raw correlation because a standardized drought index responds to departures from the seasonal cycle rather than to the cycle itself. Agreement was rated with the classes of Moriasi et al. (2007).

*Dependence of SPEI on PET.* The compilation holds only precipitation and temperature, and its temperature failed the screening, so none of the inputs of PET could be checked against stations. SPEI was therefore recomputed from ERA5 with PET replaced by its calendar-month climatology, which removes the interannual variation in evaporative demand but keeps its seasonal cycle, and the result was correlated with the full index for each division and timescale. The quantity $1 - r^2$ is the share of SPEI variance attributable to interannual PET variation (Table S5).

*Agreement of the index.* SPEI was computed separately from station precipitation and from ERA5 precipitation with the same code, the same distribution fitting and the same PET series. Because the compilation lacks the inputs of the Penman-Monteith method, both versions used PET from the method of Thornthwaite (1948) applied to ERA5 temperature, so any difference between the two indices comes from the source of precipitation alone. With the station-derived index as the reference and drought defined as SPEI $\leq -1$, agreement was scored with r, RMSE, NSE, the probability of detection, the false alarm ratio and Cohen's $\kappa$ (Cohen, 1960). The $\kappa$ values were read with the bands of Landis and Koch (1977), in which 0.21-0.40 is fair, 0.41-0.60 moderate and 0.61-0.80 substantial agreement (Table S6). The number of pairs falls from 4,128 at SPEI-1 to 4,040 at SPEI-12 because the first $k - 1$ months of each divisional series have no k-month value. The SPEI-3 series of two divisions are shown in Fig. S1.

### Text S3. Predictor screening and model inputs

*Correlation.* Pearson correlations among the eight candidate predictors, PET, the water balance and SPEI at the four timescales were computed on raw same-month values pooled over the eight divisions for 1980-2024 (4,320 division-months, pairwise complete), as in the concurrent screening of Al Mamun et al. (2024). Because monthly climate series are autocorrelated, an effective sample size based on the lag-1 autocorrelation was used when judging the significance of the coefficients. The matrix is shown in Fig. 6a of the main text.

*Multicollinearity.* The VIF (Eq. (7) of the main text) was computed on de-seasonalized anomalies, and two procedures were applied. The first removed variables on physical grounds and recomputed the factors. Removing each temperature variable in turn showed that only the removal of mean temperature brings every remaining VIF below 10, whereas the removal of maximum or minimum temperature does not. The second, a greedy procedure that repeatedly drops the variable with the highest VIF (Dormann et al., 2013), checked that a purely numerical rule reaches the same deletion (Table S7). Maximum and minimum temperature remain moderately inflated after the deletion but were kept, because they are physically distinct and are separate inputs to the Penman-Monteith equation. PET and the water balance were excluded from the start, because they are constructed from the other variables.

*Lead-lag analysis.* Three tests were applied to the de-seasonalized anomalies of each predictor for leads of 1 to 12 months. The lagged Pearson correlation (Eq. (8) of the main text) locates the lead at which the linear relationship is strongest. The lagged mutual information, estimated with the k-nearest-neighbor method of Kraskov et al. (2004), measures the same dependence without assuming that it is linear. Granger causality (Granger, 1969) asks whether the past of a predictor improves the prediction of the index beyond what its own past provides. For each division it compares an autoregression of SPEI on its own p lags (the restricted model) with one that also contains p lags of the predictor (the unrestricted model) through

$$F = \frac{(\mathrm{RSS}_r - \mathrm{RSS}_u)/p}{\mathrm{RSS}_u/(n - 2p - 1)} \tag{S12}$$

where $RSS_r$ and $RSS_u$ are the residual sums of squares of the two models and n is the number of observations. The eight divisional p-values $p_i$ were combined with the method of Fisher (1932),

$$X^2 = -2\sum_{i=1}^{8} \ln p_i \tag{S13}$$

which follows a $\chi^2$ distribution with 16 degrees of freedom when no division has a causal link. All three tests used anomalies, because SPEI has no seasonal cycle of its own and a relationship between raw series would be inflated by the shared annual cycle. The lagged correlations are shown in Fig. 6b-d of the main text.

*Autocorrelation.* The ACF (Eq. (9) of the main text) and the PACF of SPEI were computed for each division over gap-free segments and then averaged across divisions (Fig. 7 of the main text). The PACF at lag k is the last coefficient of the order-k Yule-Walker solution (Box et al., 2015), which the Durbin-Levinson recursion gives as

$$\phi_{kk} = \frac{\rho_k - \sum_{j=1}^{k-1} \phi_{k-1,j}\ \rho_{k-j}}{1 - \sum_{j=1}^{k-1} \phi_{k-1,j}\ \rho_j}, \qquad \phi_{kj} = \phi_{k-1,j} - \phi_{kk}\, \phi_{k-1,k-j} \tag{S14}$$

where $\rho_k$ is the autocorrelation at lag k. The mean ACF stays above the 95% white-noise bound up to lag 2, 5 and 10 at SPEI-3, SPEI-6 and SPEI-12. The mean PACF is significant at lags 1, 2, 3, 4, 6 and 7 at SPEI-3, at lags 1, 5, 6, 7 and 13 at SPEI-6 and at lags 1, 2, 4, 11 and 13 at SPEI-12. The largest significant lag (7, 13 and 13 months) set the depth of the lagged index in the inputs. At SPEI-6 and SPEI-12 the PACF has a distinct peak at the lag one month beyond the accumulation window (about 0.27 at lag 7 and 0.40 at lag 13), where the month that leaves the window becomes visible.

*Model inputs.* Each row of the feature matrix combines four blocks. The meteorological block holds the seven retained predictors over the three most recent months up to the forecast origin t, de-seasonalized with the calendar-month mean of each division (21 values). The drought-history block holds SPEI(t) and its lags, back to SPEI(t − 7) at SPEI-3 and to SPEI(t − 13) at SPEI-6 and SPEI-12 (8 or 14 values). The calendar month enters as a sine and cosine pair, which keeps December next to January, and the division enters as eight indicator columns, which let one pooled model learn a division-specific offset instead of eight separate models being trained.

XGBoost therefore receives a flat vector of 39 values at SPEI-3 and 45 values at SPEI-6 and SPEI-12. The deep-learning models receive sequences of ten channels per month (the seven anomalies, SPEI and the two calendar terms) over T = 7, 6 and 13 months, with the division indicators as a separate static input, since they do not change within a window (Fig. 4 of the main text). Windows were formed only within contiguous runs of months of one division, so no window spans a gap.

*Scaling and partition.* The anomaly columns were standardized with a scaler fitted on the training rows only. The lagged index was left on its native scale, because it is already standardized, which also keeps the SHAP attributions in index units. The calendar-month means used for de-seasonalization cover all years in the main pipeline and only the training years in the reference-period test (Text S6). Rows were assigned to a partition by the date of the target month. A row dated October 2018 that forecasts January 2019 at a three-month lead therefore belongs to the test set, although its own date lies in the training years. For hyperparameter search and early stopping, the deep-learning models were fitted on target months from 1980 to 2013 and validated on 2014-2018, and XGBoost was fitted on 1980-2016 and validated on 2017-2018.

**Text S4. Model architectures and training**

*Recurrent layers.* The long short-term memory (LSTM) unit (Hochreiter & Schmidhuber, 1997) keeps a cell state that is updated through a forget gate, an input gate and an output gate (Gers et al., 2000),

$$f_t = \sigma\left(W_f[h_{t-1}; x_t] + b_f\right), \quad i_t = \sigma(W_i[h_{t-1}; x_t] + b_i), \quad o_t = \sigma(W_o[h_{t-1}; x_t] + b_o) \tag{S15}$$

$$c_t = f_t \odot c_{t-1} + i_t \odot \tanh(W_c[h_{t-1}; x_t] + b_c), \qquad h_t = o_t \odot \tanh(c_t) \tag{S16}$$

where $x_t$ is the input at month t of the window, $h_t$ the hidden state, $c_t$ the cell state, W and b the weights and biases of each gate, σ the logistic sigmoid, ⊙ the element-wise product and [·;·] concatenation. In the bidirectional form (Schuster & Paliwal, 1997), one layer reads the window forwards and another reads it backwards. The Bi-LSTM passes the final states of the two directions to the dense head,

$$\mathbf{h} = [\mathbf{h}_T^{\rightarrow}; \mathbf{h}_1^{\leftarrow}] \tag{S17}$$

where T is the window length. The whole window lies before the forecast month, so the backward pass uses no future information.

*CNN and CNN-BiLSTM.* The CNN applies two blocks of one-dimensional convolution, batch normalization and ReLU activation, followed by spatial dropout and global average pooling over time (Eq. (13) of the main text). Spatial dropout removes whole feature maps rather than single values. Global average pooling reduces the time axis to a vector of fixed length in the same way for every window, so the same design serves windows of 7, 6 and 13 months. The CNN-BiLSTM uses the first convolutional block of the CNN as a front end and passes its output sequence to a bidirectional LSTM layer,

$$Z = \phi\big(\mathrm{BN}(W_1 * X)\big), \qquad \mathbf{h} = \mathrm{BiLSTM}(Z) \tag{S18}$$

where Z is the convolutional feature sequence and **h** the concatenated final states of the two directions.

*CEEMDAN-BiLSTM.* CEEMDAN (Torres et al., 2011) extends empirical mode decomposition (Huang et al., 1998) and its ensemble form (Wu & Huang, 2009). It splits a series into intrinsic mode functions (IMFs) and a residual by adding controlled white noise over an ensemble and averaging the results, which reduces the mixing of modes. It was applied to the SPEI channel of each input window,

$$X^{\mathrm{IMF}} = \mathrm{CEEMDAN}(x^{\mathrm{SPEI}}) \tag{S19}$$

with 50 noise realizations and a fixed seed, so the decomposition is identical across the five training seeds. The output was fixed at four channels. Where fewer IMFs were returned, the remaining channels were padded with zeros, and where more were returned, the surplus was summed into the last channel, so the channels still add up to the original series. The windows are short, so the decomposition gives only two or three IMFs. The raw sequence and the IMF sequence each pass through their own bidirectional LSTM, which returns the hidden state at every month,

$$H^{(s)} = \mathrm{BiLSTM}_s\big(X^{(s)}\big), \qquad s \in \{\mathrm{raw}, \mathrm{IMF}\} \tag{S20}$$

and each stream is pooled over time by the additive attention and layer normalization of Eq. (14) of the main text (Bahdanau et al., 2014).

*Dense head.* All four deep-learning models end in the same head, which joins the pooled representation with the static division vector **s** and outputs the predicted change over the lead L,

$$\Delta\hat{y}_{t+L} = \mathbf{w}_y^{\top}\,\phi(W_d[\mathbf{h};\mathbf{s}]) + b_y \tag{S21}$$

where $W_d$ holds the weights of the hidden dense layer, φ is the ReLU activation, and $\mathbf{w}_y$ and $b_y$ are the weights and bias of the output unit. For the CEEMDAN-BiLSTM, **h** is replaced by the two pooled streams $[\mathbf{p}^{(raw)}; \mathbf{p}^{(IMF)}]$. The forecast is recovered by adding the persistence forecast $y_t$ (Eq. (11) of the main text).

*Training.* The deep-learning models were trained with the Adam optimizer (Kingma & Ba, 2014) on the Huber loss (Huber, 1964) with threshold δ = 1,

$$L_\delta(e) = \begin{cases} 1/2\, e^2, & |e| \le \delta \\ \delta(|e| - 1/2\,\delta), & |e| > \delta \end{cases} \tag{S22}$$

where e is the error of the predicted change. The loss is quadratic for small errors and linear for large ones, which limits the influence of a few extreme months. Dropout was applied between layers (Srivastava et al., 2014). XGBoost was trained on the index level with a squared-error objective. Hyperparameters were searched with the tree-structured Parzen estimator of Optuna (Akiba et al., 2019), with 20 trials per model and timescale, minimizing the loss on the validation years, and with a median pruner that could stop a trial after ten epochs. The tuned values and the fixed training settings are listed in Tables S8 and S9.

*Batch order.* Each model was retrained exactly as it had been tuned. Training batches were shuffled for the Bi-LSTM and kept in temporal order for the CNN, the CNN-BiLSTM and the CEEMDAN-BiLSTM (Table S8). The match matters. In preliminary runs, the SPEI-12 CNN had been retrained with shuffled batches, unlike its tuning, and its mean lead-1 skill over five seeds was −0.017 ± 0.065. Retrained as tuned, it is 0.148 ± 0.042 (Table 2 of the main text).

**Text S5. Seeds, model selection and uncertainty estimates**

*Seeds and reproducibility.* Every configuration was trained with the seeds 42 to 46. The seed sets the weight initialization, the dropout masks and the batch order of the deep-learning models and the subsampling of XGBoost, whereas the CEEMDAN decomposition was kept fixed. Deterministic operations were enabled, so that a seed reproduces its run, and all runs were made on the same workstation. The run log holds 135 runs, 75 at lead 1 (five models, three timescales, five seeds) and 60 at leads 2 and 3 (the selected deep-learning model and XGBoost at each of the three timescales, two leads, five seeds). The five runs of the reference-period test (Text S6) are logged separately. Seed 42 reproduces the earlier single-seed results of the four deep-learning models to the precision at which those results were stored, and those of XGBoost to within 0.011, because its tuned values had been stored rounded to four decimals (Table S12). The scores of every run are listed in Table S14.

*Tuning check and selection.* The tuned Bi-LSTM was compared with an untuned configuration that had been trained with three seeds in the earlier single-seed study. The gain in mean lead-1 skill is of the same order as the seed standard deviation of the untuned model at every timescale (Table S10A). The selection rule of Section 2.7 of the main text was applied once (Table S10B). For each challenger, the margin is its mean lead-1 skill minus that of the incumbent, and the bar is the larger of the two seed standard deviations. A challenger would have replaced the incumbent only if its margin exceeded the bar, which no challenger did. A challenger whose margin is negative but smaller in size than the bar is reported as indistinguishable from the incumbent. This applies to the CNN at SPEI-6 and to the Bi-LSTM and the CNN-BiLSTM at SPEI-12.

*Median-skill seed.* Panels that show a single forecast series use the median-skill seed of each model, which is the third of its five seeds ranked by lead-1 skill. At SPEI-3, SPEI-6 and SPEI-12 these are seeds 45, 45 and 46 for XGBoost, 44, 42 and 44 for the CNN, 42, 42 and 46 for the Bi-LSTM, 44, 45 and 42 for the CNN-BiLSTM and 44, 42 and 42 for the CEEMDAN-BiLSTM.

*Circular block bootstrap.* The 576 test pairs are 72 months × 8 divisions and are correlated both in time and across divisions, so resampling single pairs would give intervals that are too narrow. The test months were therefore resampled in blocks of 12 consecutive months (Politis & Romano, 1992), with all eight divisions of a month kept together and six blocks forming one resample of 72 months. The months were treated as a circle, so that a block starting near the end of 2024 continues from January 2019. Every month then has the same chance of being drawn, whereas ordinary moving blocks draw the first and last months of a short record much less often than the middle months. In total, 2,000 resamples were drawn, and the same resampled months were used for every model and for persistence. For each model, every score was computed for each seed on each resample, and the 5 × 2,000 values were pooled, so that the interval covers both sampling and seed variation. The 95% interval is given by the 2.5th and 97.5th percentiles of the pooled values. Differences were computed resample by resample, as the model minus persistence for CSI and POD, and as the model minus the seed-mean skill of XGBoost for skill (Table S13).

*Diebold-Mariano test.* The loss differential $d_t$ of test month t is the mean over the eight divisions of the difference between the squared errors of the model and of persistence, which gives T = 72 values. Its long-run variance was estimated with the Newey-West estimator (Newey & West, 1987) and Bartlett weights,

$$\hat{\sigma}^2_{\mathrm{LR}} = \hat{\gamma}_0 + 2\sum_{k=1}^{b}\left(1 - \frac{k}{b+1}\right)\hat{\gamma}_k, \qquad b = \max\left(L - 1, \left\lfloor 4\,(T/100)^{2/9}\right\rfloor\right) \tag{S23}$$

where $\hat{\gamma}_k$ is the sample autocovariance of $d_t$ at lag k and L is the lead. With T = 72 the second term of the maximum equals 3, which exceeds L − 1 at every lead, so b = 3. The statistic of Eq. (20) of the main text was multiplied by the small-sample correction of Harvey et al. (1997),

$$DM^* = DM\sqrt{\frac{T + 1 - 2L + L(L-1)/T}{T}} \tag{S24}$$

and compared with Student's t distribution with T − 1 degrees of freedom (two-sided). A negative statistic means that the model has the smaller squared error. The test was applied to each seed and to the seed-ensemble mean forecast (Table S13C).

**Text S6. Sensitivity to the reference period of the SPEI fit**

In the main pipeline, the SPEI distributions were fitted on the full 1980-2024 record, which includes the test years, and the means used to de-seasonalize the predictors also cover all years. To test whether this use of the test years affects the results, a strict pipeline was built in two steps. First, the generalized logistic distributions were refitted on 1980-2018 only, and the index of the whole record, including 2019-2024, was recomputed with these parameters. Second, the predictor anomalies were recomputed with calendar-month means taken from the training years. The Bi-LSTM at SPEI-6 and lead 1 was then retrained with its tuned hyperparameters and the same five seeds and scored against the refitted index. The decision rule was fixed before the run: the main pipeline would be kept if the mean skill changed by less than one seed standard deviation of the main runs (0.045).

Over the test period, the refitted index is almost identical to the main index, with correlations of 0.993-0.997 at the four timescales (Table S18A; Fig. S5). It is slightly drier, with mean shifts from −0.017 at SPEI-1 to −0.110 at SPEI-12, so the number of drought months rises from 119 to 135 at SPEI-3, from 156 to 168 at SPEI-6 and from 204 to 236 at SPEI-12. The strict pipeline gives a skill of 0.354 ± 0.059, against 0.350 ± 0.045 in the main pipeline (Table S18B). The change of 0.004 is about one tenth of the bar, so the main pipeline was kept. RMSE and CSI differ more between the two pipelines (0.516 against 0.491 and 0.648 against 0.615), because the strict pipeline is scored against a different index with more drought months, so these two scores are not compared like for like.

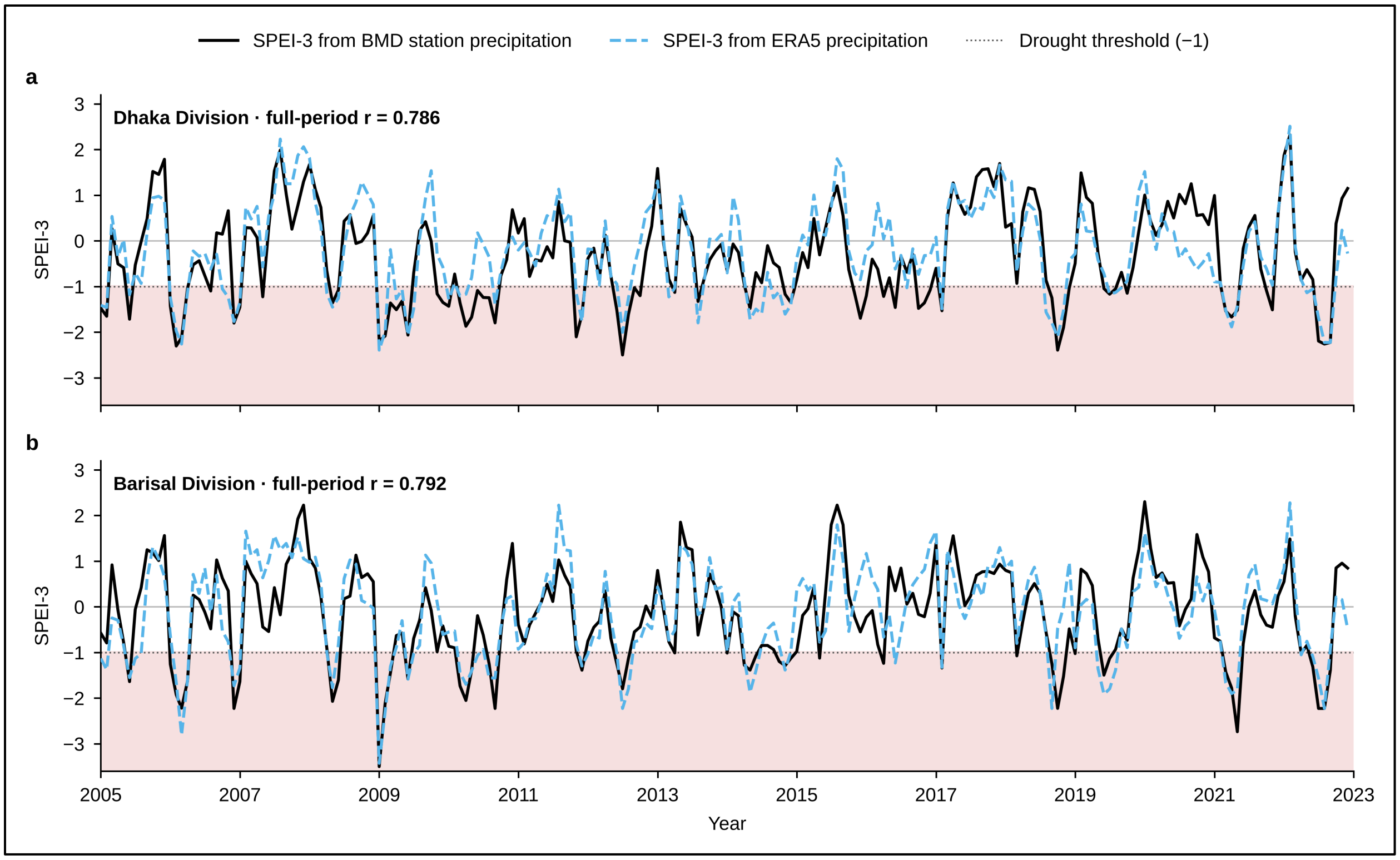

**Fig. S1.** SPEI-3 computed from BMD station precipitation (black, solid) and from ERA5 precipitation (blue, dashed) for (a) Dhaka and (b) Barisal divisions, shown for 2005-2022. Both indices use the same Thornthwaite PET derived from ERA5 temperature. The shaded band marks drought (SPEI ≤ −1). The correlation printed in each panel is computed over the full common record (n = 514 months).

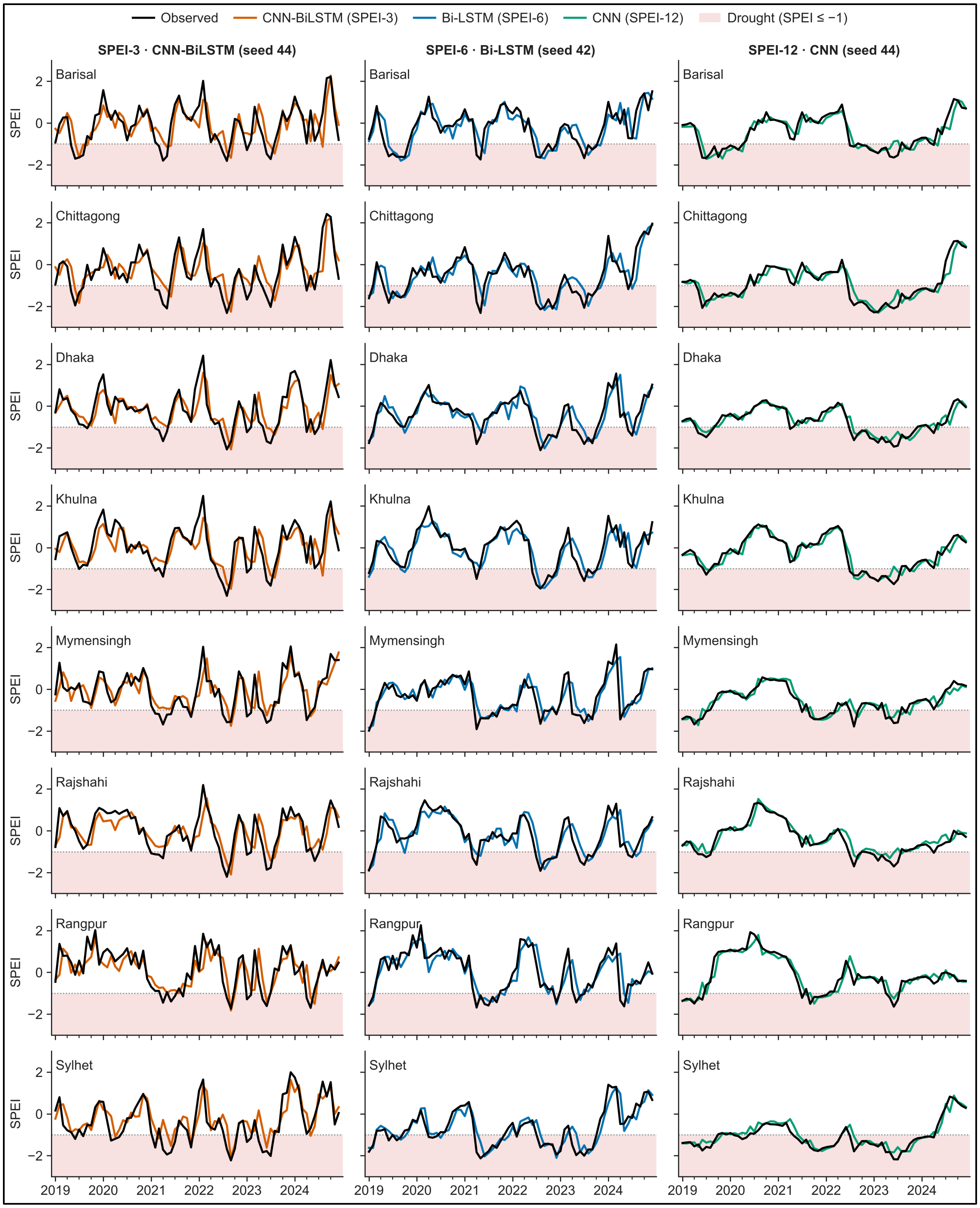

**Fig. S2.** Observed SPEI (black) and the lead-1 forecast of the selected model in each division over the test period: CNN-BiLSTM at SPEI-3 (seed 44), Bi-LSTM at SPEI-6 (seed 42) and CNN at SPEI-12 (seed 44), the median-skill seed of each. All panels share the range −3 to 3, and the shaded band marks drought (SPEI ≤ −1).

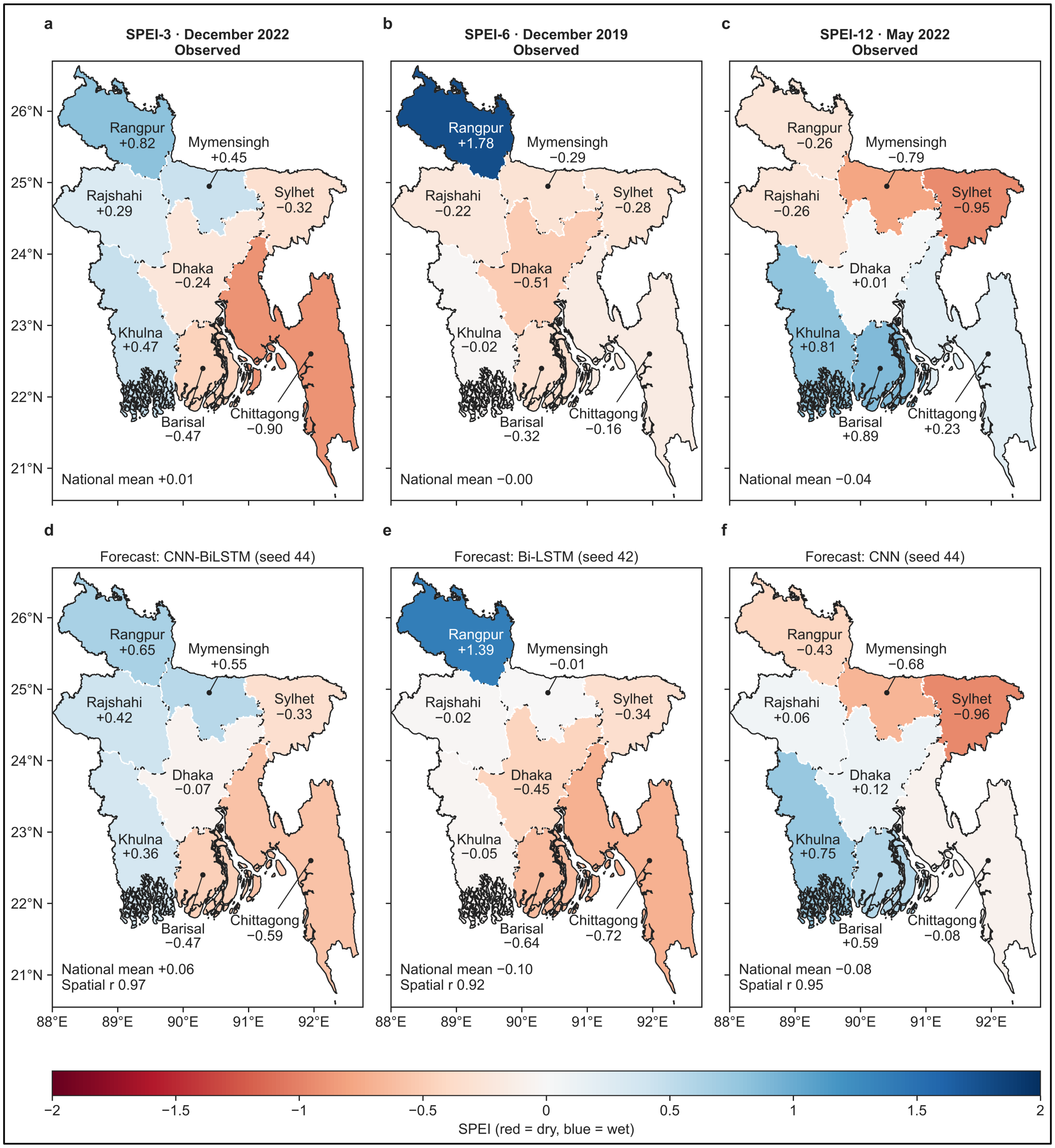

**Fig. S3.** As Fig. 14 of the main text, for a near-normal month of the test period: (a, d) SPEI-3, December 2022; (b, e) SPEI-6, December 2019; (c, f) SPEI-12, May 2022.

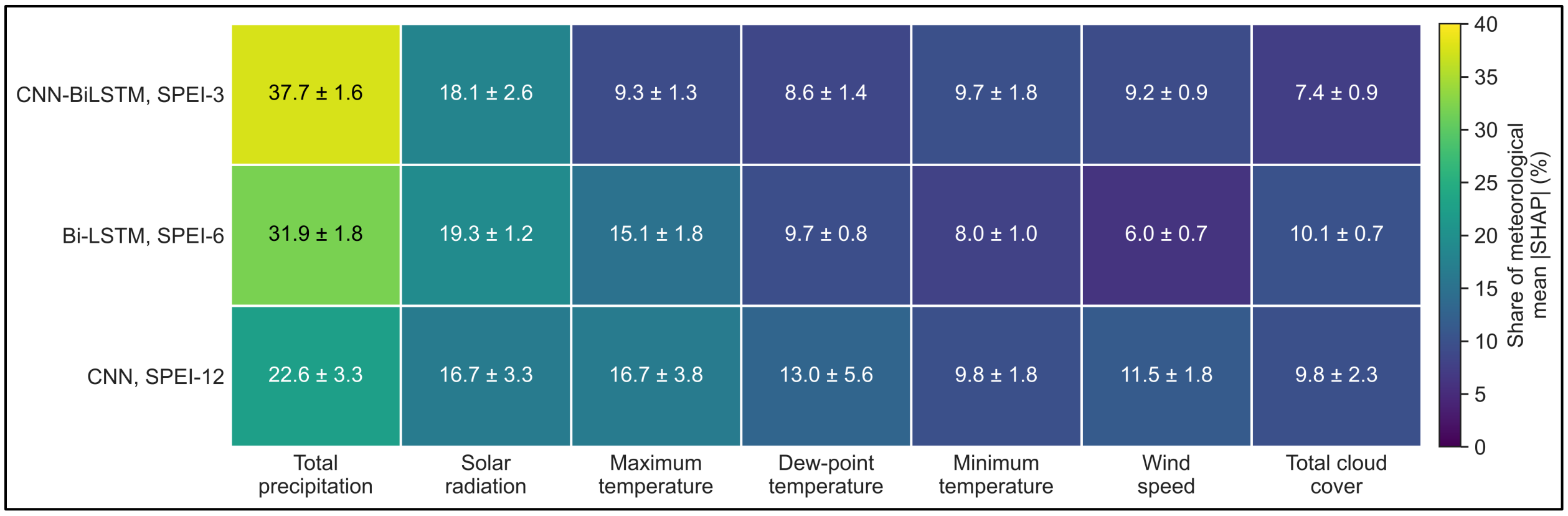

**Fig. S4.** Share (%) of each meteorological predictor in the summed mean absolute SHAP value of the meteorological predictors (lagged SPEI excluded) for the three selected models, as the mean ± SD over five seeds. The values are listed in Table S17.

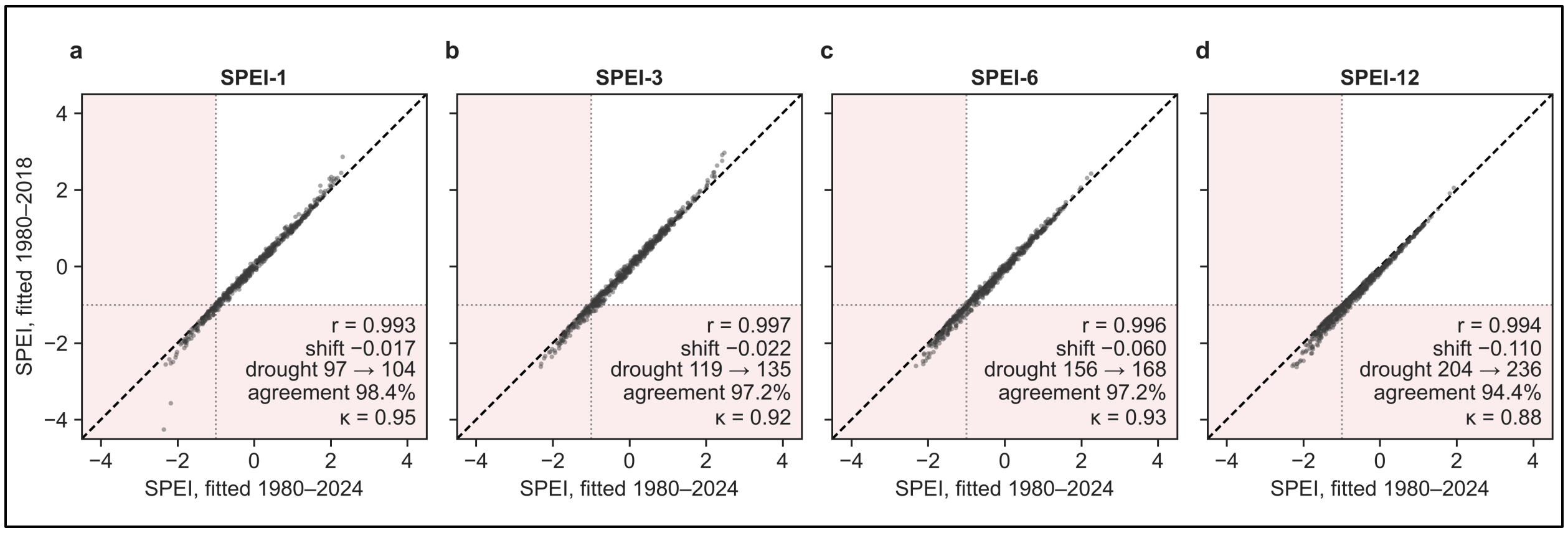

**Fig. S5.** SPEI over the test period computed with the distribution fitted on 1980-2024 (x axis, used by every model) and refitted on 1980-2018 (y axis) at (a) SPEI-1, (b) SPEI-3, (c) SPEI-6 and (d) SPEI-12. The dashed line is 1:1, and the dotted lines mark the drought threshold (SPEI = −1). Each panel gives the correlation, the mean shift (refit minus main), the number of drought months (main → refit), the agreement in drought classification and Cohen's κ. Points below the 1:1 line are drier in the refit (Text S6; Table S18).

**Table S1.** ERA5 monthly averaged variables on single levels (Hersbach et al., 2023), retrieved at 0.25° for January 1980 to December 2024, with their native units, the units after conversion and the conversion applied.

| Variable | Short name | Native unit | Converted unit | Conversion applied |
|---|---|---|---|---|
| Total precipitation | tp | m | mm | Multiplied by 1,000 |
| 2 m air temperature | t2m | K | °C | Subtracted 273.15 |
| Maximum 2 m temperature | mx2t | K | °C | Subtracted 273.15 |
| Minimum 2 m temperature | mn2t | K | °C | Subtracted 273.15 |
| 2 m dew-point temperature | d2m | K | °C | Subtracted 273.15 |
| Surface downwelling solar radiation | ssrd | $J\ m^{-2}$ | $MJ\ m^{-2}\ day^{-1}$ | Divided by 1,000,000 |
| 10 m zonal wind component | u10 | $m\ s^{-1}$ | $m\ s^{-1}$ | None required |
| 10 m meridional wind component | v10 | $m\ s^{-1}$ | $m\ s^{-1}$ | None required |
| Total cloud cover | tcc | Fraction (0-1) | Fraction (0-1) | None required |

Wind speed at 10 m was computed from the two wind components (Eq. S1) and reduced to 2 m (Eq. S6). Mean 2 m temperature was removed from the predictors after the VIF test (Table S7) but enters the computation of PET.

**Table S2.** SPEI classes (McKee et al., 1993). A month with SPEI ≤ −1 is treated as a drought month throughout the study.

| SPEI range | Class | Drought month |
|---|---|---|
| ≥ 2.00 | Extremely wet | No |
| 1.50 to 1.99 | Very wet | No |
| 1.00 to 1.49 | Moderately wet | No |
| −0.99 to 0.99 | Near normal | No |
| −1.00 to −1.49 | Moderate drought | Yes |
| −1.50 to −1.99 | Severe drought | Yes |
| ≤ −2.00 | Extreme drought | Yes |

**Table S3.** Agreement between ERA5 and BMD station precipitation for each division and for the pooled record, 1980-2022 (516 months per division; 4,128 division-months pooled).

| Division | r | $r_a$ | PBIAS (%) | RMSE (mm) | NSE | KGE |
|---|---|---|---|---|---|---|
| Barisal | 0.934 | 0.726 | −7.2 | 75.2 | 0.865 | 0.844 |
| Chittagong | 0.939 | 0.733 | −24.1 | 131.3 | 0.766 | 0.597 |
| Dhaka | 0.904 | 0.699 | +7.0 | 68.8 | 0.797 | 0.878 |
| Khulna | 0.933 | 0.756 | +5.5 | 54.8 | 0.861 | 0.913 |
| Mymensingh | 0.851 | 0.594 | +17.5 | 116.9 | 0.658 | 0.765 |
| Rajshahi | 0.898 | 0.663 | +18.0 | 69.7 | 0.733 | 0.767 |
| Rangpur | 0.869 | 0.564 | +18.0 | 109.7 | 0.698 | 0.773 |
| Sylhet | 0.926 | 0.707 | +7.2 | 102.1 | 0.839 | 0.889 |
| **All divisions (pooled)** | **0.891** | **0.696** | **+3.4** | **94.6** | **0.790** | **0.875** |

r, correlation of the raw monthly series; $r_a$, correlation of the de-seasonalized anomalies; PBIAS, percent bias of ERA5 relative to the stations; KGE, Kling-Gupta efficiency. The pooled values are reported in Table 1A of the main text. Station values stored as zero for missing data were excluded (Text S2).

**Table S4.** Optimal circular month shift (k, months) and the correlation (r) between the station climatology and the corresponding ERA5 climatology at that shift, for each of the 35 BMD stations and for maximum temperature ($T_{max}$), minimum temperature ($T_{min}$) and precipitation (P).

| Station (index) | Division | $T_{max}$ k | $T_{max}$ r | $T_{min}$ k | $T_{min}$ r | P k | P r |
|---|---|---|---|---|---|---|---|
| Barisal (11704) | Barisal | 9 | 0.604 | 10 | 0.778 | 0 | 0.999 |
| Bhola (11706) | Barisal | 7 | 0.568 | 9 | 0.784 | 0 | 0.997 |
| Khepupara (12110) | Barisal | 6 | 0.598 | 7 | 0.775 | 0 | 0.985 |
| Patuakhali (12103) | Barisal | 11 | 0.633 | 0 | 0.852 | 0 | 0.993 |
| Ambagan (41977) | Chittagong | 10 | 0.677 | 11 | 0.772 | 0 | 0.960 |
| Chandpur (11316) | Chittagong | 6 | 0.641 | 7 | 0.771 | 0 | 0.990 |
| Chittagong Airport (11921) | Chittagong | 5 | 0.659 | 6 | 0.785 | 0 | 0.980 |
| Comilla (11313) | Chittagong | 3 | 0.603 | 4 | 0.802 | 0 | 0.980 |
| Cox's Bazar (11927) | Chittagong | 2 | 0.637 | 3 | 0.796 | 0 | 0.971 |
| Feni (11805) | Chittagong | 10 | 0.627 | 11 | 0.779 | 0 | 0.984 |
| Hatiya (11814) | Chittagong | 9 | 0.625 | 10 | 0.766 | 0 | 0.984 |
| Kutubdia (11925) | Chittagong | 4 | 0.596 | 5 | 0.772 | 0 | 0.971 |
| M. Court (11809) | Chittagong | 3 | 0.607 | 4 | 0.785 | 0 | 0.985 |
| Rangamati (12007) | Chittagong | 9 | 0.649 | 10 | 0.755 | 0 | 0.995 |
| Sandwip (11916) | Chittagong | 7 | 0.650 | 8 | 0.766 | 0 | 0.978 |
| Sitakunda (11912) | Chittagong | 5 | 0.637 | 6 | 0.800 | 0 | 0.990 |
| Teknaf (11929) | Chittagong | 1 | 0.614 | 2 | 0.795 | 0 | 0.955 |
| Dhaka (11111) | Dhaka | 1 | 0.553 | 2 | 0.775 | 0 | 0.982 |
| Faridpur (11505) | Dhaka | 10 | 0.562 | 0 | 0.776 | 0 | 0.966 |
| Madaripur (11513) | Dhaka | 1 | 0.576 | 3 | 0.776 | 0 | 0.968 |
| Tangail (41909) | Dhaka | 1 | 0.544 | 3 | 0.770 | 0 | 0.984 |
| Chuadanga (41926) | Khulna | 4 | 0.586 | 5 | 0.791 | 0 | 0.969 |
| Jessore (11407) | Khulna | 7 | 0.633 | 8 | 0.753 | 0 | 0.996 |
| Khulna (11604) | Khulna | 5 | 0.570 | 6 | 0.786 | 0 | 0.996 |
| Mongla (41958) | Khulna | 0 | 0.511 | 2 | 0.811 | 0 | 0.989 |
| Satkhira (11610) | Khulna | 6 | 0.568 | 7 | 0.793 | 0 | 0.991 |
| Mymensingh (10609) | Mymensingh | 11 | 0.567 | 1 | 0.735 | 0 | 0.959 |
| Bogra (10408) | Rajshahi | 6 | 0.550 | 8 | 0.691 | 0 | 0.988 |
| Ishurdi (10910) | Rajshahi | 8 | 0.561 | 9 | 0.699 | 0 | 0.970 |
| Rajshahi (10320) | Rajshahi | 10 | 0.564 | 11 | 0.718 | 0 | 0.961 |
| Dinajpur (10120) | Rangpur | 11 | 0.562 | 1 | 0.738 | 0 | 0.965 |
| Rangpur (10208) | Rangpur | 7 | 0.629 | 9 | 0.635 | 0 | 0.981 |
| Syedpur (41858) | Rangpur | 11 | 0.589 | 0 | 0.719 | 0 | 0.985 |
| Srimongal (10724) | Sylhet | 3 | 0.539 | 5 | 0.730 | 0 | 0.972 |
| Sylhet (10705) | Sylhet | 2 | 0.527 | 4 | 0.727 | 0 | 0.989 |

A record without defects needs k = 0 and a high r. The precipitation shift is 0 at every station (r = 0.955-0.999). The temperature shifts vary between stations, with r = 0.511-0.677 for maximum and 0.635-0.852 for minimum temperature, so only precipitation was used as a validation reference (Text S2).

**Table S5.** Correlation between SPEI computed from ERA5 with time-varying PET and SPEI computed with PET held at its calendar-month climatology, by division and timescale, 1980-2024.

| Division | SPEI-1 | SPEI-3 | SPEI-6 | SPEI-12 |
|---|---|---|---|---|
| Barisal | 0.940 | 0.966 | 0.981 | 0.978 |
| Chittagong | 0.951 | 0.974 | 0.979 | 0.977 |
| Dhaka | 0.946 | 0.975 | 0.976 | 0.971 |
| Khulna | 0.946 | 0.971 | 0.976 | 0.965 |
| Mymensingh | 0.939 | 0.981 | 0.984 | 0.988 |
| Rajshahi | 0.929 | 0.967 | 0.968 | 0.967 |
| Rangpur | 0.865 | 0.964 | 0.976 | 0.977 |
| Sylhet | 0.961 | 0.988 | 0.990 | 0.991 |
| **Mean** | **0.935** | **0.973** | **0.979** | **0.977** |
| **1 − $r^2$ of the mean (%)** | **12.6** | **5.3** | **4.2** | **4.6** |

The last row gives the share of SPEI variance attributable to interannual PET variation, as in Table 1B of the main text.

**Table S6.** Agreement between SPEI computed from BMD station precipitation and SPEI computed from ERA5 precipitation, pooled over the eight divisions, 1980-2022.

| Timescale | n | r | RMSE | NSE | POD | FAR | Cohen's κ | Agreement (Landis & Koch, 1977) |
|---|---|---|---|---|---|---|---|---|
| SPEI-1 | 4,128 | 0.809 | 0.635 | 0.622 | 0.713 | 0.270 | 0.660 | Substantial |
| SPEI-3 | 4,112 | 0.766 | 0.689 | 0.534 | 0.652 | 0.345 | 0.579 | Moderate |
| SPEI-6 | 4,088 | 0.712 | 0.755 | 0.427 | 0.557 | 0.436 | 0.468 | Moderate |
| SPEI-12 | 4,040 | 0.647 | 0.825 | 0.294 | 0.522 | 0.482 | 0.422 | Moderate |

Both indices use the same Thornthwaite PET derived from ERA5 temperature, so differences arise from precipitation alone. The station index is the reference, and drought is SPEI ≤ −1. n is the number of division-months with a value at that timescale. The full-record correlations of SPEI-3 for Dhaka (0.786) and Barisal (0.792) are printed in Fig. S1.

**Table S7.** Variance inflation factors of the eight candidate predictors, computed on de-seasonalized anomalies, before and after the removal of mean 2 m temperature.

| Predictor | VIF, eight predictors | VIF after removing mean temperature |
|---|---|---|
| Maximum temperature (mx2t) | 11.5 | 8.4 |
| Minimum temperature (mn2t) | 10.9 | 8.8 |
| Mean 2 m temperature (t2m) | 11.3 | Removed |
| Solar radiation (ssrd) | 7.2 | 7.2 |
| Dewpoint temperature (d2m) | 3.7 | 3.5 |
| Precipitation (tp) | 2.9 | 2.8 |
| Cloud cover (tcc) | 2.7 | 2.7 |
| Wind speed (wind) | 1.0 | 1.0 |

A VIF above 10 was treated as serious multicollinearity. The greedy procedure of Dormann et al. (2013) removes the same variable (Text S3).

**Table S8.** Tuned hyperparameters of the four deep-learning models (A) and the training settings common to them (B).

**(A) Tuned values**

| Model | SPEI | T (months) | Conv. filters | Kernel | Second conv. filters | LSTM units per direction | Dense units | Dropout | Learning rate | Batch order |
|---|---|---|---|---|---|---|---|---|---|---|
| CNN | 3 | 7 | 16 | 5 | 80 | n/a | 48 | 0.108 | 0.008706 | Temporal |
| CNN | 6 | 6 | 96 | 3 | 64 | n/a | 32 | 0.280 | 0.000359 | Temporal |
| CNN | 12 | 13 | 32 | 4 | 112 | n/a | 24 | 0.110 | 0.001660 | Temporal |
| Bi-LSTM | 3 | 7 | n/a | n/a | n/a | 16 | 48 | 0.429 | 0.003810 | Shuffled |
| Bi-LSTM | 6 | 6 | n/a | n/a | n/a | 16 | 64 | 0.495 | 0.002713 | Shuffled |
| Bi-LSTM | 12 | 13 | n/a | n/a | n/a | 128 | 32 | 0.488 | 0.001904 | Shuffled |
| CNN-BiLSTM | 3 | 7 | 112 | 4 | n/a | 96 | 40 | 0.431 | 0.000173 | Temporal |
| CNN-BiLSTM | 6 | 6 | 80 | 2 | n/a | 16 | 64 | 0.486 | 0.004138 | Temporal |
| CNN-BiLSTM | 12 | 13 | 96 | 4 | n/a | 112 | 40 | 0.419 | 0.000205 | Temporal |
| CEEMDAN-BiLSTM | 3 | 7 | n/a | n/a | n/a | 32 | 48 | 0.434 | 0.005893 | Temporal |
| CEEMDAN-BiLSTM | 6 | 6 | n/a | n/a | n/a | 48 | 40 | 0.273 | 0.000382 | Temporal |
| CEEMDAN-BiLSTM | 12 | 13 | n/a | n/a | n/a | 32 | 48 | 0.429 | 0.004019 | Temporal |

**(B) Common settings**

| Setting | Value |
|---|---|
| Optimizer and loss | Adam (Kingma & Ba, 2014); Huber loss with threshold 1.0 (Huber, 1964) |
| Hyperparameter search | Optuna tree-structured Parzen estimator (Akiba et al., 2019), 20 trials per model and timescale, at lead 1 |
| Pruning | Median pruner, active after ten epochs |
| Fitting and validation years | 1980-2013 and 2014-2018 (target months) |
| Batch size and epochs | 32; up to 200 epochs |
| Early stopping | Patience of 15 epochs in the search and 40 epochs in the final training |
| Learning-rate schedule | Halved after 15 epochs without improvement, down to $1 \times 10^{-5}$ |
| Seeds | 42-46; the same tuned values are used at every seed and lead |

T is the length of the input window. The values are read from the run log and are identical in every seed and lead; they equal those of the earlier single-seed study. For the CNN, the dropout rate applies to the spatial dropout after the second block and to the dense layer.

**Table S9.** Tuned hyperparameters of XGBoost at each timescale.

| SPEI | Maximum depth | Learning rate | Subsample ratio | Column sample by tree | Minimum child weight | γ (minimum split loss) | λ (L2 penalty) | α (L1 penalty) | Input features |
|---|---|---|---|---|---|---|---|---|---|
| 3 | 3 | 0.0441 | 0.6138 | 0.9637 | 3 | 2.6003 | 0.4468 | 0.0177 | 39 |
| 6 | 3 | 0.1716 | 0.9863 | 0.9234 | 3 | 2.2008 | 0.0025 | 0.5457 | 45 |
| 12 | 3 | 0.1151 | 0.8827 | 0.8916 | 7 | 0.5793 | 0.0020 | 0.0272 | 45 |

XGBoost used a squared-error objective on the SPEI level, up to 3,000 trees and early stopping after 50 rounds without improvement on the 2017-2018 validation years (fitting years 1980-2016). The search used the same Optuna settings as Table S8. The values are read from the run log and are identical in every seed and lead.

**Table S10.** Tuning check of the Bi-LSTM (A) and application of the model-selection rule (B). Values are mean lead-1 skill against persistence ± seed SD.

**(A) Untuned and tuned Bi-LSTM**

| Timescale | Untuned (three seeds) | Tuned (five seeds) | Gain |
|---|---|---|---|
| SPEI-3 | 0.335 ± 0.032 | 0.375 ± 0.012 | +0.040 |
| SPEI-6 | 0.254 ± 0.082 | 0.350 ± 0.045 | +0.096 |
| SPEI-12 | 0.008 ± 0.102 | 0.141 ± 0.118 | +0.133 |

**(B) Selection rule**

| Timescale | Incumbent | Incumbent skill | Challenger | Challenger skill | Margin | Bar | Outcome |
|---|---|---|---|---|---|---|---|
| SPEI-3 | CNN-BiLSTM | 0.408 ± 0.016 | XGBoost | 0.373 ± 0.006 | −0.035 | 0.016 | Below incumbent |
| | | | CNN | 0.255 ± 0.031 | −0.154 | 0.031 | Below incumbent |
| | | | Bi-LSTM | 0.375 ± 0.012 | −0.034 | 0.016 | Below incumbent |
| | | | CEEMDAN-BiLSTM | 0.280 ± 0.059 | −0.128 | 0.059 | Below incumbent |
| SPEI-6 | Bi-LSTM | 0.350 ± 0.045 | XGBoost | 0.144 ± 0.019 | −0.206 | 0.045 | Below incumbent |
| | | | CNN | 0.327 ± 0.059 | −0.022 | 0.059 | Indistinguishable |
| | | | CNN-BiLSTM | 0.274 ± 0.044 | −0.075 | 0.045 | Below incumbent |
| | | | CEEMDAN-BiLSTM | 0.111 ± 0.060 | −0.238 | 0.060 | Below incumbent |
| SPEI-12 | CNN | 0.148 ± 0.042 | XGBoost | −0.093 ± 0.038 | −0.241 | 0.042 | Below incumbent |
| | | | Bi-LSTM | 0.141 ± 0.118 | −0.007 | 0.118 | Indistinguishable |
| | | | CNN-BiLSTM | 0.110 ± 0.059 | −0.038 | 0.059 | Indistinguishable |
| | | | CEEMDAN-BiLSTM | −0.034 ± 0.105 | −0.182 | 0.105 | Below incumbent |

(A) The untuned configuration was trained with three seeds in the earlier single-seed study; the tuned values are those of Table 2 of the main text. (B) Margin is the challenger mean minus the incumbent mean; bar is the larger of the two seed SDs. A challenger replaces the incumbent only if the margin exceeds the bar, and a challenger whose shortfall is smaller than the bar is reported as indistinguishable (Text S5).

**Table S11.** Drought-event contingency counts and scores over the test period (576 division-months), for all models at lead 1 and for the selected model, XGBoost and persistence at leads 2 and 3. Model values are the mean ± SD over five seeds.

| SPEI | Lead | Model | Hits | Misses | False alarms | Correct negatives | POD | FAR | CSI | SR |
|---|---|---|---|---|---|---|---|---|---|---|
| 3 | 1 | XGBoost | 33.8 ± 2.9 | 85.2 ± 2.9 | 15.2 ± 2.5 | 441.8 ± 2.5 | 0.284 ± 0.025 | 0.310 ± 0.043 | 0.252 ± 0.022 | 0.690 ± 0.043 |
| 3 | 1 | CNN | 40.8 ± 9.1 | 78.2 ± 9.1 | 25.0 ± 6.9 | 432.0 ± 6.9 | 0.343 ± 0.076 | 0.377 ± 0.018 | 0.281 ± 0.051 | 0.623 ± 0.018 |
| 3 | 1 | Bi-LSTM | 50.6 ± 7.9 | 68.4 ± 7.9 | 26.8 ± 5.5 | 430.2 ± 5.5 | 0.425 ± 0.066 | 0.345 ± 0.015 | 0.346 ± 0.041 | 0.655 ± 0.015 |
| **3** | **1** | **CNN-BiLSTM** | **42.4 ± 6.7** | **76.6 ± 6.7** | **23.2 ± 3.6** | **433.8 ± 3.6** | **0.356 ± 0.056** | **0.354 ± 0.042** | **0.298 ± 0.044** | **0.646 ± 0.042** |
| 3 | 1 | CEEMDAN-BiLSTM | 47.2 ± 3.3 | 71.8 ± 3.3 | 31.4 ± 3.1 | 425.6 ± 3.1 | 0.397 ± 0.027 | 0.399 ± 0.026 | 0.314 ± 0.020 | 0.601 ± 0.026 |
| 3 | 1 | Persistence | 66 | 53 | 55 | 402 | 0.555 | 0.455 | 0.379 | 0.545 |
| 3 | 2 | XGBoost | 0.6 ± 0.5 | 118.4 ± 0.5 | 0.6 ± 1.3 | 456.4 ± 1.3 | 0.005 ± 0.005 | 0.250 ± 0.433 (3) | 0.005 ± 0.005 | 0.750 ± 0.433 (3) |
| **3** | **2** | **CNN-BiLSTM** | **7.2 ± 3.4** | **111.8 ± 3.4** | **12.6 ± 4.4** | **444.4 ± 4.4** | **0.061 ± 0.029** | **0.641 ± 0.118** | **0.054 ± 0.025** | **0.359 ± 0.118** |
| 3 | 2 | Persistence | 38 | 81 | 89 | 368 | 0.319 | 0.701 | 0.183 | 0.299 |
| 3 | 3 | XGBoost | 0.0 ± 0.0 | 119.0 ± 0.0 | 0.2 ± 0.4 | 456.8 ± 0.4 | 0.000 ± 0.000 | 1.000 (1) | 0.000 ± 0.000 | 0.000 (1) |
| **3** | **3** | **CNN-BiLSTM** | **0.2 ± 0.4** | **118.8 ± 0.4** | **1.2 ± 1.3** | **455.8 ± 1.3** | **0.002 ± 0.004** | **0.889 ± 0.192 (3)** | **0.002 ± 0.004** | **0.111 ± 0.192 (3)** |
| 3 | 3 | Persistence | 22 | 97 | 112 | 345 | 0.185 | 0.836 | 0.095 | 0.164 |
| 6 | 1 | XGBoost | 94.6 ± 2.2 | 61.4 ± 2.2 | 27.6 ± 1.7 | 392.4 ± 1.7 | 0.606 ± 0.014 | 0.226 ± 0.014 | 0.515 ± 0.016 | 0.774 ± 0.014 |
| 6 | 1 | CNN | 104.4 ± 7.4 | 51.6 ± 7.4 | 23.6 ± 4.2 | 396.4 ± 4.2 | 0.669 ± 0.048 | 0.183 ± 0.020 | 0.581 ± 0.032 | 0.817 ± 0.020 |
| **6** | **1** | **Bi-LSTM** | **111.4 ± 3.1** | **44.6 ± 3.1** | **25.2 ± 4.3** | **394.8 ± 4.3** | **0.714 ± 0.020** | **0.184 ± 0.026** | **0.615 ± 0.023** | **0.816 ± 0.026** |
| 6 | 1 | CNN-BiLSTM | 97.4 ± 3.4 | 58.6 ± 3.4 | 20.2 ± 1.6 | 399.8 ± 1.6 | 0.624 ± 0.022 | 0.172 ± 0.009 | 0.553 ± 0.016 | 0.828 ± 0.009 |
| 6 | 1 | CEEMDAN-BiLSTM | 102.0 ± 2.9 | 54.0 ± 2.9 | 23.8 ± 6.7 | 396.2 ± 6.7 | 0.654 ± 0.019 | 0.187 ± 0.043 | 0.568 ± 0.013 | 0.813 ± 0.043 |
| 6 | 1 | Persistence | 120 | 36 | 42 | 378 | 0.769 | 0.259 | 0.606 | 0.741 |
| 6 | 2 | XGBoost | 67.6 ± 2.6 | 88.4 ± 2.6 | 28.6 ± 2.6 | 391.4 ± 2.6 | 0.433 ± 0.017 | 0.297 ± 0.014 | 0.366 ± 0.011 | 0.703 ± 0.014 |
| **6** | **2** | **Bi-LSTM** | **79.0 ± 7.2** | **77.0 ± 7.2** | **30.6 ± 8.7** | **389.4 ± 8.7** | **0.506 ± 0.046** | **0.276 ± 0.041** | **0.423 ± 0.031** | **0.724 ± 0.041** |
| 6 | 2 | Persistence | 91 | 65 | 76 | 344 | 0.583 | 0.455 | 0.392 | 0.545 |
| 6 | 3 | XGBoost | 42.6 ± 5.3 | 113.4 ± 5.3 | 20.4 ± 8.0 | 399.6 ± 8.0 | 0.273 ± 0.034 | 0.314 ± 0.071 | 0.241 ± 0.022 | 0.686 ± 0.071 |
| **6** | **3** | **Bi-LSTM** | **56.2 ± 2.8** | **99.8 ± 2.8** | **23.6 ± 4.3** | **396.4 ± 4.3** | **0.360 ± 0.018** | **0.294 ± 0.035** | **0.313 ± 0.015** | **0.706 ± 0.035** |
| 6 | 3 | Persistence | 70 | 86 | 101 | 319 | 0.449 | 0.591 | 0.272 | 0.409 |
| 12 | 1 | XGBoost | 145.8 ± 3.4 | 58.2 ± 3.4 | 15.6 ± 1.8 | 356.4 ± 1.8 | 0.715 ± 0.017 | 0.096 ± 0.009 | 0.664 ± 0.012 | 0.904 ± 0.009 |
| **12** | **1** | **CNN** | **165.2 ± 3.3** | **38.8 ± 3.3** | **20.8 ± 4.1** | **351.2 ± 4.1** | **0.810 ± 0.016** | **0.111 ± 0.018** | **0.735 ± 0.014** | **0.889 ± 0.018** |
| 12 | 1 | Bi-LSTM | 160.8 ± 9.2 | 43.2 ± 9.2 | 15.8 ± 3.6 | 356.2 ± 3.6 | 0.788 ± 0.045 | 0.089 ± 0.016 | 0.731 ± 0.037 | 0.911 ± 0.016 |
| 12 | 1 | CNN-BiLSTM | 155.2 ± 2.9 | 48.8 ± 2.9 | 13.0 ± 2.1 | 359.0 ± 2.1 | 0.761 ± 0.014 | 0.077 ± 0.011 | 0.715 ± 0.013 | 0.923 ± 0.011 |
| 12 | 1 | CEEMDAN-BiLSTM | 159.8 ± 4.7 | 44.2 ± 4.7 | 15.0 ± 3.2 | 357.0 ± 3.2 | 0.783 ± 0.023 | 0.085 ± 0.016 | 0.730 ± 0.019 | 0.915 ± 0.016 |
| 12 | 1 | Persistence | 174 | 30 | 33 | 339 | 0.853 | 0.159 | 0.734 | 0.841 |
| 12 | 2 | XGBoost | 109.4 ± 7.9 | 94.6 ± 7.9 | 15.0 ± 3.0 | 357.0 ± 3.0 | 0.536 ± 0.039 | 0.120 ± 0.017 | 0.499 ± 0.033 | 0.880 ± 0.017 |
| **12** | **2** | **CNN** | **130.2 ± 12.8** | **73.8 ± 12.8** | **14.0 ± 7.1** | **358.0 ± 7.1** | **0.638 ± 0.063** | **0.093 ± 0.039** | **0.596 ± 0.043** | **0.907 ± 0.039** |
| 12 | 2 | Persistence | 159 | 45 | 51 | 321 | 0.779 | 0.243 | 0.624 | 0.757 |
| 12 | 3 | XGBoost | 90.4 ± 2.7 | 113.6 ± 2.7 | 13.4 ± 2.6 | 358.6 ± 2.6 | 0.443 ± 0.013 | 0.129 ± 0.023 | 0.416 ± 0.015 | 0.871 ± 0.023 |
| **12** | **3** | **CNN** | **96.4 ± 28.3** | **107.6 ± 28.3** | **14.0 ± 10.5** | **358.0 ± 10.5** | **0.473 ± 0.139** | **0.112 ± 0.055** | **0.438 ± 0.110** | **0.888 ± 0.055** |
| 12 | 3 | Persistence | 146 | 58 | 67 | 305 | 0.716 | 0.315 | 0.539 | 0.685 |

Drought is SPEI ≤ −1; the test period holds 119, 156 and 204 observed drought months at SPEI-3, SPEI-6 and SPEI-12. Bold rows mark the selected model. FAR and SR are undefined for a seed that issues no drought forecast; a number in parentheses gives the seeds over which they are averaged.

**Table S12.** Reproduction check at lead 1: the seed-42 run of this study against the earlier single-seed result of the same configuration.

| Model | SPEI | RMSE, this study | RMSE, earlier | NSE, this study | NSE, earlier | Skill, this study | Skill, earlier | Largest absolute difference |
|---|---|---|---|---|---|---|---|---|
| XGBoost | 3 | 0.647178 | 0.6447 | 0.567618 | 0.5710 | 0.368813 | 0.3737 | 0.004887 |
| XGBoost | 6 | 0.557340 | 0.5610 | 0.635910 | 0.6311 | 0.163615 | 0.1525 | 0.011115 |
| XGBoost | 12 | 0.323378 | 0.3232 | 0.835722 | 0.8359 | −0.093216 | −0.0922 | 0.001016 |
| CNN | 3 | 0.705634 | 0.705634 | 0.485982 | 0.485982 | 0.249642 | 0.249642 | 0.000000 |
| CNN | 6 | 0.506858 | 0.506858 | 0.698880 | 0.698880 | 0.308269 | 0.308269 | 0.000000 |
| CNN | 12 | 0.281777 | 0.281777 | 0.875270 | 0.875270 | 0.169962 | 0.169962 | 0.000000 |
| Bi-LSTM | 3 | 0.645677 | 0.6457 | 0.569622 | 0.5696 | 0.371739 | 0.3717 | 0.000039 |
| Bi-LSTM | 6 | 0.496631 | 0.4966 | 0.710908 | 0.7109 | 0.335900 | 0.3359 | 0.000031 |
| Bi-LSTM | 12 | 0.310854 | 0.3109 | 0.848200 | 0.8482 | −0.010178 | −0.0102 | 0.000046 |
| CNN-BiLSTM | 3 | 0.621006 | 0.621006 | 0.601883 | 0.601883 | 0.418833 | 0.418833 | 0.000000 |
| CNN-BiLSTM | 6 | 0.535428 | 0.535428 | 0.663976 | 0.663976 | 0.228089 | 0.228089 | 0.000000 |
| CNN-BiLSTM | 12 | 0.291397 | 0.291397 | 0.866608 | 0.866608 | 0.112320 | 0.112320 | 0.000000 |
| CEEMDAN-BiLSTM | 3 | 0.680010 | 0.680010 | 0.522636 | 0.522636 | 0.303148 | 0.303148 | 0.000000 |
| CEEMDAN-BiLSTM | 6 | 0.569016 | 0.569016 | 0.620496 | 0.620496 | 0.128206 | 0.128206 | 0.000000 |
| CEEMDAN-BiLSTM | 12 | 0.314559 | 0.314559 | 0.844560 | 0.844560 | −0.034400 | −0.0344 | 0.000000 |

Earlier values are shown with the precision at which they were stored. The four deep-learning models reproduce to that precision. XGBoost differs by up to 0.011 because its tuned values had been stored rounded to four decimals.

**Table S13.** Sampling uncertainty over the test period: 95% circular block bootstrap intervals of the scores (A), paired differences (B) and Diebold-Mariano tests against persistence (C).

**(A) Scores with 95% intervals**

| SPEI | Lead | Model | RMSE | NSE | Skill | POD | CSI |
|---|---|---|---|---|---|---|---|
| 3 | 1 | XGBoost | 0.645 [0.543, 0.732] | 0.571 [0.481, 0.634] | 0.373 [0.312, 0.433] | 0.284 [0.107, 0.418] | 0.252 [0.095, 0.365] |
| 3 | 1 | CNN | 0.703 [0.615, 0.789] | 0.489 [0.387, 0.560] | 0.255 [0.118, 0.365] | 0.343 [0.127, 0.527] | 0.281 [0.115, 0.414] |
| 3 | 1 | Bi-LSTM | 0.644 [0.553, 0.727] | 0.572 [0.475, 0.640] | 0.374 [0.282, 0.441] | 0.425 [0.243, 0.589] | 0.346 [0.204, 0.469] |
| 3 | 1 | CNN-BiLSTM | 0.627 [0.540, 0.707] | 0.595 [0.509, 0.653] | 0.408 [0.340, 0.473] | 0.356 [0.186, 0.497] | 0.298 [0.152, 0.420] |
| 3 | 1 | CEEMDAN-BiLSTM | 0.691 [0.590, 0.780] | 0.507 [0.366, 0.589] | 0.280 [0.063, 0.378] | 0.397 [0.248, 0.505] | 0.314 [0.189, 0.402] |
| 3 | 1 | Persistence | 0.815 [0.697, 0.918] | 0.315 [0.177, 0.423] | 0 | 0.555 [0.451, 0.616] | 0.379 [0.288, 0.433] |
| 3 | 2 | XGBoost | 0.923 [0.775, 1.042] | 0.120 [−0.009, 0.231] | 0.365 [0.273, 0.445] | 0.005 [0.000, 0.019] | 0.005 [0.000, 0.019] |
| 3 | 2 | CNN-BiLSTM | 0.902 [0.781, 1.014] | 0.159 [0.000, 0.256] | 0.394 [0.282, 0.482] | 0.060 [0.006, 0.149] | 0.054 [0.006, 0.127] |
| 3 | 2 | Persistence | 1.159 [0.968, 1.313] | −0.387 [−0.702, −0.147] | 0 | 0.319 [0.207, 0.466] | 0.183 [0.110, 0.287] |
| 3 | 3 | XGBoost | 1.039 [0.901, 1.164] | −0.115 [−0.284, −0.025] | 0.439 [0.350, 0.528] | 0.000 [0.000, 0.000] | 0.000 [0.000, 0.000] |
| 3 | 3 | CNN-BiLSTM | 1.115 [0.971, 1.248] | −0.283 [−0.503, −0.180] | 0.355 [0.213, 0.474] | 0.002 [0.000, 0.017] | 0.002 [0.000, 0.017] |
| 3 | 3 | Persistence | 1.388 [1.194, 1.549] | −0.989 [−1.386, −0.727] | 0 | 0.185 [0.095, 0.270] | 0.095 [0.046, 0.136] |
| 6 | 1 | XGBoost | 0.564 [0.474, 0.642] | 0.627 [0.524, 0.714] | 0.144 [0.016, 0.237] | 0.606 [0.462, 0.692] | 0.515 [0.362, 0.609] |
| 6 | 1 | CNN | 0.499 [0.402, 0.591] | 0.707 [0.593, 0.795] | 0.327 [0.177, 0.424] | 0.669 [0.433, 0.787] | 0.581 [0.361, 0.705] |
| 6 | 1 | Bi-LSTM | 0.491 [0.393, 0.598] | 0.717 [0.592, 0.806] | 0.350 [0.265, 0.455] | 0.714 [0.578, 0.805] | 0.615 [0.447, 0.723] |
| 6 | 1 | CNN-BiLSTM | 0.519 [0.402, 0.623] | 0.684 [0.557, 0.794] | 0.275 [0.126, 0.401] | 0.624 [0.474, 0.724] | 0.553 [0.392, 0.662] |
| 6 | 1 | CEEMDAN-BiLSTM | 0.574 [0.481, 0.674] | 0.613 [0.478, 0.706] | 0.111 [−0.051, 0.246] | 0.654 [0.495, 0.742] | 0.568 [0.406, 0.669] |
| 6 | 1 | Persistence | 0.609 [0.501, 0.721] | 0.565 [0.405, 0.682] | 0 | 0.769 [0.670, 0.831] | 0.606 [0.492, 0.676] |
| 6 | 2 | XGBoost | 0.712 [0.604, 0.808] | 0.405 [0.255, 0.526] | 0.322 [0.228, 0.385] | 0.433 [0.300, 0.532] | 0.366 [0.244, 0.455] |
| 6 | 2 | Bi-LSTM | 0.688 [0.560, 0.806] | 0.445 [0.239, 0.605] | 0.367 [0.228, 0.513] | 0.506 [0.338, 0.617] | 0.423 [0.277, 0.519] |
| 6 | 2 | Persistence | 0.865 [0.728, 0.988] | 0.123 [−0.144, 0.329] | 0 | 0.583 [0.433, 0.679] | 0.392 [0.270, 0.479] |
| 6 | 3 | XGBoost | 0.847 [0.719, 0.959] | 0.159 [−0.049, 0.308] | 0.372 [0.275, 0.453] | 0.273 [0.148, 0.392] | 0.241 [0.132, 0.333] |
| 6 | 3 | Bi-LSTM | 0.851 [0.708, 0.986] | 0.151 [−0.133, 0.343] | 0.367 [0.190, 0.497] | 0.360 [0.230, 0.485] | 0.313 [0.195, 0.419] |
| 6 | 3 | Persistence | 1.069 [0.908, 1.227] | −0.340 [−0.728, −0.051] | 0 | 0.449 [0.311, 0.549] | 0.272 [0.175, 0.346] |
| 12 | 1 | XGBoost | 0.323 [0.284, 0.363] | 0.836 [0.741, 0.884] | −0.093 [−0.258, 0.043] | 0.715 [0.564, 0.813] | 0.664 [0.503, 0.759] |
| 12 | 1 | CNN | 0.285 [0.252, 0.321] | 0.872 [0.796, 0.910] | 0.148 [−0.045, 0.304] | 0.810 [0.691, 0.883] | 0.735 [0.608, 0.809] |
| 12 | 1 | Bi-LSTM | 0.286 [0.247, 0.333] | 0.871 [0.783, 0.913] | 0.141 [−0.229, 0.363] | 0.788 [0.648, 0.892] | 0.731 [0.600, 0.825] |
| 12 | 1 | CNN-BiLSTM | 0.292 [0.252, 0.332] | 0.866 [0.778, 0.907] | 0.110 [−0.179, 0.302] | 0.761 [0.649, 0.823] | 0.715 [0.596, 0.785] |
| 12 | 1 | CEEMDAN-BiLSTM | 0.314 [0.268, 0.355] | 0.845 [0.750, 0.893] | −0.034 [−0.390, 0.218] | 0.783 [0.661, 0.860] | 0.730 [0.615, 0.803] |
| 12 | 1 | Persistence | 0.309 [0.277, 0.345] | 0.850 [0.773, 0.891] | 0 | 0.853 [0.765, 0.905] | 0.734 [0.610, 0.807] |
| 12 | 2 | XGBoost | 0.474 [0.399, 0.557] | 0.647 [0.414, 0.765] | −0.042 [−0.270, 0.151] | 0.536 [0.405, 0.655] | 0.499 [0.371, 0.608] |
| 12 | 2 | CNN | 0.468 [0.376, 0.629] | 0.652 [0.282, 0.782] | −0.025 [−0.945, 0.269] | 0.638 [0.479, 0.742] | 0.596 [0.465, 0.692] |
| 12 | 2 | Persistence | 0.465 [0.408, 0.531] | 0.661 [0.471, 0.763] | 0 | 0.779 [0.677, 0.861] | 0.624 [0.512, 0.718] |
| 12 | 3 | XGBoost | 0.586 [0.487, 0.693] | 0.460 [0.080, 0.642] | −0.004 [−0.343, 0.251] | 0.443 [0.324, 0.538] | 0.416 [0.299, 0.510] |
| 12 | 3 | CNN | 0.617 [0.485, 0.820] | 0.393 [−0.201, 0.638] | −0.129 [−1.156, 0.259] | 0.472 [0.210, 0.673] | 0.438 [0.207, 0.590] |
| 12 | 3 | Persistence | 0.585 [0.513, 0.658] | 0.462 [0.190, 0.625] | 0 | 0.716 [0.578, 0.836] | 0.539 [0.416, 0.645] |

### (B) Paired differences with 95% intervals

| SPEI | Lead | Model | ΔPOD (model − persistence) | ΔCSI (model − persistence) | ΔSkill (model − XGBoost) |
|---|---|---|---|---|---|
| 3 | 1 | XGBoost | −0.271 [−0.440, −0.135] | −0.127 [−0.274, −0.016] | n/a |
| 3 | 1 | CNN | −0.212 [−0.433, −0.021] | −0.098 [−0.273, +0.038] | −0.118 [−0.240, −0.018] |
| 3 | 1 | Bi-LSTM | −0.129 [−0.303, +0.010] | −0.033 [−0.168, +0.071] | +0.002 [−0.095, +0.073] |
| 3 | 1 | CNN-BiLSTM | −0.198 [−0.371, −0.061] | −0.081 [−0.231, +0.038] | +0.035 [−0.029, +0.090] |
| 3 | 1 | CEEMDAN-BiLSTM | −0.158 [−0.297, −0.055] | −0.066 [−0.181, +0.021] | −0.093 [−0.307, −0.002] |
| 3 | 2 | XGBoost | −0.314 [−0.466, −0.198] | −0.178 [−0.287, −0.102] | n/a |
| 3 | 2 | CNN-BiLSTM | −0.259 [−0.434, −0.107] | −0.128 [−0.259, −0.018] | +0.028 [−0.049, +0.096] |
| 3 | 3 | XGBoost | −0.185 [−0.270, −0.095] | −0.095 [−0.136, −0.046] | n/a |
| 3 | 3 | CNN-BiLSTM | −0.183 [−0.269, −0.095] | −0.094 [−0.134, −0.045] | −0.085 [−0.166, −0.015] |
| 6 | 1 | XGBoost | −0.163 [−0.243, −0.092] | −0.091 [−0.169, −0.027] | n/a |
| 6 | 1 | CNN | −0.100 [−0.277, −0.007] | −0.025 [−0.171, +0.063] | +0.184 [+0.066, +0.343] |
| 6 | 1 | Bi-LSTM | −0.055 [−0.151, +0.025] | +0.009 [−0.087, +0.085] | +0.206 [+0.063, +0.368] |
| 6 | 1 | CNN-BiLSTM | −0.145 [−0.252, −0.058] | −0.053 [−0.156, +0.031] | +0.131 [−0.005, +0.286] |
| 6 | 1 | CEEMDAN-BiLSTM | −0.115 [−0.221, −0.045] | −0.038 [−0.125, +0.034] | −0.032 [−0.180, +0.121] |
| 6 | 2 | XGBoost | −0.150 [−0.228, −0.066] | −0.026 [−0.101, +0.056] | n/a |
| 6 | 2 | Bi-LSTM | −0.077 [−0.208, +0.018] | +0.031 [−0.067, +0.111] | +0.045 [−0.085, +0.194] |
| 6 | 3 | XGBoost | −0.176 [−0.281, −0.070] | −0.031 [−0.119, +0.061] | n/a |
| 6 | 3 | Bi-LSTM | −0.088 [−0.203, +0.037] | +0.041 [−0.045, +0.149] | −0.005 [−0.139, +0.095] |
| 12 | 1 | XGBoost | −0.138 [−0.244, −0.067] | −0.070 [−0.138, −0.021] | n/a |
| 12 | 1 | CNN | −0.043 [−0.118, +0.013] | +0.001 [−0.043, +0.049] | +0.241 [+0.042, +0.392] |
| 12 | 1 | Bi-LSTM | −0.065 [−0.192, +0.023] | −0.003 [−0.098, +0.067] | +0.234 [−0.113, +0.475] |
| 12 | 1 | CNN-BiLSTM | −0.092 [−0.166, −0.033] | −0.019 [−0.075, +0.034] | +0.203 [−0.042, +0.383] |
| 12 | 1 | CEEMDAN-BiLSTM | −0.070 [−0.164, −0.004] | −0.004 [−0.070, +0.065] | +0.059 [−0.322, +0.295] |
| 12 | 2 | XGBoost | −0.243 [−0.361, −0.128] | −0.124 [−0.208, −0.040] | n/a |
| 12 | 2 | CNN | −0.141 [−0.302, −0.038] | −0.027 [−0.151, +0.063] | +0.017 [−0.839, +0.289] |
| 12 | 3 | XGBoost | −0.273 [−0.391, −0.153] | −0.123 [−0.196, −0.029] | n/a |
| 12 | 3 | CNN | −0.243 [−0.561, −0.049] | −0.101 [−0.343, +0.068] | −0.125 [−1.076, +0.271] |

## (C) Diebold-Mariano tests against persistence

| SPEI | Lead | Model | DM*, ensemble | p, ensemble | Mean loss differential | p, seed 42 | p, seed 43 | p, seed 44 | p, seed 45 | p, seed 46 | Seeds with p < 0.05 |
|---|---|---|---|---|---|---|---|---|---|---|---|
| 3 | 1 | XGBoost | −4.427 | <0.0001 | −0.2511 | 0.0001 | 0.0001 | <0.0001 | 0.0001 | <0.0001 | 5/5 |
| 3 | 1 | CNN | −3.452 | 0.0009 | −0.2131 | 0.0073 | 0.0772 | 0.0080 | 0.0004 | 0.0256 | 4/5 |
| 3 | 1 | Bi-LSTM | −5.056 | <0.0001 | −0.2638 | <0.0001 | <0.0001 | 0.0001 | <0.0001 | <0.0001 | 5/5 |
| 3 | 1 | CNN-BiLSTM | −4.967 | <0.0001 | −0.2770 | <0.0001 | <0.0001 | <0.0001 | <0.0001 | <0.0001 | 5/5 |
| 3 | 1 | CEEMDAN-BiLSTM | −3.864 | 0.0002 | −0.2239 | 0.0008 | 0.1567 | 0.0014 | 0.0001 | 0.0003 | 4/5 |
| 3 | 2 | XGBoost | −4.037 | 0.0001 | −0.4961 | 0.0001 | 0.0003 | 0.0001 | 0.0002 | 0.0002 | 5/5 |
| 3 | 2 | CNN-BiLSTM | −4.112 | 0.0001 | −0.5434 | 0.0002 | 0.0001 | 0.0001 | 0.0003 | 0.0003 | 5/5 |
| 3 | 3 | XGBoost | −4.737 | <0.0001 | −0.8530 | <0.0001 | <0.0001 | <0.0001 | <0.0001 | <0.0001 | 5/5 |
| 3 | 3 | CNN-BiLSTM | −3.241 | 0.0018 | −0.7012 | 0.0025 | 0.0012 | 0.0038 | 0.0020 | 0.0040 | 5/5 |
| 6 | 1 | XGBoost | −1.820 | 0.0730 | −0.0581 | 0.0462 | 0.0566 | 0.2043 | 0.0643 | 0.1919 | 1/5 |
| 6 | 1 | CNN | −3.513 | 0.0008 | −0.1334 | 0.0072 | 0.0070 | 0.0004 | 0.0232 | 0.0002 | 5/5 |
| 6 | 1 | Bi-LSTM | −4.089 | 0.0001 | −0.1383 | 0.0002 | 0.0001 | 0.0003 | 0.0002 | 0.0012 | 5/5 |
| 6 | 1 | CNN-BiLSTM | −2.888 | 0.0051 | −0.1166 | 0.0534 | 0.0520 | 0.0082 | 0.0139 | 0.0012 | 3/5 |
| 6 | 1 | CEEMDAN-BiLSTM | −2.142 | 0.0356 | −0.0668 | 0.2559 | 0.5898 | 0.0738 | 0.0694 | 0.5262 | 0/5 |
| 6 | 2 | XGBoost | −3.154 | 0.0024 | −0.2467 | 0.0028 | 0.0042 | 0.0036 | 0.0043 | 0.0013 | 5/5 |
| 6 | 2 | Bi-LSTM | −3.844 | 0.0003 | −0.2968 | 0.0002 | 0.0035 | 0.0058 | <0.0001 | 0.0022 | 5/5 |
| 6 | 3 | XGBoost | −3.713 | 0.0004 | −0.4363 | 0.0021 | 0.0004 | 0.0004 | 0.0005 | 0.0005 | 5/5 |
| 6 | 3 | Bi-LSTM | −3.487 | 0.0008 | −0.4504 | 0.0025 | 0.0004 | 0.0006 | 0.0233 | 0.0012 | 5/5 |
| 12 | 1 | XGBoost | 0.761 | 0.4489 | 0.0059 | 0.2504 | 0.7094 | 0.1853 | 0.1729 | 0.2570 | 0/5 |
| 12 | 1 | CNN | −2.821 | 0.0062 | −0.0214 | 0.1021 | 0.0360 | 0.1193 | 0.0303 | 0.2974 | 2/5 |
| 12 | 1 | Bi-LSTM | −1.827 | 0.0720 | −0.0184 | 0.9463 | 0.0209 | 0.7301 | 0.0346 | 0.0503 | 2/5 |
| 12 | 1 | CNN-BiLSTM | −1.146 | 0.2555 | −0.0127 | 0.4087 | 0.4223 | 0.2648 | 0.0673 | 0.8450 | 0/5 |
| 12 | 1 | CEEMDAN-BiLSTM | −1.251 | 0.2152 | −0.0118 | 0.8051 | 0.3729 | 0.4234 | 0.4681 | 0.4059 | 0/5 |
| 12 | 2 | XGBoost | 0.151 | 0.8805 | 0.0038 | 0.9462 | 0.6752 | 0.3800 | 0.7510 | 0.9717 | 0/5 |
| 12 | 2 | CNN | −1.160 | 0.2500 | −0.0278 | 0.6926 | 0.1173 | 0.1899 | 0.2181 | 0.6478 | 0/5 |
| 12 | 3 | XGBoost | −0.118 | 0.9067 | −0.0061 | 0.7442 | 0.6335 | 0.6975 | 0.8511 | 0.9366 | 0/5 |
| 12 | 3 | CNN | −0.372 | 0.7110 | −0.0195 | 0.8580 | 0.4395 | 0.0627 | 0.9617 | 0.0180 | 1/5 |

Intervals come from 2,000 resamples of the 72 test months in circular blocks of 12 months, with all divisions of a month kept together and the scores of the five seeds pooled (Text S5). Point values are seed means. (B) Differences are paired by resample; the skill difference is taken against the seed-mean XGBoost skill. (C) DM* is the Diebold-Mariano statistic with the correction of Harvey et al. (1997), computed on the monthly mean loss differential (model minus persistence squared error, $T = 72$); a negative value favors the model.

**Table S14.** Scores of every run over the test period (135 runs: five models at lead 1, and the selected model and XGBoost at leads 2 and 3, each with seeds 42-46).

| Model | SPEI | Lead | Seed | RMSE | MAE | NSE | WI | R² | Skill | POD | FAR | CSI | Hits | Misses | False alarms |
|---|---|---|---|---|---|---|---|---|---|---|---|---|---|---|---|
| XGBoost | 3 | 1 | 42 | 0.647 | 0.518 | 0.568 | 0.838 | 0.572 | 0.369 | 0.286 | 0.244 | 0.262 | 34 | 85 | 11 |
| XGBoost | 3 | 1 | 43 | 0.648 | 0.517 | 0.566 | 0.838 | 0.569 | 0.366 | 0.294 | 0.300 | 0.261 | 35 | 84 | 15 |
| XGBoost | 3 | 1 | 44 | 0.641 | 0.516 | 0.576 | 0.836 | 0.585 | 0.381 | 0.261 | 0.340 | 0.230 | 31 | 88 | 16 |
| XGBoost | 3 | 1 | 45 | 0.645 | 0.520 | 0.570 | 0.836 | 0.577 | 0.372 | 0.261 | 0.354 | 0.228 | 31 | 88 | 17 |
| XGBoost | 3 | 1 | 46 | 0.643 | 0.518 | 0.573 | 0.839 | 0.577 | 0.376 | 0.319 | 0.309 | 0.279 | 38 | 81 | 17 |
| XGBoost | 3 | 2 | 42 | 0.920 | 0.752 | 0.127 | 0.486 | 0.128 | 0.371 | 0.000 | n/a | 0.000 | 0 | 119 | 0 |
| XGBoost | 3 | 2 | 43 | 0.929 | 0.756 | 0.108 | 0.486 | 0.113 | 0.357 | 0.008 | 0.000 | 0.008 | 1 | 118 | 0 |
| XGBoost | 3 | 2 | 44 | 0.923 | 0.756 | 0.120 | 0.468 | 0.122 | 0.366 | 0.000 | n/a | 0.000 | 0 | 119 | 0 |
| XGBoost | 3 | 2 | 45 | 0.928 | 0.752 | 0.112 | 0.497 | 0.118 | 0.360 | 0.008 | 0.750 | 0.008 | 1 | 118 | 3 |
| XGBoost | 3 | 2 | 46 | 0.917 | 0.749 | 0.133 | 0.500 | 0.134 | 0.375 | 0.008 | 0.000 | 0.008 | 1 | 118 | 0 |
| XGBoost | 3 | 3 | 42 | 1.031 | 0.854 | −0.097 | 0.122 | 0.067 | 0.448 | 0.000 | n/a | 0.000 | 0 | 119 | 0 |
| XGBoost | 3 | 3 | 43 | 1.053 | 0.867 | −0.144 | 0.145 | 0.057 | 0.425 | 0.000 | n/a | 0.000 | 0 | 119 | 0 |
| XGBoost | 3 | 3 | 44 | 1.005 | 0.835 | −0.042 | 0.151 | 0.038 | 0.476 | 0.000 | n/a | 0.000 | 0 | 119 | 0 |
| XGBoost | 3 | 3 | 45 | 1.043 | 0.865 | −0.123 | 0.135 | 0.055 | 0.435 | 0.000 | n/a | 0.000 | 0 | 119 | 0 |
| XGBoost | 3 | 3 | 46 | 1.065 | 0.875 | −0.170 | 0.155 | 0.058 | 0.412 | 0.000 | 1.000 | 0.000 | 0 | 119 | 1 |
| XGBoost | 6 | 1 | 42 | 0.557 | 0.423 | 0.636 | 0.882 | 0.641 | 0.164 | 0.628 | 0.210 | 0.538 | 98 | 58 | 26 |
| XGBoost | 6 | 1 | 43 | 0.560 | 0.424 | 0.632 | 0.883 | 0.640 | 0.155 | 0.603 | 0.230 | 0.511 | 94 | 62 | 28 |
| XGBoost | 6 | 1 | 44 | 0.573 | 0.436 | 0.615 | 0.875 | 0.622 | 0.115 | 0.590 | 0.246 | 0.495 | 92 | 64 | 30 |
| XGBoost | 6 | 1 | 45 | 0.562 | 0.423 | 0.630 | 0.883 | 0.638 | 0.150 | 0.609 | 0.215 | 0.522 | 95 | 61 | 26 |
| XGBoost | 6 | 1 | 46 | 0.567 | 0.432 | 0.623 | 0.879 | 0.631 | 0.135 | 0.603 | 0.230 | 0.511 | 94 | 62 | 28 |
| XGBoost | 6 | 2 | 42 | 0.714 | 0.553 | 0.402 | 0.786 | 0.425 | 0.319 | 0.429 | 0.316 | 0.358 | 67 | 89 | 31 |
| XGBoost | 6 | 2 | 43 | 0.716 | 0.550 | 0.398 | 0.785 | 0.422 | 0.314 | 0.410 | 0.281 | 0.354 | 64 | 92 | 25 |
| XGBoost | 6 | 2 | 44 | 0.715 | 0.551 | 0.401 | 0.784 | 0.420 | 0.317 | 0.442 | 0.296 | 0.373 | 69 | 87 | 29 |
| XGBoost | 6 | 2 | 45 | 0.715 | 0.549 | 0.400 | 0.785 | 0.421 | 0.316 | 0.429 | 0.287 | 0.366 | 67 | 89 | 27 |
| XGBoost | 6 | 2 | 46 | 0.701 | 0.539 | 0.424 | 0.794 | 0.438 | 0.343 | 0.455 | 0.304 | 0.380 | 71 | 85 | 31 |
| XGBoost | 6 | 3 | 42 | 0.861 | 0.689 | 0.131 | 0.662 | 0.208 | 0.352 | 0.301 | 0.382 | 0.254 | 47 | 109 | 29 |
| XGBoost | 6 | 3 | 43 | 0.838 | 0.668 | 0.178 | 0.628 | 0.216 | 0.386 | 0.250 | 0.204 | 0.235 | 39 | 117 | 10 |
| XGBoost | 6 | 3 | 44 | 0.840 | 0.675 | 0.174 | 0.633 | 0.215 | 0.383 | 0.224 | 0.300 | 0.205 | 35 | 121 | 15 |
| XGBoost | 6 | 3 | 45 | 0.845 | 0.676 | 0.162 | 0.650 | 0.212 | 0.375 | 0.295 | 0.313 | 0.260 | 46 | 110 | 21 |
| XGBoost | 6 | 3 | 46 | 0.853 | 0.678 | 0.147 | 0.654 | 0.207 | 0.363 | 0.295 | 0.370 | 0.251 | 46 | 110 | 27 |
| XGBoost | 12 | 1 | 42 | 0.323 | 0.246 | 0.836 | 0.954 | 0.859 | −0.093 | 0.706 | 0.089 | 0.661 | 144 | 60 | 14 |
| XGBoost | 12 | 1 | 43 | 0.314 | 0.240 | 0.845 | 0.957 | 0.867 | −0.030 | 0.716 | 0.087 | 0.670 | 146 | 58 | 14 |
| XGBoost | 12 | 1 | 44 | 0.327 | 0.248 | 0.832 | 0.953 | 0.853 | −0.115 | 0.730 | 0.102 | 0.674 | 149 | 55 | 17 |
| XGBoost | 12 | 1 | 45 | 0.329 | 0.252 | 0.830 | 0.952 | 0.856 | −0.131 | 0.691 | 0.096 | 0.644 | 141 | 63 | 15 |
| XGBoost | 12 | 1 | 46 | 0.324 | 0.249 | 0.836 | 0.954 | 0.859 | −0.094 | 0.730 | 0.108 | 0.671 | 149 | 55 | 18 |
| XGBoost | 12 | 2 | 42 | 0.467 | 0.347 | 0.658 | 0.899 | 0.720 | −0.008 | 0.529 | 0.100 | 0.500 | 108 | 96 | 12 |
| XGBoost | 12 | 2 | 43 | 0.476 | 0.350 | 0.644 | 0.895 | 0.697 | −0.049 | 0.583 | 0.144 | 0.531 | 119 | 85 | 20 |
| XGBoost | 12 | 2 | 44 | 0.490 | 0.369 | 0.623 | 0.887 | 0.686 | −0.111 | 0.495 | 0.129 | 0.461 | 101 | 103 | 15 |
| XGBoost | 12 | 2 | 45 | 0.473 | 0.359 | 0.648 | 0.895 | 0.707 | −0.036 | 0.505 | 0.120 | 0.472 | 103 | 101 | 14 |
| XGBoost | 12 | 2 | 46 | 0.466 | 0.349 | 0.659 | 0.900 | 0.719 | −0.004 | 0.569 | 0.108 | 0.532 | 116 | 88 | 14 |
| XGBoost | 12 | 3 | 42 | 0.602 | 0.453 | 0.431 | 0.824 | 0.545 | −0.057 | 0.431 | 0.137 | 0.404 | 88 | 116 | 14 |
| XGBoost | 12 | 3 | 43 | 0.566 | 0.428 | 0.497 | 0.840 | 0.583 | 0.066 | 0.466 | 0.112 | 0.440 | 95 | 109 | 12 |

| Model | SPEI | Lead | Seed | RMSE | MAE | NSE | WI | $R^2$ | Skill | POD | FAR | CSI | Hits | Misses | False alarms |
|---|---|---|---|---|---|---|---|---|---|---|---|---|---|---|---|
| XGBoost | 12 | 3 | 44 | 0.604 | 0.456 | 0.427 | 0.826 | 0.549 | −0.065 | 0.441 | 0.100 | 0.421 | 90 | 114 | 10 |
| XGBoost | 12 | 3 | 45 | 0.577 | 0.439 | 0.476 | 0.832 | 0.570 | 0.026 | 0.436 | 0.136 | 0.408 | 89 | 115 | 14 |
| XGBoost | 12 | 3 | 46 | 0.582 | 0.444 | 0.469 | 0.835 | 0.567 | 0.012 | 0.441 | 0.159 | 0.407 | 90 | 114 | 17 |
| CNN | 3 | 1 | 42 | 0.706 | 0.565 | 0.486 | 0.820 | 0.493 | 0.250 | 0.378 | 0.400 | 0.302 | 45 | 74 | 30 |
| CNN | 3 | 1 | 43 | 0.722 | 0.573 | 0.462 | 0.801 | 0.472 | 0.215 | 0.261 | 0.354 | 0.228 | 31 | 88 | 17 |
| CNN | 3 | 1 | 44 | 0.704 | 0.566 | 0.488 | 0.825 | 0.498 | 0.252 | 0.412 | 0.388 | 0.327 | 49 | 70 | 31 |
| CNN | 3 | 1 | 45 | 0.680 | 0.549 | 0.523 | 0.830 | 0.529 | 0.303 | 0.403 | 0.377 | 0.324 | 48 | 71 | 29 |
| CNN | 3 | 1 | 46 | 0.704 | 0.566 | 0.489 | 0.814 | 0.500 | 0.254 | 0.261 | 0.367 | 0.226 | 31 | 88 | 18 |
| CNN | 6 | 1 | 42 | 0.507 | 0.383 | 0.699 | 0.909 | 0.700 | 0.308 | 0.660 | 0.202 | 0.566 | 103 | 53 | 26 |
| CNN | 6 | 1 | 43 | 0.521 | 0.384 | 0.682 | 0.904 | 0.686 | 0.269 | 0.628 | 0.183 | 0.551 | 98 | 58 | 22 |
| CNN | 6 | 1 | 44 | 0.475 | 0.354 | 0.735 | 0.922 | 0.737 | 0.392 | 0.731 | 0.203 | 0.616 | 114 | 42 | 29 |
| CNN | 6 | 1 | 45 | 0.517 | 0.373 | 0.686 | 0.904 | 0.697 | 0.279 | 0.622 | 0.157 | 0.557 | 97 | 59 | 18 |
| CNN | 6 | 1 | 46 | 0.477 | 0.355 | 0.734 | 0.920 | 0.734 | 0.389 | 0.705 | 0.173 | 0.615 | 110 | 46 | 23 |
| CNN | 12 | 1 | 42 | 0.282 | 0.205 | 0.875 | 0.965 | 0.877 | 0.170 | 0.819 | 0.097 | 0.752 | 167 | 37 | 18 |
| CNN | 12 | 1 | 43 | 0.289 | 0.214 | 0.869 | 0.965 | 0.874 | 0.129 | 0.784 | 0.111 | 0.714 | 160 | 44 | 20 |
| CNN | 12 | 1 | 44 | 0.282 | 0.202 | 0.875 | 0.966 | 0.876 | 0.168 | 0.809 | 0.098 | 0.743 | 165 | 39 | 18 |
| CNN | 12 | 1 | 45 | 0.278 | 0.200 | 0.878 | 0.967 | 0.880 | 0.190 | 0.809 | 0.108 | 0.737 | 165 | 39 | 20 |
| CNN | 12 | 1 | 46 | 0.296 | 0.216 | 0.863 | 0.963 | 0.863 | 0.085 | 0.828 | 0.142 | 0.728 | 169 | 35 | 28 |
| CNN | 12 | 2 | 42 | 0.456 | 0.322 | 0.674 | 0.907 | 0.720 | 0.038 | 0.632 | 0.065 | 0.606 | 129 | 75 | 9 |
| CNN | 12 | 2 | 43 | 0.566 | 0.404 | 0.497 | 0.849 | 0.587 | −0.483 | 0.534 | 0.044 | 0.522 | 109 | 95 | 5 |
| CNN | 12 | 2 | 44 | 0.432 | 0.308 | 0.706 | 0.915 | 0.726 | 0.134 | 0.652 | 0.101 | 0.607 | 133 | 71 | 15 |
| CNN | 12 | 2 | 45 | 0.431 | 0.320 | 0.708 | 0.921 | 0.741 | 0.140 | 0.691 | 0.113 | 0.635 | 141 | 63 | 18 |
| CNN | 12 | 2 | 46 | 0.453 | 0.330 | 0.677 | 0.897 | 0.679 | 0.048 | 0.681 | 0.142 | 0.612 | 139 | 65 | 23 |
| CNN | 12 | 3 | 42 | 0.578 | 0.425 | 0.475 | 0.845 | 0.556 | 0.024 | 0.529 | 0.122 | 0.493 | 108 | 96 | 15 |
| CNN | 12 | 3 | 43 | 0.626 | 0.490 | 0.385 | 0.802 | 0.542 | −0.143 | 0.309 | 0.060 | 0.303 | 63 | 141 | 4 |
| CNN | 12 | 3 | 44 | 0.533 | 0.397 | 0.553 | 0.860 | 0.568 | 0.170 | 0.623 | 0.186 | 0.545 | 127 | 77 | 29 |
| CNN | 12 | 3 | 45 | 0.583 | 0.437 | 0.466 | 0.844 | 0.538 | 0.007 | 0.559 | 0.136 | 0.514 | 114 | 90 | 18 |
| CNN | 12 | 3 | 46 | 0.763 | 0.594 | 0.084 | 0.764 | 0.501 | −0.702 | 0.343 | 0.054 | 0.337 | 70 | 134 | 4 |
| Bi-LSTM | 3 | 1 | 42 | 0.646 | 0.520 | 0.570 | 0.855 | 0.576 | 0.372 | 0.345 | 0.328 | 0.295 | 41 | 78 | 20 |
| Bi-LSTM | 3 | 1 | 43 | 0.649 | 0.515 | 0.565 | 0.858 | 0.571 | 0.365 | 0.420 | 0.351 | 0.342 | 50 | 69 | 27 |
| Bi-LSTM | 3 | 1 | 44 | 0.651 | 0.513 | 0.563 | 0.854 | 0.576 | 0.362 | 0.529 | 0.357 | 0.409 | 63 | 56 | 35 |
| Bi-LSTM | 3 | 1 | 45 | 0.636 | 0.507 | 0.582 | 0.860 | 0.583 | 0.390 | 0.412 | 0.329 | 0.343 | 49 | 70 | 24 |
| Bi-LSTM | 3 | 1 | 46 | 0.639 | 0.505 | 0.578 | 0.857 | 0.579 | 0.384 | 0.420 | 0.359 | 0.340 | 50 | 69 | 28 |
| Bi-LSTM | 6 | 1 | 42 | 0.497 | 0.365 | 0.711 | 0.917 | 0.716 | 0.336 | 0.731 | 0.214 | 0.610 | 114 | 42 | 31 |
| Bi-LSTM | 6 | 1 | 43 | 0.501 | 0.358 | 0.706 | 0.920 | 0.722 | 0.324 | 0.718 | 0.182 | 0.619 | 112 | 44 | 25 |
| Bi-LSTM | 6 | 1 | 44 | 0.481 | 0.347 | 0.729 | 0.925 | 0.745 | 0.378 | 0.679 | 0.197 | 0.582 | 106 | 50 | 26 |
| Bi-LSTM | 6 | 1 | 45 | 0.467 | 0.340 | 0.744 | 0.928 | 0.753 | 0.413 | 0.724 | 0.144 | 0.646 | 113 | 43 | 19 |
| Bi-LSTM | 6 | 1 | 46 | 0.511 | 0.369 | 0.694 | 0.915 | 0.710 | 0.298 | 0.718 | 0.182 | 0.619 | 112 | 44 | 25 |
| Bi-LSTM | 6 | 2 | 42 | 0.682 | 0.505 | 0.454 | 0.827 | 0.486 | 0.378 | 0.526 | 0.255 | 0.446 | 82 | 74 | 28 |
| Bi-LSTM | 6 | 2 | 43 | 0.705 | 0.517 | 0.418 | 0.819 | 0.464 | 0.336 | 0.506 | 0.255 | 0.432 | 79 | 77 | 27 |
| Bi-LSTM | 6 | 2 | 44 | 0.715 | 0.524 | 0.402 | 0.811 | 0.451 | 0.318 | 0.429 | 0.272 | 0.370 | 67 | 89 | 25 |
| Bi-LSTM | 6 | 2 | 45 | 0.622 | 0.462 | 0.547 | 0.851 | 0.556 | 0.484 | 0.519 | 0.250 | 0.443 | 81 | 75 | 27 |
| Bi-LSTM | 6 | 2 | 46 | 0.714 | 0.524 | 0.402 | 0.819 | 0.455 | 0.319 | 0.551 | 0.348 | 0.426 | 86 | 70 | 46 |

| Model | SPEI | Lead | Seed | RMSE | MAE | NSE | WI | R² | Skill | POD | FAR | CSI | Hits | Misses | False alarms |
|---|---|---|---|---|---|---|---|---|---|---|---|---|---|---|---|
| Bi-LSTM | 6 | 3 | 42 | 0.838 | 0.639 | 0.177 | 0.714 | 0.269 | 0.386 | 0.353 | 0.353 | 0.296 | 55 | 101 | 30 |
| Bi-LSTM | 6 | 3 | 43 | 0.814 | 0.627 | 0.223 | 0.730 | 0.302 | 0.420 | 0.391 | 0.299 | 0.335 | 61 | 95 | 26 |
| Bi-LSTM | 6 | 3 | 44 | 0.854 | 0.656 | 0.145 | 0.684 | 0.233 | 0.362 | 0.353 | 0.267 | 0.312 | 55 | 101 | 20 |
| Bi-LSTM | 6 | 3 | 45 | 0.879 | 0.653 | 0.095 | 0.697 | 0.234 | 0.324 | 0.346 | 0.280 | 0.305 | 54 | 102 | 21 |
| Bi-LSTM | 6 | 3 | 46 | 0.868 | 0.669 | 0.118 | 0.673 | 0.216 | 0.341 | 0.359 | 0.273 | 0.316 | 56 | 100 | 21 |
| Bi-LSTM | 12 | 1 | 42 | 0.311 | 0.217 | 0.848 | 0.958 | 0.862 | −0.010 | 0.716 | 0.082 | 0.673 | 146 | 58 | 13 |
| Bi-LSTM | 12 | 1 | 43 | 0.265 | 0.190 | 0.890 | 0.970 | 0.890 | 0.266 | 0.833 | 0.091 | 0.769 | 170 | 34 | 17 |
| Bi-LSTM | 12 | 1 | 44 | 0.302 | 0.207 | 0.857 | 0.961 | 0.862 | 0.047 | 0.799 | 0.114 | 0.724 | 163 | 41 | 21 |
| Bi-LSTM | 12 | 1 | 45 | 0.273 | 0.195 | 0.883 | 0.969 | 0.889 | 0.221 | 0.814 | 0.088 | 0.755 | 166 | 38 | 16 |
| Bi-LSTM | 12 | 1 | 46 | 0.280 | 0.199 | 0.877 | 0.967 | 0.889 | 0.182 | 0.779 | 0.070 | 0.736 | 159 | 45 | 12 |
| CNN-BiLSTM | 3 | 1 | 42 | 0.621 | 0.495 | 0.602 | 0.863 | 0.602 | 0.419 | 0.303 | 0.429 | 0.247 | 36 | 83 | 27 |
| CNN-BiLSTM | 3 | 1 | 43 | 0.617 | 0.493 | 0.606 | 0.867 | 0.607 | 0.426 | 0.420 | 0.342 | 0.345 | 50 | 69 | 26 |
| CNN-BiLSTM | 3 | 1 | 44 | 0.623 | 0.493 | 0.600 | 0.861 | 0.600 | 0.416 | 0.311 | 0.339 | 0.268 | 37 | 82 | 19 |
| CNN-BiLSTM | 3 | 1 | 45 | 0.635 | 0.503 | 0.583 | 0.857 | 0.584 | 0.391 | 0.412 | 0.329 | 0.343 | 49 | 70 | 24 |
| CNN-BiLSTM | 3 | 1 | 46 | 0.636 | 0.511 | 0.582 | 0.858 | 0.584 | 0.390 | 0.336 | 0.333 | 0.288 | 40 | 79 | 20 |
| CNN-BiLSTM | 3 | 2 | 42 | 0.904 | 0.737 | 0.156 | 0.575 | 0.173 | 0.392 | 0.042 | 0.688 | 0.038 | 5 | 114 | 11 |
| CNN-BiLSTM | 3 | 2 | 43 | 0.902 | 0.733 | 0.160 | 0.588 | 0.175 | 0.394 | 0.101 | 0.600 | 0.088 | 12 | 107 | 18 |
| CNN-BiLSTM | 3 | 2 | 44 | 0.890 | 0.720 | 0.183 | 0.580 | 0.190 | 0.411 | 0.067 | 0.467 | 0.063 | 8 | 111 | 7 |
| CNN-BiLSTM | 3 | 2 | 45 | 0.909 | 0.742 | 0.147 | 0.562 | 0.158 | 0.385 | 0.067 | 0.667 | 0.059 | 8 | 111 | 16 |
| CNN-BiLSTM | 3 | 2 | 46 | 0.907 | 0.731 | 0.151 | 0.583 | 0.171 | 0.388 | 0.025 | 0.786 | 0.023 | 3 | 116 | 11 |
| CNN-BiLSTM | 3 | 3 | 42 | 1.110 | 0.909 | −0.272 | 0.249 | 0.015 | 0.360 | 0.000 | 1.000 | 0.000 | 0 | 119 | 3 |
| CNN-BiLSTM | 3 | 3 | 43 | 1.115 | 0.911 | −0.283 | 0.229 | 0.027 | 0.355 | 0.000 | n/a | 0.000 | 0 | 119 | 0 |
| CNN-BiLSTM | 3 | 3 | 44 | 1.115 | 0.908 | −0.282 | 0.223 | 0.028 | 0.355 | 0.000 | 1.000 | 0.000 | 0 | 119 | 1 |
| CNN-BiLSTM | 3 | 3 | 45 | 1.115 | 0.914 | −0.284 | 0.225 | 0.026 | 0.355 | 0.000 | n/a | 0.000 | 0 | 119 | 0 |
| CNN-BiLSTM | 3 | 3 | 46 | 1.120 | 0.909 | −0.295 | 0.268 | 0.012 | 0.349 | 0.008 | 0.667 | 0.008 | 1 | 118 | 2 |
| CNN-BiLSTM | 6 | 1 | 42 | 0.535 | 0.376 | 0.664 | 0.899 | 0.682 | 0.228 | 0.590 | 0.171 | 0.526 | 92 | 64 | 19 |
| CNN-BiLSTM | 6 | 1 | 43 | 0.531 | 0.377 | 0.670 | 0.906 | 0.690 | 0.242 | 0.622 | 0.171 | 0.551 | 97 | 59 | 20 |
| CNN-BiLSTM | 6 | 1 | 44 | 0.515 | 0.369 | 0.689 | 0.910 | 0.703 | 0.286 | 0.647 | 0.185 | 0.564 | 101 | 55 | 23 |
| CNN-BiLSTM | 6 | 1 | 45 | 0.518 | 0.366 | 0.685 | 0.911 | 0.708 | 0.277 | 0.635 | 0.161 | 0.566 | 99 | 57 | 19 |
| CNN-BiLSTM | 6 | 1 | 46 | 0.495 | 0.351 | 0.713 | 0.918 | 0.725 | 0.340 | 0.628 | 0.169 | 0.557 | 98 | 58 | 20 |
| CNN-BiLSTM | 12 | 1 | 42 | 0.291 | 0.204 | 0.867 | 0.963 | 0.878 | 0.112 | 0.745 | 0.084 | 0.697 | 152 | 52 | 14 |
| CNN-BiLSTM | 12 | 1 | 43 | 0.293 | 0.203 | 0.865 | 0.962 | 0.874 | 0.101 | 0.770 | 0.092 | 0.714 | 157 | 47 | 16 |
| CNN-BiLSTM | 12 | 1 | 44 | 0.290 | 0.206 | 0.868 | 0.964 | 0.884 | 0.120 | 0.760 | 0.066 | 0.721 | 155 | 49 | 11 |
| CNN-BiLSTM | 12 | 1 | 45 | 0.278 | 0.195 | 0.878 | 0.966 | 0.885 | 0.191 | 0.779 | 0.076 | 0.733 | 159 | 45 | 13 |
| CNN-BiLSTM | 12 | 1 | 46 | 0.305 | 0.213 | 0.854 | 0.960 | 0.874 | 0.025 | 0.750 | 0.067 | 0.712 | 153 | 51 | 11 |
| CEEMDAN-BiLSTM | 3 | 1 | 42 | 0.680 | 0.547 | 0.523 | 0.845 | 0.534 | 0.303 | 0.412 | 0.395 | 0.325 | 49 | 70 | 32 |
| CEEMDAN-BiLSTM | 3 | 1 | 43 | 0.738 | 0.585 | 0.438 | 0.813 | 0.458 | 0.179 | 0.353 | 0.432 | 0.278 | 42 | 77 | 32 |
| CEEMDAN-BiLSTM | 3 | 1 | 44 | 0.685 | 0.553 | 0.515 | 0.841 | 0.527 | 0.293 | 0.420 | 0.405 | 0.327 | 50 | 69 | 34 |
| CEEMDAN-BiLSTM | 3 | 1 | 45 | 0.664 | 0.537 | 0.545 | 0.845 | 0.548 | 0.336 | 0.387 | 0.361 | 0.317 | 46 | 73 | 26 |
| CEEMDAN-BiLSTM | 3 | 1 | 46 | 0.686 | 0.551 | 0.514 | 0.840 | 0.525 | 0.291 | 0.412 | 0.402 | 0.322 | 49 | 70 | 33 |
| CEEMDAN-BiLSTM | 6 | 1 | 42 | 0.569 | 0.404 | 0.620 | 0.889 | 0.642 | 0.128 | 0.622 | 0.118 | 0.574 | 97 | 59 | 13 |
| CEEMDAN-BiLSTM | 6 | 1 | 43 | 0.599 | 0.443 | 0.579 | 0.881 | 0.613 | 0.033 | 0.654 | 0.227 | 0.548 | 102 | 54 | 30 |
| CEEMDAN-BiLSTM | 6 | 1 | 44 | 0.551 | 0.412 | 0.644 | 0.899 | 0.666 | 0.183 | 0.667 | 0.212 | 0.565 | 104 | 52 | 28 |

| Model | SPEI | Lead | Seed | RMSE | MAE | NSE | WI | $R^2$ | Skill | POD | FAR | CSI | Hits | Misses | False alarms |
|---|---|---|---|---|---|---|---|---|---|---|---|---|---|---|---|
| CEEMDAN-BiLSTM | 6 | 1 | 45 | 0.564 | 0.413 | 0.627 | 0.892 | 0.645 | 0.144 | 0.667 | 0.175 | 0.584 | 104 | 52 | 22 |
| CEEMDAN-BiLSTM | 6 | 1 | 46 | 0.588 | 0.420 | 0.595 | 0.889 | 0.635 | 0.069 | 0.660 | 0.202 | 0.566 | 103 | 53 | 26 |
| CEEMDAN-BiLSTM | 12 | 1 | 42 | 0.315 | 0.211 | 0.845 | 0.959 | 0.857 | −0.034 | 0.770 | 0.082 | 0.720 | 157 | 47 | 14 |
| CEEMDAN-BiLSTM | 12 | 1 | 43 | 0.332 | 0.226 | 0.827 | 0.955 | 0.848 | −0.154 | 0.809 | 0.083 | 0.753 | 165 | 39 | 15 |
| CEEMDAN-BiLSTM | 12 | 1 | 44 | 0.296 | 0.207 | 0.863 | 0.963 | 0.867 | 0.087 | 0.804 | 0.109 | 0.732 | 164 | 40 | 20 |
| CEEMDAN-BiLSTM | 12 | 1 | 45 | 0.301 | 0.215 | 0.858 | 0.962 | 0.872 | 0.053 | 0.755 | 0.089 | 0.703 | 154 | 50 | 15 |
| CEEMDAN-BiLSTM | 12 | 1 | 46 | 0.328 | 0.224 | 0.831 | 0.955 | 0.846 | −0.122 | 0.779 | 0.065 | 0.740 | 159 | 45 | 11 |

FAR is undefined (n/a) for a run that issues no drought forecast. The five runs of the reference-period test are reported in Table S18.

**Table S15.** Dispersion and increment diagnostics of the lead-1 forecasts over the test period. Model values are the mean ± SD over five seeds.

| SPEI | Model | SD observed | SD forecast | SD ratio | r (forecast, observed) | Slope of forecast on observed | r (forecast, persistence) | r (Δ observed, Δ forecast) | Δ amplitude ratio |
|---|---|---|---|---|---|---|---|---|---|
| 3 | XGBoost | 0.985 | 0.679 ± 0.012 | 0.689 ± 0.013 | 0.759 ± 0.004 | 0.523 ± 0.007 | 0.846 ± 0.011 | 0.614 ± 0.004 | 0.671 ± 0.015 |
| 3 | CNN | 0.985 | 0.746 ± 0.028 | 0.757 ± 0.029 | 0.706 ± 0.014 | 0.535 ± 0.025 | 0.844 ± 0.041 | 0.531 ± 0.019 | 0.655 ± 0.069 |
| 3 | Bi-LSTM | 0.985 | 0.787 ± 0.012 | 0.799 ± 0.012 | 0.760 ± 0.003 | 0.607 ± 0.007 | 0.888 ± 0.028 | 0.622 ± 0.006 | 0.563 ± 0.059 |
| **3** | **CNN-BiLSTM** | **0.985** | **0.764 ± 0.009** | **0.775 ± 0.010** | **0.772 ± 0.007** | **0.598 ± 0.009** | **0.883 ± 0.021** | **0.644 ± 0.016** | **0.581 ± 0.044** |
| 3 | CEEMDAN-BiLSTM | 0.985 | 0.808 ± 0.014 | 0.820 ± 0.014 | 0.720 ± 0.025 | 0.590 ± 0.022 | 0.848 ± 0.060 | 0.550 ± 0.029 | 0.634 ± 0.110 |
| 3 | Persistence | 0.985 | n/a | 1.000 | 0.657 | n/a | n/a | n/a | n/a |
| 6 | XGBoost | 0.924 | 0.764 ± 0.009 | 0.826 ± 0.010 | 0.797 ± 0.005 | 0.658 ± 0.010 | 0.956 ± 0.006 | 0.403 ± 0.015 | 0.482 ± 0.021 |
| 6 | CNN | 0.924 | 0.808 ± 0.013 | 0.874 ± 0.014 | 0.843 ± 0.014 | 0.737 ± 0.021 | 0.925 ± 0.013 | 0.579 ± 0.050 | 0.574 ± 0.047 |
| **6** | **Bi-LSTM** | **0.924** | **0.870 ± 0.016** | **0.941 ± 0.018** | **0.854 ± 0.011** | **0.804 ± 0.017** | **0.930 ± 0.008** | **0.599 ± 0.042** | **0.556 ± 0.030** |
| 6 | CNN-BiLSTM | 0.924 | 0.840 ± 0.019 | 0.908 ± 0.020 | 0.838 ± 0.010 | 0.761 ± 0.024 | 0.913 ± 0.006 | 0.552 ± 0.031 | 0.615 ± 0.022 |
| 6 | CEEMDAN-BiLSTM | 0.924 | 0.864 ± 0.024 | 0.934 ± 0.026 | 0.800 ± 0.012 | 0.747 ± 0.021 | 0.936 ± 0.030 | 0.380 ± 0.093 | 0.515 ± 0.148 |
| 6 | Persistence | 0.924 | n/a | 0.993 | 0.781 | n/a | n/a | n/a | n/a |
| 12 | XGBoost | 0.799 | 0.725 ± 0.001 | 0.909 ± 0.002 | 0.927 ± 0.003 | 0.842 ± 0.002 | 0.977 ± 0.003 | 0.337 ± 0.018 | 0.574 ± 0.032 |
| **12** | **CNN** | **0.799** | **0.750 ± 0.014** | **0.939 ± 0.018** | **0.935 ± 0.003** | **0.878 ± 0.016** | **0.971 ± 0.006** | **0.435 ± 0.053** | **0.618 ± 0.065** |
| 12 | Bi-LSTM | 0.799 | 0.754 ± 0.009 | 0.944 ± 0.012 | 0.937 ± 0.008 | 0.885 ± 0.015 | 0.954 ± 0.009 | 0.512 ± 0.041 | 0.769 ± 0.072 |
| 12 | CNN-BiLSTM | 0.799 | 0.735 ± 0.007 | 0.920 ± 0.008 | 0.938 ± 0.003 | 0.862 ± 0.009 | 0.957 ± 0.004 | 0.502 ± 0.015 | 0.749 ± 0.033 |
| 12 | CEEMDAN-BiLSTM | 0.799 | 0.769 ± 0.014 | 0.963 ± 0.017 | 0.926 ± 0.006 | 0.892 ± 0.013 | 0.940 ± 0.028 | 0.470 ± 0.048 | 0.857 ± 0.242 |
| 12 | Persistence | 0.799 | n/a | 0.995 | 0.925 | n/a | n/a | n/a | n/a |

SD is the sample standard deviation over the 576 test pairs. Δ is the change from the last observed month (value minus persistence), and the Δ amplitude ratio is the SD of the forecast change divided by the SD of the observed change. Bold rows mark the selected models. For persistence, the SD ratio and correlation are those plotted in the Taylor diagrams of Fig. 9.

**Table S16.** Spatial verification of the selected models at lead 1 in the worst observed drought month and in a near-normal month of the test period, summarized over five seeds.

| SPEI (model) | Case | Month | Observed national mean | Forecast national mean | Forecast / observed | Spatial r | Spatial SD ratio | Driest division, observed | Driest division, forecast (seeds) | Sign reversals (seeds) |
|---|---|---|---|---|---|---|---|---|---|---|
| SPEI-3 (CNN-BiLSTM) | Worst drought | Aug 2022 | −1.899 | −1.149 | 0.60 | 0.932 ± 0.007 | 1.36 | Chittagong | Khulna 4/5, Chittagong 1/5 | None |
| SPEI-6 (Bi-LSTM) | Worst drought | Aug 2022 | −1.718 | −1.243 | 0.72 | 0.912 ± 0.016 | 1.44 | Chittagong | Khulna 4/5, Chittagong 1/5 | Rangpur 4/5 |
| SPEI-12 (CNN) | Worst drought | Jun 2023 | −1.821 | −1.365 | 0.75 | 0.854 ± 0.054 | 1.48 | Sylhet | Sylhet 4/5, Chittagong 1/5 | None |
| SPEI-3 (CNN-BiLSTM) | Near-normal | Dec 2022 | +0.013 | +0.050 | n/a | 0.970 ± 0.007 | 0.84 | Chittagong | Chittagong 5/5 | Dhaka 2/5 |
| SPEI-6 (Bi-LSTM) | Near-normal | Dec 2019 | −0.003 | +0.004 | n/a | 0.932 ± 0.027 | 0.86 | Dhaka | Barisal 3/5, Chittagong 2/5 | Mymensingh 4/5, Rajshahi 4/5, Khulna 3/5 |
| SPEI-12 (CNN) | Near-normal | May 2022 | −0.039 | +0.019 | n/a | 0.927 ± 0.029 | 0.90 | Sylhet | Sylhet 5/5 | Chittagong 4/5, Rajshahi 4/5 |

Forecast national means, spatial correlations and spatial SD ratios are seed means (spatial r ± SD). The months are those with the most negative and the nearest-to-zero observed national mean. In the worst month the observed driest division is a near-tie at SPEI-3 (Chittagong −2.32, Khulna −2.31) and SPEI-6 (Chittagong −2.14, Dhaka −2.11); at SPEI-12 it is Sylhet (−2.17). The only sign reversal in a worst month is Rangpur at SPEI-6, which is observed at −0.71 and forecast at +0.23 by the seed shown in Fig. 14e of the main text. Fig. 14 of the main text and Fig. S3 show the median-skill seeds (44, 42 and 44).

**Table S17.** SHAP values of the selected models at lead 1 over the test windows, as the mean ± SD over five seeds.

| SPEI (model) | Predictor | Mean \|SHAP\| | Mean signed SHAP | Share of meteorological SHAP (%) | Mean rank | Seeds with positive signed SHAP |
|---|---|---|---|---|---|---|
| SPEI-3 (CNN-BiLSTM) | Total precipitation | 0.176 ± 0.013 | 0.015 ± 0.005 | 37.7 ± 1.6 | 1.4 | 5/5 |
| | Lagged SPEI | 0.173 ± 0.017 | 0.010 ± 0.003 | n/a | 1.6 | 5/5 |
| | Solar radiation | 0.084 ± 0.012 | 0.009 ± 0.005 | 18.1 ± 2.6 | 3.0 | 4/5 |
| | Minimum temperature | 0.046 ± 0.012 | 0.012 ± 0.011 | 9.7 ± 1.8 | 5.4 | 4/5 |
| | Maximum temperature | 0.044 ± 0.009 | 0.005 ± 0.008 | 9.3 ± 1.3 | 5.6 | 3/5 |
| | Wind speed | 0.043 ± 0.007 | −0.011 ± 0.006 | 9.2 ± 0.9 | 4.8 | 0/5 |
| | Dew-point temperature | 0.040 ± 0.009 | −0.008 ± 0.013 | 8.6 ± 1.4 | 6.8 | 1/5 |
| | Total cloud cover | 0.035 ± 0.006 | −0.002 ± 0.005 | 7.4 ± 0.9 | 7.4 | 1/5 |
| SPEI-6 (Bi-LSTM) | Total precipitation | 0.114 ± 0.008 | 0.045 ± 0.005 | 31.9 ± 1.8 | 1.0 | 5/5 |
| | Lagged SPEI | 0.070 ± 0.009 | 0.041 ± 0.007 | n/a | 2.6 | 5/5 |
| | Solar radiation | 0.069 ± 0.003 | 0.021 ± 0.004 | 19.3 ± 1.2 | 2.4 | 5/5 |
| | Maximum temperature | 0.054 ± 0.008 | 0.000 ± 0.006 | 15.1 ± 1.8 | 4.0 | 3/5 |
| | Total cloud cover | 0.036 ± 0.004 | −0.001 ± 0.002 | 10.1 ± 0.7 | 5.4 | 2/5 |
| | Dew-point temperature | 0.035 ± 0.003 | −0.008 ± 0.012 | 9.7 ± 0.8 | 5.8 | 1/5 |
| | Minimum temperature | 0.029 ± 0.005 | −0.002 ± 0.004 | 8.0 ± 1.0 | 6.8 | 3/5 |
| | Wind speed | 0.022 ± 0.004 | 0.002 ± 0.007 | 6.0 ± 0.7 | 8.0 | 3/5 |
| SPEI-12 (CNN) | Lagged SPEI | 0.128 ± 0.043 | 0.107 ± 0.036 | n/a | 1.0 | 5/5 |
| | Total precipitation | 0.066 ± 0.013 | 0.018 ± 0.013 | 22.6 ± 3.3 | 2.2 | 5/5 |
| | Solar radiation | 0.049 ± 0.013 | 0.024 ± 0.013 | 16.7 ± 3.3 | 4.0 | 5/5 |
| | Maximum temperature | 0.048 ± 0.012 | −0.033 ± 0.013 | 16.7 ± 3.8 | 3.6 | 0/5 |
| | Dew-point temperature | 0.037 ± 0.014 | 0.014 ± 0.033 | 13.0 ± 5.6 | 5.2 | 3/5 |
| | Wind speed | 0.033 ± 0.007 | −0.017 ± 0.004 | 11.5 ± 1.8 | 6.4 | 0/5 |
| | Minimum temperature | 0.028 ± 0.006 | 0.008 ± 0.015 | 9.8 ± 1.8 | 6.8 | 2/5 |
| | Total cloud cover | 0.028 ± 0.006 | −0.017 ± 0.008 | 9.8 ± 2.3 | 6.8 | 0/5 |

SHAP values were computed with the gradient-based explainer (background of 150 training windows) and summed over the time steps of each predictor. The share excludes lagged SPEI. A positive signed value raises the predicted SPEI (wetter). The sign is interpreted only where all five seeds agree (0/5 or 5/5), and only the top three ranks are stable across seeds.

**Table S18.** Sensitivity to the reference period of the SPEI fit (Text S6): agreement between the main index and the index refitted on 1980-2018 over the test period (A), and lead-1 scores of the Bi-LSTM at SPEI-6 in the main and strict pipelines (B).

**(A) Agreement of the refitted index, 2019-2024**

| Timescale | n | r | Mean shift (refit − main) | Largest absolute difference | Drought months, main | Drought months, refit | Agreement (%) | Cohen's κ |
|---|---|---|---|---|---|---|---|---|
| SPEI-1 | 576 | 0.993 | −0.017 | 1.893 | 97 | 104 | 98.4 | 0.95 |
| SPEI-3 | 576 | 0.997 | −0.022 | 0.494 | 119 | 135 | 97.2 | 0.92 |
| SPEI-6 | 576 | 0.996 | −0.060 | 0.413 | 156 | 168 | 97.2 | 0.93 |
| SPEI-12 | 576 | 0.994 | −0.110 | 0.476 | 204 | 236 | 94.4 | 0.88 |

**(B) Bi-LSTM, SPEI-6, lead 1**

| Pipeline | Seeds | Skill | NSE | RMSE | CSI | POD |
|---|---|---|---|---|---|---|
| Main (fit 1980-2024) | 5 | 0.350 ± 0.045 | 0.717 ± 0.020 | 0.491 ± 0.017 | 0.615 ± 0.023 | 0.714 ± 0.020 |
| Strict (fit 1980-2018) | 5 | 0.354 ± 0.059 | 0.715 ± 0.026 | 0.516 ± 0.023 | 0.648 ± 0.022 | 0.751 ± 0.032 |

(B) Model values are the mean ± SD over seeds 42-46; each pipeline is scored against its own index. The decision rule, fixed before the run, kept the main pipeline if the mean skill changed by less than one seed SD of the main runs (0.045); the change is 0.004.